\RequirePackage{fix-cm}
\documentclass[epjc3]{svjour3} \usepackage{amsmath,amssymb}

\usepackage{upgreek}
\usepackage{subfigure}
\usepackage{multirow}
\usepackage{microtype}
\usepackage{color}

\usepackage[english]{babel}
\usepackage[utf8x]{inputenc}
\usepackage[T1]{fontenc}

\usepackage{amsmath,amssymb} 
\usepackage{graphicx}
\usepackage{upgreek}
\usepackage{multirow}
\usepackage{subfigure}
\usepackage{epstopdf}
\usepackage[colorlinks=true, allcolors=red]{hyperref}
\usepackage{verbatim}
\usepackage{tikz}
\usepackage{lineno}
\usetikzlibrary{shapes}
\usetikzlibrary{arrows}
\usetikzlibrary{calc,positioning}
\usetikzlibrary{decorations.markings}

\begin{document}

\title{The Double Chooz antineutrino detectors}

\authorrunning{The Double Chooz Collaboration}

\author{H.~de~Kerret\thanksref{e1,addr3,addr38} \and Y.~Abe\thanksref{addr32} \and C.~Aberle\thanksref{addr23} \and T.~Abrah\~{a}o\thanksref{addr6,addr3} \and J.M.~Ahijado\thanksref{addr9} \and T.~Akiri\thanksref{addr3,addr5} \and J.M.~Alarc\'on\thanksref{addr9} \and J.~Alba\thanksref{addr9} \and H.~Almazan\thanksref{addr23} \and J.C.~dos Anjos\thanksref{addr6}
% ...
} 

\thankstext{e1}{deceased in 2017}

\institute{
% ...
APC, CNRS/IN2P3, CEA/IRFU, Observatoire de Paris, Sorbonne Paris Cit\'{e} Universit\'{e}, 75205 Paris Cedex 13, France \label{addr3}
% ...
\and
DRF/IRFU, CEA, Universit\'{e} Paris-Saclay, 91191 Gif-sur-Yvette, France \label{addr5}\and
Centro Brasileiro de Pesquisas F\'{i}sicas, Rio de Janeiro, RJ, 22290-180, Brazil \label{addr6}
% ...
\and
Centro de Investigaciones Energ\'{e}ticas, Medioambientales y Tecnol\'{o}gicas, CIEMAT, 28040, Madrid, Spain \label{addr9}
% ...
\and
Max-Planck-Institut f\"{u}r Kernphysik, 69117 Heidelberg, Germany \label{addr23}
% ...
\and
Department of Physics, Tokyo Institute of Technology, Tokyo, 152-8551, Japan \label{addr32}
% ...
\and
LNCA Underground Laboratory, IN2P3/CNRS - CEA, Chooz, France \label{addr38}
% ...
}

\maketitle

\newpage

\begin{abstract}
This article describes the setup and performance of the near and far detectors in the Double Chooz experiment. The electron antineutrinos of the Chooz nuclear power plant were measured in two identically designed detectors with different average baselines of about 400~m and 1050~m from the two reactor cores. Over many years of data taking the neutrino signals were extracted from interactions in the detectors with the goal of measuring a fundamental parameter in the context of neutrino oscillation, the mixing angle $\theta_{13}$. The central part of the Double Chooz detectors was a main detector comprising four cylindrical volumes filled with organic liquids. From the inside towards the outside there were volumes containing gadolinium-loaded scintillator,  gadolinium-free scintillator, a buffer oil and, optically separated, another liquid scintillator acting as veto system. Above this main detector an additional outer veto system using plastic scintillator strips was installed. The technologies developed in Double Chooz were inspiration for several other antineutrino detectors in the field. The detector design allowed implementation of efficient background rejection techniques including use of pulse shape information provided by the data acquisition system. The Double Chooz detectors featured remarkable stability, in particular for the detected photons, as well as high radiopurity of the detector components.
\end{abstract}

\tableofcontents

\section{Introduction}

The Double Chooz (DC) experiment was built to study neutrino oscillation properties in the vicinity of nuclear reactors~\cite{Ardellier:2004ui}\cite{Ardellier:2006mn}. In particular it was designed to search for the smallest mixing angle in the three neutrino framework, known as $\theta_{13}$. The knowledge of this fundamental parameter in particle physics is critical for the determination of other oscillation parameters such as the CP violating phase~\cite{T2K:2019bcf}. Among all km-baseline reactor neutrino experiments, DC provided the first indication for a nonzero value of $\theta_{13}$~\cite{Abe:2011fz} in 2011. Today $\theta_{13}$ is determined with the highest precision of all mixing angles by the three reactor neutrino experiments Daya Bay~\cite{Adey:2018zwh}, DC~\cite{DC_Nature} and RENO~\cite{Seo:2016uom}.   

The DC concept is to have two identical detectors at different distances from the reactor cores. A nuclear reactor is a localized, pure source of electron antineutrinos which have less than 10\,MeV energy. Whereas the near detector monitors the reactor neutrino flux almost without any oscillation effect, the second, far detector is placed close to the first oscillation minimum. In this way the disappearance of the electron antineutrinos in the far detector is observed in order to extract $\theta_{13}$. 

Both detectors were in underground laboratories with shielding against cosmic radiation of 120~m\,w.e.\ in the near and 300~m\,w.e.\ in the far detector. The neutrinos were detected via inverse beta decay (IBD) on free protons in organic liquid scintillator. 
Since the far detector could be installed in the existing laboratory of the original CHOOZ experiment~\cite{Apollonio:2002gd}, its data taking started first, in April 2011. After construction of the new near laboratory and detector, data taking with the second detector started at the end of 2014. Both detectors ran in parallel for about three years, until the end of 2017, when neutrino data taking was stopped. With the near (far) detector about 800 (100) neutrinos could be detected per day in the active detector volume at a signal to noise ratio of more than 20 (10).

The most recent published result on $\theta_{13}$ from DC is sin$^22\theta_{13}=0.105\pm0.014$~\cite{DC_Nature}. Besides the $\theta_{13}$ measurement, DC found in 2014 distortions in the measured neutrino spectrum with respect to the prediction, in particular around 5~MeV energy~\cite{Abe:2014bwa}, subsequently corroborated by Daya Bay~\cite{DayaBay:2015lja} and RENO~\cite{RENO:2015ksa}. Moreover, world leading results were obtained for the mean cross section per fission~\cite{DC_Nature} and unique background studies were done, including periods with both reactors turned off~\cite{Abrahao:2020ztg}. 

This article is structured as follows: In section 2, a general detector description is provided. Section 3 includes a characterization of the liquids used in the DC detector. The proton number estimation of the target liquid is detailed in section 4. In section 5 the gas and liquid system used for detector filling are discussed. The light produced in the liquid scintillator was observed by photomultiplier tubes, which are the topic of section 6. Section 7 reports on the electronics and the data acquisition system. The Geant4 based DC Monte Carlo simulation is reviewed in section 8. The manifold calibration strategy of the experiment is illustrated in section 9. Section 10 highlights results on radiopurity investigations and background studies. In section 11 the stability of the detector response over 7 years of data taking is discussed and a final summary is given in section 12.  

\section{Detector description}

\subsection{Underground laboratories and detector overview}

\begin{figure}
    \centering
    \includegraphics[width=0.5\textwidth]{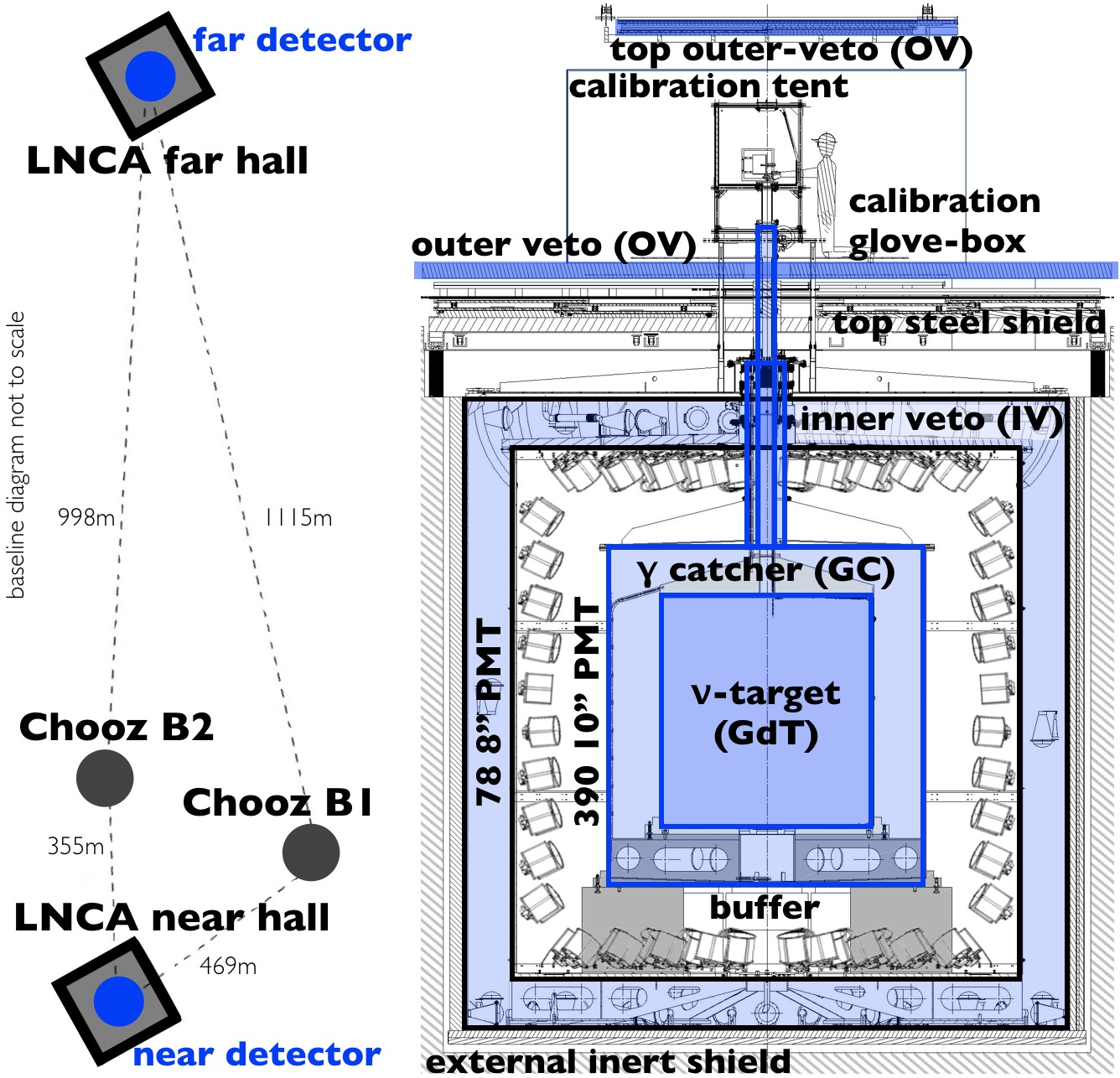}
    \caption{
    Arrangement of the two DC detectors with respect to the nuclear reactors (left) and sketch of the detector design (right)~\cite{DC_Nature}.}
    \label{fig:site}
\end{figure}

Antineutrino detection in DC was based on two almost identical detectors located at the Chooz-B nuclear power plant operated by the EDF company~\cite{EDF}. The far detector was positioned at an average distance of 1.05\,km from the two nuclear cores of the plant with a thermal power of up to 4.25\,GW each. The near detector was installed in a new underground laboratory at around 400\,m from the nuclear cores. One of the unique features of the DC site configuration was the effective isoflux position. This means that both detectors were exposed to  similar ratios of the antineutrino flux coming from the two reactor cores. Therefore, the systematic uncertainty in the neutrino flux calculation was strongly reduced in the ratio of measurements with far and near detectors. The detector positions with respect to the reactor cores and the main detector components are depicted in fig.~\ref{fig:site}. The location of the experiment at a nuclear power plant resulted in security issues regarding access to the labs and detectors. These issues did not compromise any of the physics capabilities of the detectors.

The far detector was built in an existing experimental hall, therefore there were geometrical constraints on the size and shape of the DC detector. The environmental conditions in the laboratory are a stable temperature around 15$^{\circ}$C and relative humidity close to 100\,\%. The access tunnel is about 150\,m long with a mean slope of 10\,\% from the main entrance until the lab entrance. Its shape is more or less that of a half cylinder with a radius around 3.8\,m setting constraints on the maximal size of pieces entering the lab.

At the near underground laboratory, which was newly constructed for the DC experiment, a 90\,m ramp (14\,\% slope) gives access to the 150\,m long tunnel (12\,\% slope). The near laboratory was constructed larger than the far laboratory to simplify handling of large components. It is about 12\,m wide and 29\,m long. Both detectors were installed inside cylindrical pits at the centers of the laboratories with the detector top lid below the ground floor. Both laboratories are equipped with dedicated ventilation systems and a crane (5\,t capacity in the far and 12\,t in the near laboratory) passing in one direction along the central axis. The muon flux in the laboratories is $(3.64 \pm 0.04)\cdot10^{-4}\,\text{cm}^{-2}$\,s$^{-1}$ in the near and $(7.00 \pm 0.05) \cdot 10^{-5}$\,cm$^{-2}$\,s$^{-1}$ in the far detector.~\cite{DoubleChooz:2016sdt}

The DC detector system can be separated into the following components: inner detector, inner veto, outer veto, shielding and calibration devices. In the inner detector, the rate and spectrum of the reactor antineutrino flux was measured. The inner detector consisted of three concentric cylindrical vessels with a central chimney at the top. These vessels were filled with liquid scintillators or mineral oil. The central part was a gadolinium-loaded liquid scintillator contained in an acrylic vessel with a volume of 10.3\,m$^3$ (Target). This volume was surrounded by the gamma catcher (GC), an unloaded liquid scintillator in a second acrylic vessel. The GC volume was 22.5\,m$^3$ and its original purpose was just to detect gamma rays escaping the Target for inclusive calorimetry. However, neutrino interactions in the GC can also be included in the neutrino analysis increasing the target volume of the detector by about a factor of 3~\cite{DC_Nature}. Outside the GC was the Buffer, more than 100\,m$^3$ of mineral oil acting as a shield against radioactivity from detector components and the surrounding rock. On the inner wall of the stainless steel Buffer tank, 390 photomultiplier tubes (PMTs) of 10-inch diameter were installed.

The inner detector was embedded in another cylindrical steel tank containing about 90\,m$^3$ of liquid scintillator, the inner veto. It was equipped with 78 PMTs (8-inch) and mainly used as a cosmic muon veto. In addition, it acted as a shield and detector for spallation neutrons and high energy gammas produced outside the detector. The detector was covered and surrounded by 15 cm of demagnetized steel (far detector) or 1~m of water (near detector) to further suppress external gamma-rays. Above the inner veto tank an outer veto system (OV) consisting of layers of plastic scintillator was arranged in strips so that both cosmic muon tagging and tracking were possible. 

\subsection{Inner detector}
\subsubsection{Double Chooz acrylic vessels} 
\begin{figure}
    \begin{center}
\subfigure{
	\includegraphics[width=0.3\textwidth]{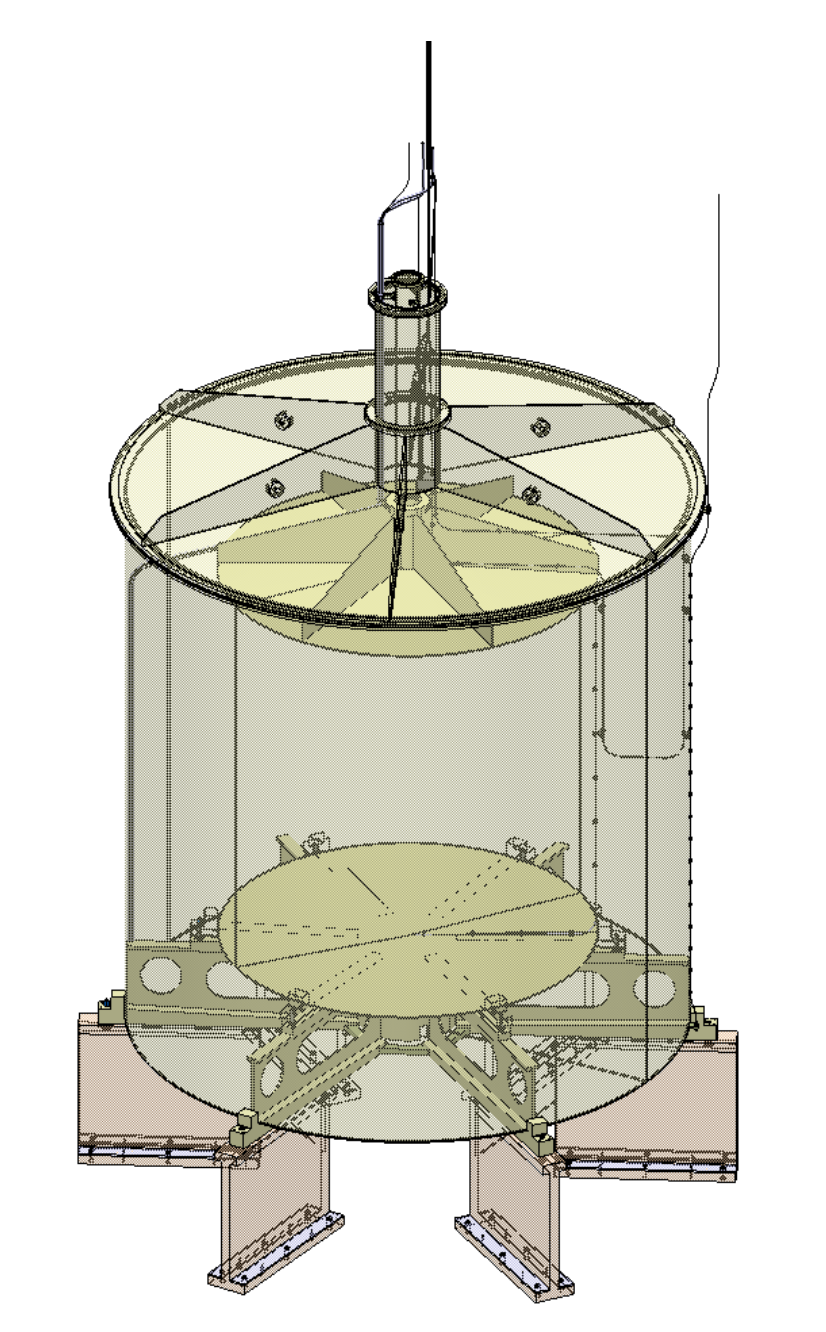}
	\label{Target_figa}}
\subfigure{
	\includegraphics[height=6.5cm]{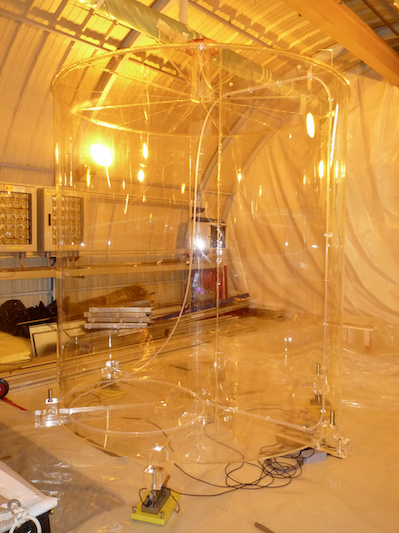}
	\label{Target_figb}}
\subfigure{
	\includegraphics[height=6.5cm]{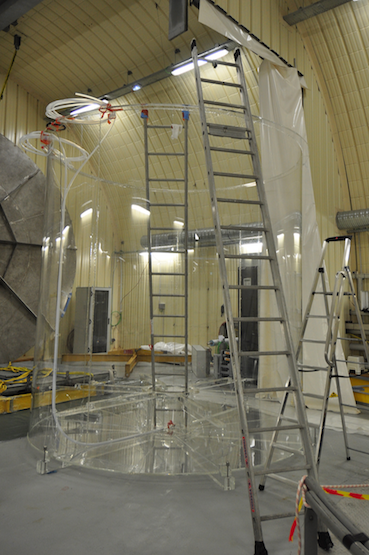}
	\label{Target_figc}}
\end{center}
    \caption{\label{fig:acrylics}
    View of the acrylic vessels (top). The 8\,mm thick Target vessel (bottom, left) contained  10.3\,m$^3$ of Gadolinium-doped scintillator. It was placed inside the 12\,mm thick Gamma Catcher vessel (bottom, right), filled with organic non-doped scintillator. The ensemble was supported by six massive acrylic legs.}
\end{figure}

The DC acrylic vessels were made of polymethyl methacrylate (PMMA). In partnership with the Degussa company, the collaboration designed a new UV-transparent PMMA material, called GSOZ18 (monomer C$_5$H$_8$O$_2$, 1.19 g$\cdot$cm$^{-3}$), maximizing the length of polymers in order to enhance the long-term resistance to organic liquid scintillators. For the support structure of the acrylic vessels, in particular the legs (see fig.~\ref{fig:acrylics}), the same transparent PMMA material was used. The free proton concentration of the GSOZ18 was measured to be $7.9\pm0.1$\,\% by weight~\cite{Queval:2010ala}. Taking into account the aging of the material in contact with organic liquid scintillators, possible defects, and glued joints, the maximum internal stress was limited to 5\,MPa, by design. Concerning backgrounds, the material was radiopure enough so that the trigger rate coming from the acrylic was negligible compared to the overall rate of single events (see section~\ref{radio_sec}). A single acrylic batch of 9\,tons of PMMA was used for the construction of three Target and two GC vessels. The radiopurity of this batch was controlled through gamma-ray spectroscopy and neutron activation analyses. The main contaminations were measured to be $\lesssim 10^{-10}$ g/g for $^{238}$U, $\lesssim 10^{-11}$ g/g for $^{232}$Th, and $\lesssim 10^{-12}$ g/g for $^{40}$K~\cite{Queval:2010ala}. The acrylic sheets were first thermoformed at the manufacture site and covered with a UV-protecting foil in order to prevent any degradation or crazing of the acrylic material. The assembly of the vessels was done by gluing together acrylic sheets, using Acryfix 190 glue. About 5 kg of glue was necessary for a single Target vessel assembly, and about 20 kg for a single GC vessel assembly. The glue radioactive contamination was measured to be negligible in comparison to the bulk materials representing $>$98\% of the vessel masses~\cite{Queval:2010ala}.

A delicate part of each acrylic vessel was the central chimney attached to the top. The chimney of the Target vessel had to pass through all the other detector volumes and connected the innermost liquid volume to a glove box (see section~\ref{glovebox}) containing calibration devices. Along these chimneys there were several pipes, cables for sensors and connections to external expansion tanks (see section~\ref{sec:LH}). There were challenging constraints related to mechanical stability, material compatibility and tightness. The chimney parts of the inner three volumes had to be machined with very high precision to assure they matched the dimensions of all the other detector parts and to minimize stresses.

A critical step in the detector construction was the integration of the acrylic vessels, in particular at the smaller far laboratory. After transport the rather fragile GC vessel had to be swivelled into a horizontal position to enter the tunnel and the laboratory. Then it was positioned above the pit and swivelled again to the vertical position using dedicated tools. After lowering the GC down the Buffer vessel it was fixed at the bottom. The more compact Target acrylic vessel was fixed inside the GC and finally the GC top lid glued to the vessel body.

Cleanliness was a major concern during the construction and integration of the acrylic vessels. To mitigate the encapsulation of non-radiopure materials, the vessel gluing was carried out inside an ISO~7 clean room at the manufacture site, and the final vessel assemblies and integration were performed inside an ISO~6 clean room built around the DC detector pits. After manufacturing and integration, stress measurements were performed using photoelasticimetry~\cite{Queval:2010ala}. It was noticed that water trapped within the acrylics might degrade the stability of the Gd-doped liquid scintillator in the long term. Therefore, radon-free dehydrating bags and nitrogen flushing were intensively used prior to the filling of the detectors. In order to fill the vessels, an ensemble of perfluoroalkoxy alkanes (PFA) tubes were glued onto custom-made GSOZ18 holders in both the Target (1/2\,inch) and GC vessels (3/4\,inch). Both Target and GC vessels are displayed in fig.~\ref{fig:acrylics}. 

\paragraph{Target acrylic vessel} 
The Target acrylic vessel was the innermost mechanical part of the detector. The vessel was a cylinder of 2300\,mm diameter and 2458\,mm height. Its thickness was limited to 8\,mm in order to minimize dead material within the active volume that could induce a spectral distortion of the measured prompt IBD spectrum. The Target vessel was filled with 10.3\,m$^3$ of Gd-loaded liquid scintillator (see section~\ref{sec:liquids}), which originally constituted the Neutrino Target, where neutron captures on gadolinium were detected. Three acrylic containers were manufactured simultaneously, using the same construction methods and templates. After a complete mechanical characterization (dimension and weight measurements), the two most similar vessels were used for integration of the detectors at the far and near sites, in 2010 and 2014 respectively. The volume difference of the two final vessels was less than 0.2\%~\cite{Queval:2010ala}. Prior to their integration in the DC detector pit, the Target vessels were annealed in a large oven, in a carefully planned temperature sequence ranging from 20 to 80$^\circ$~\cite{Queval:2010ala}. Moreover, a helium leak test was performed.  

\paragraph{Gamma Catcher acrylic vessel} 
The GC vessel was a 12\,mm thick acrylic cylinder of 3416\,mm diameter and 3572\,mm height. It surrounded the Target, providing a 55\,cm thick layer of undoped liquid scintillator (22.5\,m$^3$), originally designed to contain the gamma rays emitted from the positron annihilation and neutron capture in the Target. The detection of these energy depositions limited edge effects at the Target border, thus increasing the detection efficiency and reducing its associated uncertainty. The undoped GC liquid scintillator also served as a detection volume when performing the antineutrino analysis using neutron capture on hydrogen. After thermoforming of the acrylic sheets at the manufacture site, the GC vessels were glued and assembled at the Chooz nuclear power station, under an ISO~6 clean room environment~\cite{Queval:2010ala}.

\subsubsection{Buffer vessel} 
The DC Buffer region was a major upgrade compared to the inner detector of the former CHOOZ  experiment, which consisted only of a Target and a GC volume. The Buffer was filled with about 110\,m$^3$ of non-scintillating mineral oil. This region served multiple purposes. First, it supported the 390 low-background PMTs, which detected the light emitted upon energy depositions in the Target and the GC. Second, this 105-cm thick layer of transparent oil served to shield the Target and GC from external (e.g.,~fast neutrons) and PMT radioactivity, the latter mostly from $^{40}$K in the PMT glass. Third, the Buffer volume enabled a more uniform detector response for energy depositions across the full scintillator volumes.

\begin{figure}
\label{fig:buffer}
    \begin{center}
\subfigure{
	\includegraphics[height=7 cm]{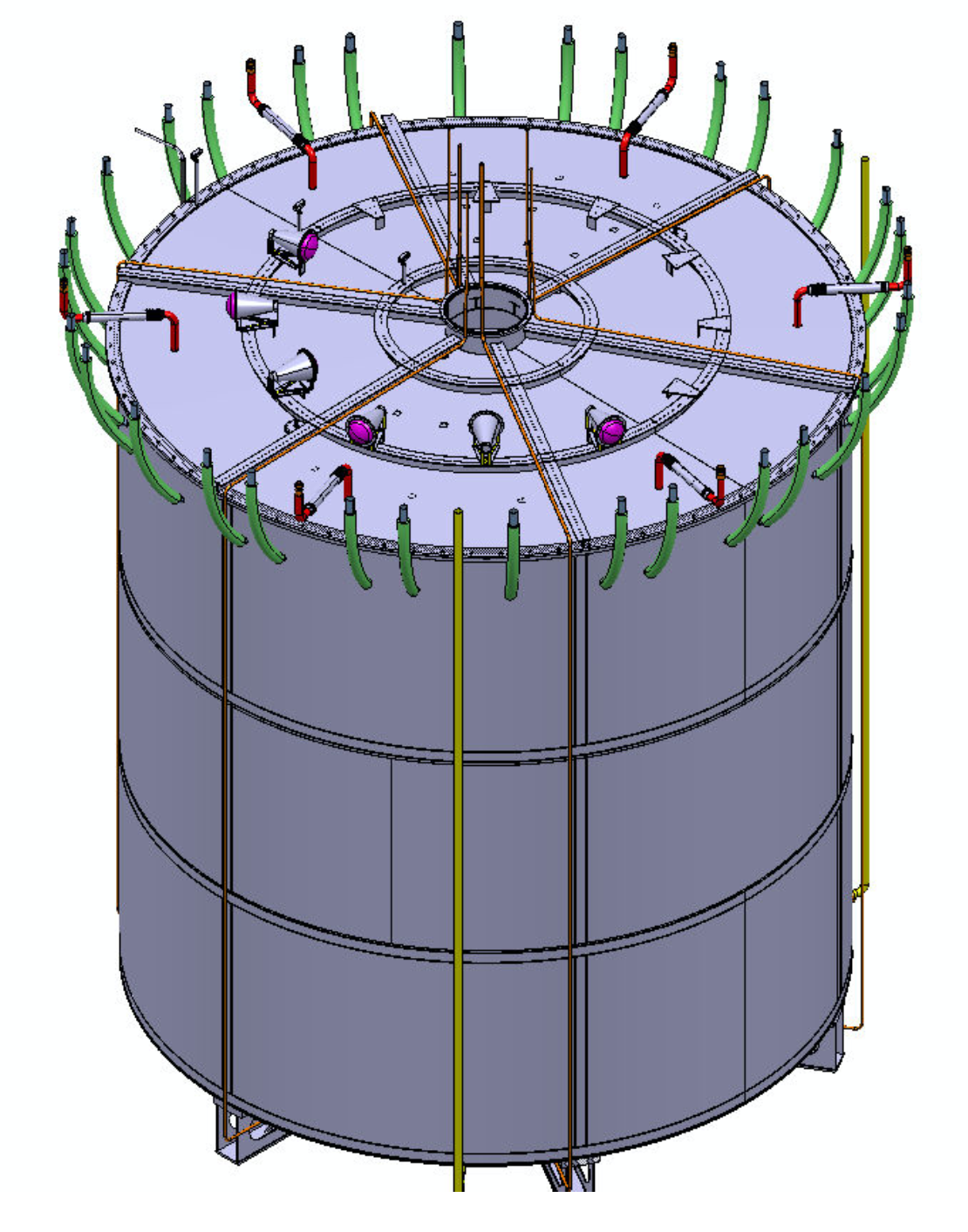}
	\label{fig:buffera}}
\subfigure{
	\includegraphics[height=7 cm]{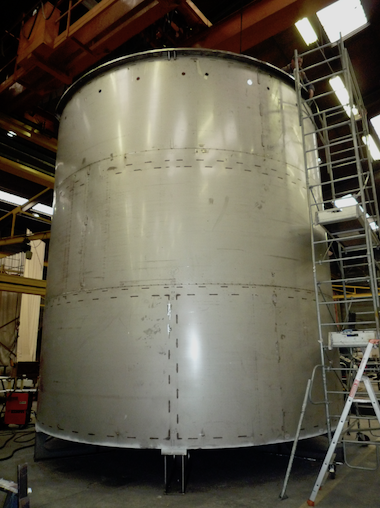}
	\label{fig:bufferb}}
\end{center}
\caption{Technical drawing (top) and photograph (bottom) of the Double Chooz Buffer vessel, composed of a frame made of stiffening profiles (1565 kg) on which a "skin" made of 3 mm thick steel sheets (4500 kg) was fixed. It rested on 6 feet (3525 kg) and supported the 390 photomultiplier tubes, drain and expansion tubes.}
\end{figure}

The Buffer vessel was the outermost cylinder of the inner detector, surrounding the GC, with a diameter of 5516\,mm and a height of 5675\,mm (see fig.~\ref{fig:buffera} and fig.~\ref{fig:bufferb}). The vessel was made of electropolished 304L and 316L stainless steel. It was mainly composed of a reinforcement consisting of thick beams (1565 kg) to which a "skin" made of 3\,mm thick steel sheets (4500 kg) was attached. It rested on six legs (3525 kg) and supported the PMTs, as well as the drainage and expansion tubes. The reflectivity coefficient at the surface of the Buffer vessel was about 40\,\% \cite{Mueller:2010mja}.

The materials required to build the Buffer vessels were identical for both detectors. The number of steel castings was minimized and the material was tested prior to acquisition. The materials were selected so that each Buffer vessel contributed a count rate less than 1\,Hz in the inner scintillator volumes, for energy depositions of more than 0.7\,MeV. For the most common radioactive isotopes, the concentrations were found to be $\lesssim 10^{-10}$ g/g for $^{238}$U, $\lesssim 10^{-9}$\,g/g for $^{232}$Th, $\lesssim 10^{-7}$\,g/g for $^{40}$K, and $\lesssim 15$\,mBq/kg for $^{60}$Co~\cite{Mueller:2010mja}. 

Due to the size of the DC laboratories, the Buffer vessel was produced in the form of eight subparts: six half-cylinders corresponding to the lower, middle and upper parts of the enclosure and two half-lids. Each sub-part was cleaned after manufacture and prepared for transport to the experiment site according to the following protocol: cleaning with mild soap and water, rinsing with deionized water, cleaning  with isopropyl alcohol. In addition, to limit the risk of corrosion and reduce the quantity of impurities or foreign bodies, each sub-part underwent a pickling and then passivation process~\cite{Mueller:2010mja}.

The manufacture of the Buffer vessels and their integration on site were carried out in a clean environment, reaching the class ISO 7 (standard ISO 14644-1), in order to guarantee that dust contamination in the enclosure not induce an additional background rate of more than 10\%, nor affect the optical transparency properties of the buffer mineral oil \cite{Mueller:2010mja}.

\subsection{Inner Veto}

\begin{figure}[h]
\label{fig:dc_pmt_layout}
    \begin{center}
\subfigure{
	\includegraphics[height=6.5 cm]{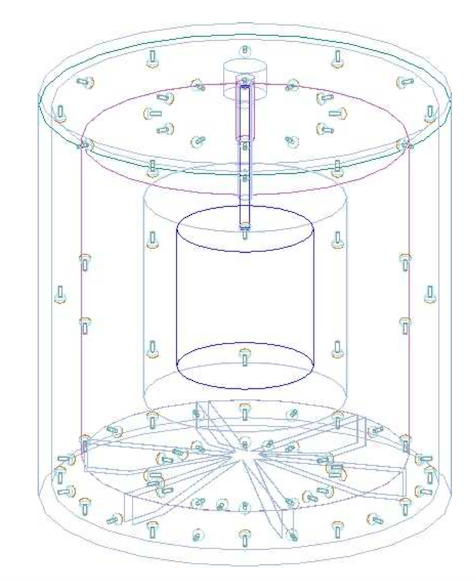}}
\subfigure{
	\includegraphics[height=6.5 cm]{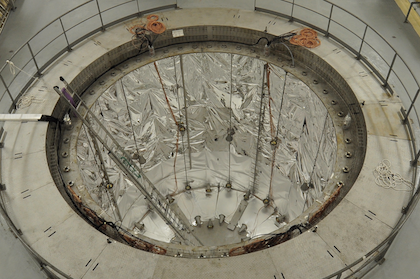}}
\end{center}
    \caption{\label{fig:veto}
    Schematic view of the PMT layout in the inner vetos (top). There were 78 PMTs arranged in five rings with alternating orientations along the rings. The open near detector veto vessel with the inner surface covered by the VM2000 reflective foil is shown at bottom.}
\end{figure}

The inner veto (IV) was optically separated from the inner detector and acted both as an extra shield against externally originating fast neutrons and gammas as well as actively identifying cosmic muons. Moreover, analysis methods were developed to tag neutron scatterings as well as high energy gammas for background suppression and flux evaluation~\cite{Abe:2014bwa}. The veto consisted of a steel tank, creating a cylindrical shell surrounding the Buffer vessel with approximately 50\,cm of space to the Buffer vessel on all sides. However, above the Buffer vessel, reaching to the IV lid, the central chimney cut into the IV's sensitive volume. Therefore, an extra OV layer was installed above the calibration tent as shown in fig.~\ref{fig:site}. The IV outer dimensions were chosen to fit within the existing pit in the far laboratory. The cylindrical IV vessel was 6810\,mm high and had a radius of 3240\,mm. It was constructed from 10\,mm thick steel (far detector) or stainless steel (near detector). The IV contained 90\,m$^3$ of LAB (linear alkyl benzene) based liquid scintillator. To increase light collection, 48 sheets of a highly reflective foil (VM2000$^{\mathrm{TM}}$ Enhanced Specular Reflector (ESR) from the 3M company)~\cite{Motta:2005ua} were mounted to cover completely the side walls of the Buffer in both the far detector and the near detector. This ESR foil was a multi-layer polymeric film with the outer layer being polyethylene-naphthalate. The specular reflectivity exceeded 95\% in the relevant wavelength range of  scintillator emission. The foils were manufactured with an adhesive coating on the back side of the polycarbonate carrier, which was removed prior to  installation to ensure compatibility with the scintillator liquid~\cite{phdroehling}. In the near detector, VM2000 was also used to cover the inner surface of the IV vessel (see fig.~\ref{fig:veto}). The inner surfaces of the far detector IV vessel were instead painted with a reflective white coating with a lower reflectivity of about 80\%~\cite{phdroehling} in the most relevant wavelength region. Several pipes crossed the inner veto,  to guide the PMT cables from the Buffer to the top flange of the IV, to fill and drain the liquids, and for level measurement devices.

The layout of the 78~PMTs (8-inch) in the IV as shown in fig.~\ref{fig:veto} was the same for the far and near detectors and was rotationally symmetric, with two rings at the bottom of the veto volume: one outer ring of 24 and one inner ring of 18~PMTs. Three rings of 12~PMTs each were arranged on top of the Buffer vessel, at the IV side wall at IV half-height, and 30 cm below the outer, upper edge of the cylindrical veto volume. PMT positions in the inner veto were optimized with regard to maximizing the efficiency in rejecting muons and their corresponding correlated backgrounds using a MC simulation \cite{vetojinst,phdgreiner}. The larger number of PMTs at the bottom was due to the non-transparent support structure of the Buffer vessel. The PMTs were attached to the IV surfaces with stainless steel support structures and their orientation alternated along each ring. The installation of the reflective foils and the PMTs was carried out with a clean tent (ISO class 7-8) erected in the DC laboratories.

\subsection{Shielding}
Different detector shielding concepts were applied for the near and far detector. The purpose in both cases was to protect the detector from natural radioactivity coming from the rock outside the detector pits. For the far detector a steel structure was installed around the IV. The steel grade S235JRG2 (similar to E24-2 or A283C) and a layer thickness of 15\,cm were chosen. The thickness was optimized taking into account simulation studies and space constraints in the tunnel and the laboratory. The shielding was installed above, below and around the IV vessel. The links between the shielding pieces were machined with a 60$^\circ$ rafter shape, very efficient for avoidance of external radiation leakage
through cracks between neighboring  shielding parts. The selection of the 300~tons of steel used in the shielding satisfied radiopurity constraints. Samples were measured by germanium spectroscopy before the approval of the batches. In all the shielding pieces the strength of the residual magnetic field was determined. In case it did not meet the specification of $<$\,0.5~gauss the pieces were demagnetized on site at Chooz. A rail system
allowed moving the shielding to the sides of the pit, thus allowing access the inner detector.

At the near site the pit housing the detector was longer and wider than that at the far laboratory. Therefore it was feasible to replace the dense steel shielding of the far detector with a more cost-effective water shielding. Simulations showed that the 1.0~m thick water shield of the near detector was about as efficient as 15~cm of steel. The water surrounded the sides and bottom of the IV vessel. Above the near detector was a steel shielding layer.

\subsection{Outer Veto}
\label{OV_sec}
The main source of correlated background events mimicking an IBD signal in the detector was due to cosmic muons, e.g.,~muons stopping and decaying inside the detector or creating neutrons crossing the surroundings of the inner detector. To further increase the rejection power against such muon events in addition to the IV an Outer Veto (OV) made of plastic scintillator panels and read by PMTs was installed above the detector vessels. Although muons could be tagged in the central detector as well, the OV was still useful, because background events could also be produced by muons passing some distance away from the central detector via secondary particle production or bremsstrahlung. Ideally the OV would have covered all sides of the detector. However, in combination with the IV system, a large OV situated above the detector provided sufficient background rejection for DC's purposes. Such a top-only design was the only possibility for the existing far laboratory due to the hall geometry. The OV information was not used as a real-time veto. Instead, it was combined with the neutrino detector data offline using timing information.

\begin{figure}[h]
\begin{center}
\subfigure[]{
	\includegraphics[width=0.5\textwidth]{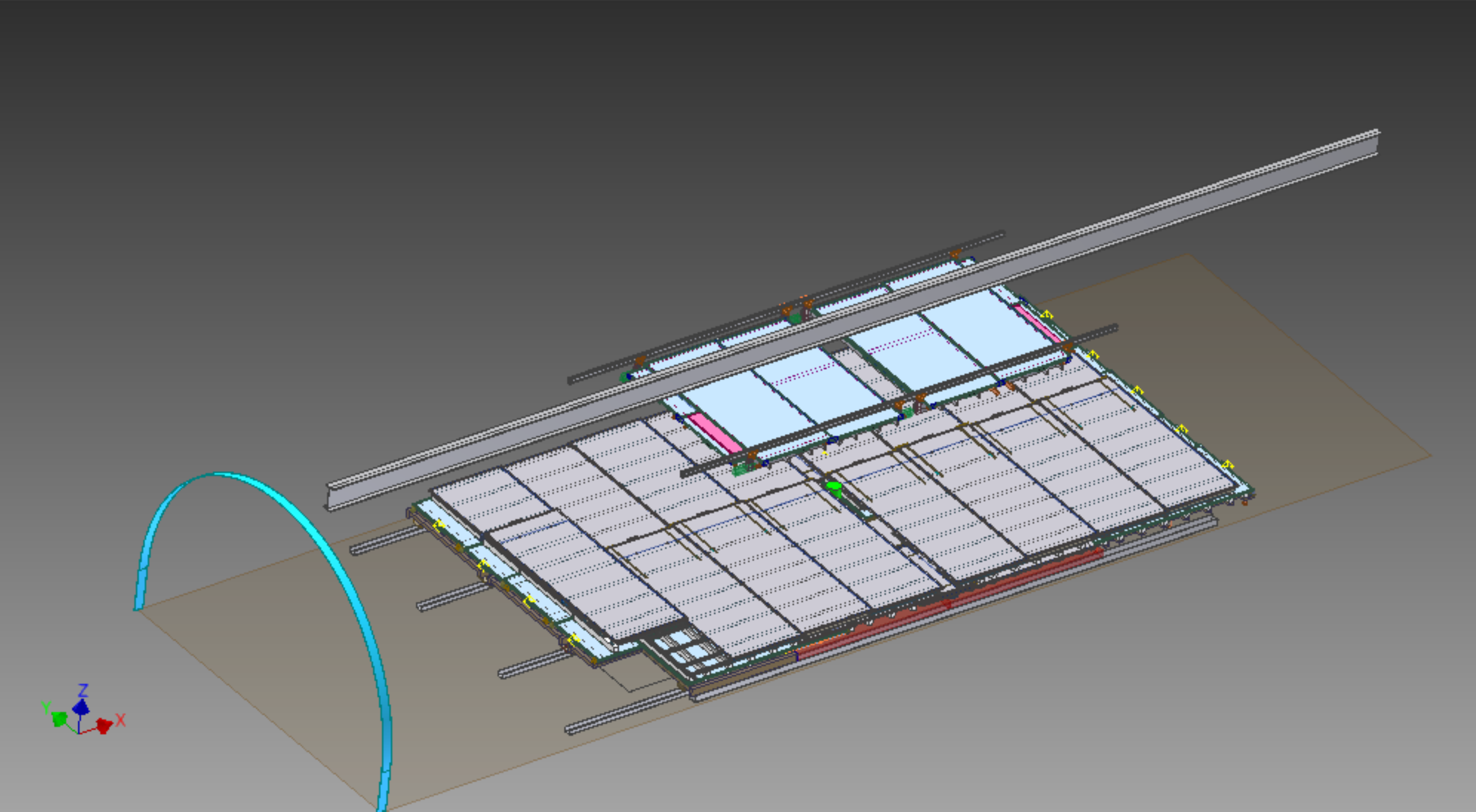}
	\label{OV_figa}}
\subfigure[]{
	\includegraphics[width=0.45\textwidth]{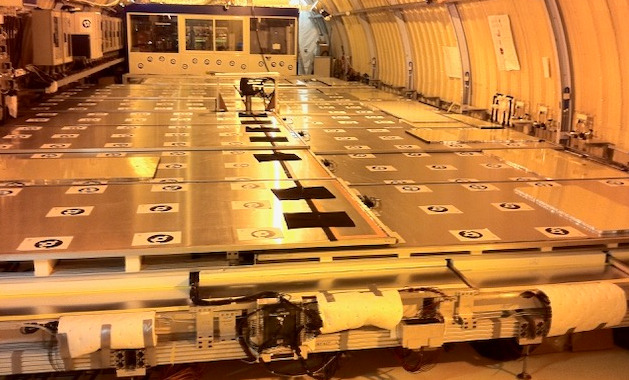}
	\label{OV_figb}}
\end{center}
\caption[]{(a) Schematic drawing of the OV in the far detector hall. The lower OV rested on the floor of the hall, the upper OV hung in two sections off of the ceiling on crane rails. Each panel was made up of two perpendicular layers, x and y. The chimney, indicated in green, required an opening in the OV coverage area. (The entrance to the hall is in the upper right of the image.) (b) A photo of the lower OV in the far detector hall is shown.  
}
\end{figure}

An OV was mounted directly above the steel shielding and provided the (x, y) coordinates of muons passing through the area centered on the chimney as shown for the far detector case in fig.~\ref{OV_figa} and fig.~\ref{OV_figb}. The coverage in the near detector laboratory was $13\times3.6$\,m$^2$, smaller than the $13\times7$\,m$^2$ coverage in the far laboratory. A small region around the chimney was left open in both the near and far detector lower outer vetos. Therefore, an additional upper OV was  mounted above the glove box. For the far detector, this upper OV included 8 modules, while the near detector upper OV used a single module.

\begin{figure}[htpb!]
	\centering
	\includegraphics[width=0.7\textwidth]{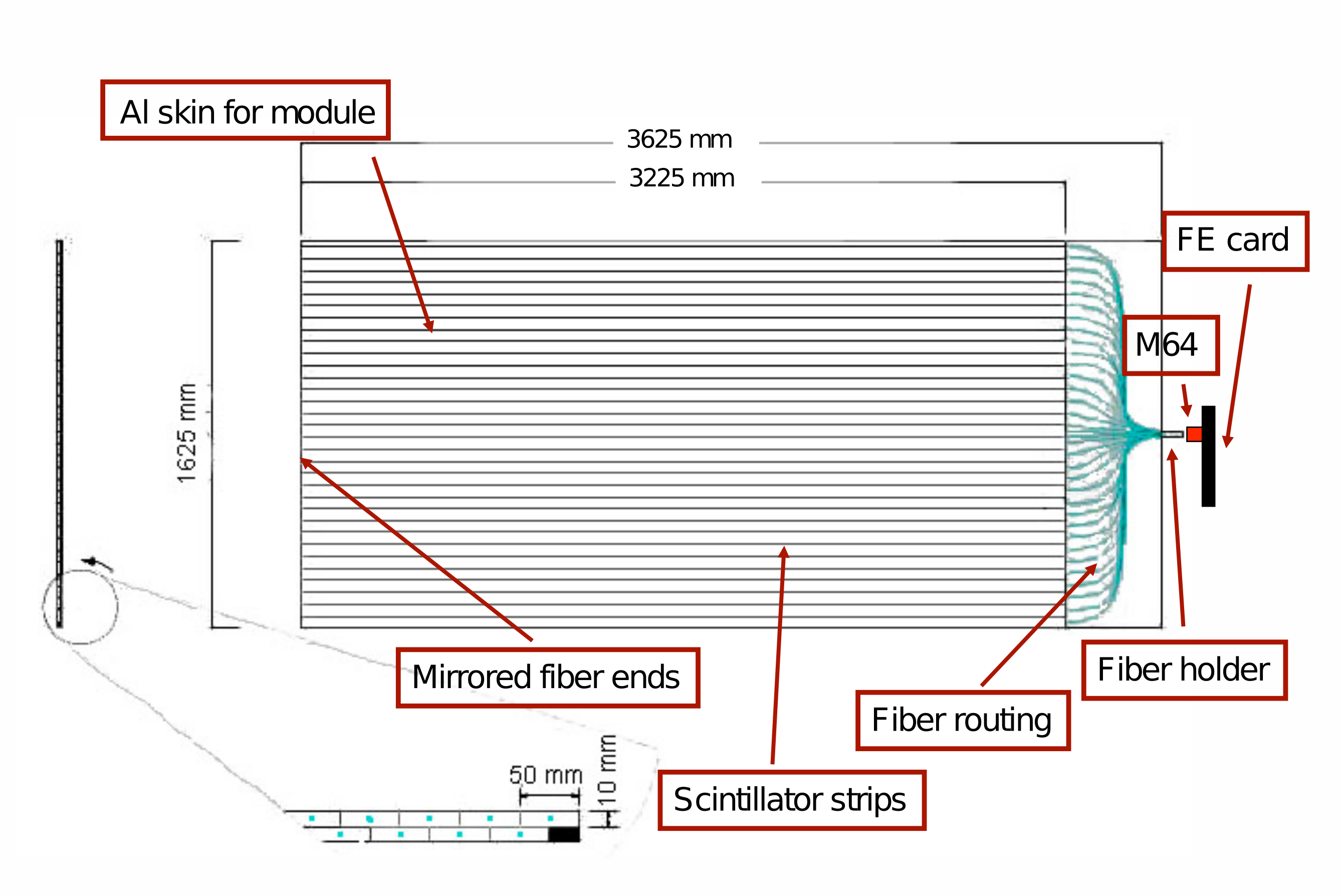}
	\caption{Schematic drawing of a short OV module. The 32 strips in the top layer are visible in the diagram, with fibers (light blue) routed to the fiber holder, which is attached to an M64 multi-anode PMT and front end (FE) board. In the inset at the bottom of the figure the two layers of strips can be seen, offset by half a strip width. Each strip was 50\,mm wide and 10\,mm thick and had a single fiber in the center. The strips at the far end of the PMT were mirrored to provide better light collection at the PMT. The whole module was covered with an aluminum sheet.}
	\label{fig:OV_module}
\end{figure}

The OV was assembled from modules containing 64 scintillator strips, each 5\,cm~×~1\,cm with a length of 320\,cm or 360\,cm (see fig.~\ref{fig:OV_module}). Each strip was extruded with a hole running through its length, through which a 1.5\,mm diameter wavelength-shifting fiber was threaded. Modules were built out of two superimposed 32-strip layers with the top layer offset by 2.5\,cm with respect to the bottom layer. These OV modules were positioned over the detectors in two levels, one with strips oriented in the x direction and one in the y direction.

The 64 fibers were coupled at one end to a Hamamatsu H8804 multi-anode photo-multiplier tube (M64). The other fiber ends were mirrored. Each M64 was connected to a custom Front-End board with a MAROC2 ASIC~\cite{Blin:2010tsa}. The MAROC2 chip allowed adjustment of the electronic gain of each of the 64 channels, as needed to correct for the factor of 2 pixel-to-pixel gain variation in the M64. Signals that exceeded a common threshold were sent to a multiplexed 12-bit ADC, providing charge and time information for hit strips. The output of the MAROC2 was fed into an FPGA, also on the board. This FPGA recorded the time-stamps of hits, buffered the data, provided control signals for the MAROC2 and a signal to the central system when data were present to be read out. A parallel to serial converter was used to transfer data to central receiver boards when the FPGA was polled by the main system. The outer veto data were read out as a separate data stream of hits and times. The data were synchronized and merged with the inner detector event records offline.

The IV and OV described above were fundamental to the precise muon reconstruction of DC~\cite{DoubleChooz:2014qeg}, which allowed the collaboration to make leading measurements on the cosmic muon characterization and annual modulation~\cite{DoubleChooz:2016sdt}, muon capture on light isotopes~\cite{DoubleChooz:2015jlf}, and yields as well as production rates of cosmogenic $^9$Li and $^8$He~\cite{DoubleChooz:2018kvj}. After each tagged muon event an 1.25\,ms wide offline veto was applied. The corresponding deadtimes in the detectors related to the muon veto were 25.5\% in the near and 5.4\% in the far detector.

\section{Organic scintillators and shielding liquids}
\label{sec:liquids}
Since several Gadolinium (Gd) liquid scintillators (Gd-LS) in reactor neutrino experiments of the past suffered from optical and chemical instabilities (e.g.~\cite{Apollonio:2002gd}\cite{Piepke:1999db}) a new type of metal loaded organic LS was developed and produced for DC based on a Gd-$\beta$-diketone (Gd-BDK) complex~\cite{Aberle:2011ar}. Among such other basic requirements as Gd solubility, optical transparency, safety considerations and material compatibility, a main focus during the design of the Gd-LS was the radiopurity of the components. A Gd concentration of 1\,g/l was chosen for the Target. 

\subsection{Composition and basic properties}

The composition of all four DC liquids is summarized in table~\ref{LS_T01} for the case of the far detector. Whereas the Target was identical in both detectors, slightly different mineral oil and paraffin concentrations had to be used in the near detector due to variations in the density of the mineral oil. This had negligible impact on the optical properties of the liquids. Some of the relevant basic properties of the DC liquids are summarized in table~\ref{LS_T02}.

\begin{table*}[h]
\caption[Composition]{Composition and volumes of the Double Chooz Target and Gamma Catcher scintillators (far detector). \label{LS_T01}}
\begin{center}
\begin{tabular}{lll}
Scintillator & Composition & CAS Number\\
\hline
Target (10.3\,m$^3$) & 80~\%$_\mathrm{vol}$ n-dodecane (99.1~\%) & 112-40-3\\
	& 20~\%$_\mathrm{vol}$ o-PXE (ortho-Phenylxylylethane) (99.2~\%) & 6196-95-8\\
	& 4.5~g/l~Gd-(thd)$_3$ (Gd(III)-tris-(2,2,6,6-tetramethyl- & 14768-15-1\\
        & heptane-3,5-dionate)) (sublimed) & \\
	& 0.5~\%$_\mathrm{wt.}$ Oxolane (tetrahydrofuran, THF, $> 99.9$~\%) & 1099-99-9\\
	& 7~g/l PPO (2,5-Diphenyloxazole, neutrino grade) & 92-71-7\\
	& 20~mg/l~bis-MSB (4-bis-(2-Methylstyryl)benzene) & 13280-61-0\\ [5px]
GC (22.4\,m$^3$) & 66~\%$_\mathrm{vol}$ Mineral oil (Shell Ondina 909) & 8042-47-5\\
	& 30~\%$_\mathrm{vol}$ n-dodecane & 112-40-3\\
	& 4~\%$_\mathrm{vol}$ o-PXE (ortho-Phenylxylylethane) & 6196-95-8\\
	& 2~g/l  PPO (2,5-Diphenyloxazole) & 92-71-7\\
	& 20~mg/l~bis-MSB (4-bis-(2-Methylstyryl)benzene) & 13280-61-0\\ [5px]
Buffer (110\,m$^3$) & 53~\%$_\mathrm{vol}$ Mineral oil (Shell Ondina 917) & 8042-47-5\\
        & 47~\%$_\mathrm{vol}$ n-paraffins (Cobersol C 70)&  64771-72-8\\ [5 px]
Inner Veto (90\,m$^3$) &  50~\%$_\mathrm{vol}$ Linear Alkyl Benzene (LAB) & 67774-74-7\\
        & 47~\%$_\mathrm{vol}$ n-paraffins (Cobersol C 70)&  64771-72-8\\
        & 2~g/l PPO (2,5-Diphenyloxazole) & 92-71-7\\
	& 20~mg/l~bis-MSB (4-bis-(2-Methylstyryl)benzene) & 13280-61-0\\
\hline
\end{tabular}
\end{center}
\end{table*}

\begin{table*}
\caption[Properties]{Main properties of the liquids used in Double Chooz: density (d), thermal expansion coefficient ($\alpha$), kinematic viscosity ($\eta$), attenuation length ($\Lambda$), light yield (LY), refractive index (n) and H fraction. As a standard for the LY measurement we used the liquid scintillator BC-505: Bicron, St. Gobain Crystals (80\% light yield anthracene). \label{LS_T02}}
\begin{center}
\begin{tabular}{lcccc}
 & Target & Gamma Catcher & Buffer & Veto\\ \hline
d [kg/l] at 15$^\circ$C & $0.8035\pm0.0010$ & $0.8041\pm0.0010$  &  $0.8040\pm0.0010$  &  $0.8040\pm0.0010$  \\
$\alpha$ [10$^{-3}$/K] & $0.6\pm0.1$ & $0.6\pm0.1$ & $0.6\pm0.1$ & $0.6\pm0.1$\\
$\eta$ [mm$^2$/s] & $2.3\pm0.1$  & $3.7\pm0.1$ &  $7.04\pm0.04$ &  $3.58\pm0.04$ \\ 
 & $(21^\circ$C) & $(21^\circ$C)  & $(15^\circ$C)  &  $(15^\circ$C) \\
$\Lambda$ at 430 nm [m]  & $7.8\pm0.5$  & $13.5\pm1.0$ & $ > 15$ & $ 10.1\pm 1.0 $\\ 
LY [\% BC-505]  &  $48.1\pm0.5$ & $46.6\pm1.0$ & -  &  $54.5\pm1.3$ \\ 
n (589~nm, 22$^\circ$C) & 1.450 & 1.445 & 1.445 & 1.450 \\ 
H fraction [wt.\%]  & $13.60\pm0.04$  &  $14.6\pm0.2$ &  $14.8\pm0.2$  & not determined\\ \hline \\
\end{tabular}
\end{center}
\end{table*}

\subsection{Light yield}
The light yield of the Target was optimized with the help of a light yield model specifically developed for Double Chooz~\cite{Aberle:2011zm}. The Target light yield was compared to several standards. Relative to an unloaded o-PXE scintillator with no n-dodecane and similar fluor amounts, the light yield was measured to be $56\pm2$~\%. The scintillation signal for highly ionizing particles is quenched. The energy dependent quenching factors were determined using small samples in the laboratory for alpha-particles~\cite{CADiss}, proton recoils~\cite{VZDiss} and low energy electrons~\cite{Aberle:2011zz}. The quenching factors found for alphas of 7.7~MeV ($^{214}$Po decay) could be confirmed with detector data~\cite{MFDiss} and were determined to be 9.6 in the Target and 12.3 in the GC.

The light yield and density of the GC scintillator had to be matched to the Target values (see table~\ref{LS_T02}). This was mandatory to assure a homogeneous detector response as well as the mechanical stability of the acrylic vessels respectively. It was important to keep the densities of all detector liquids the same within a precision of the order of 0.1\%. The densities of all liquids were matched at the detector temperature of about 15$^\circ$C to $0.804\pm0.001$\,kg/l and the thermal expansion coefficients were determined.   

\subsection{Scintillator emission spectrum}
\begin{figure}[tb]
\begin{center}
\subfigure[]{
	\includegraphics[width=0.45\textwidth]{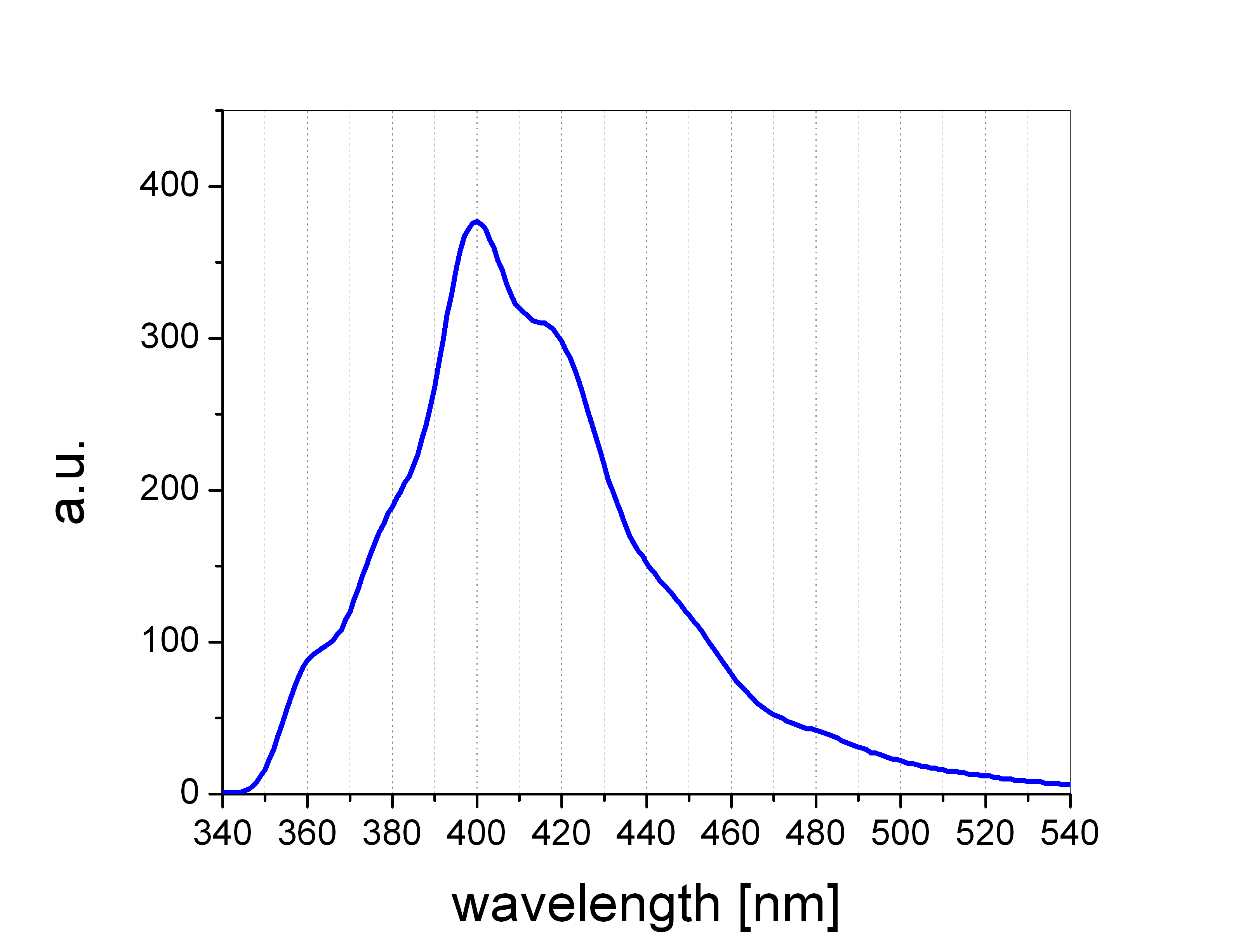}
    \label{LS_01}}
\subfigure[]{
	\includegraphics[width=0.45\textwidth]{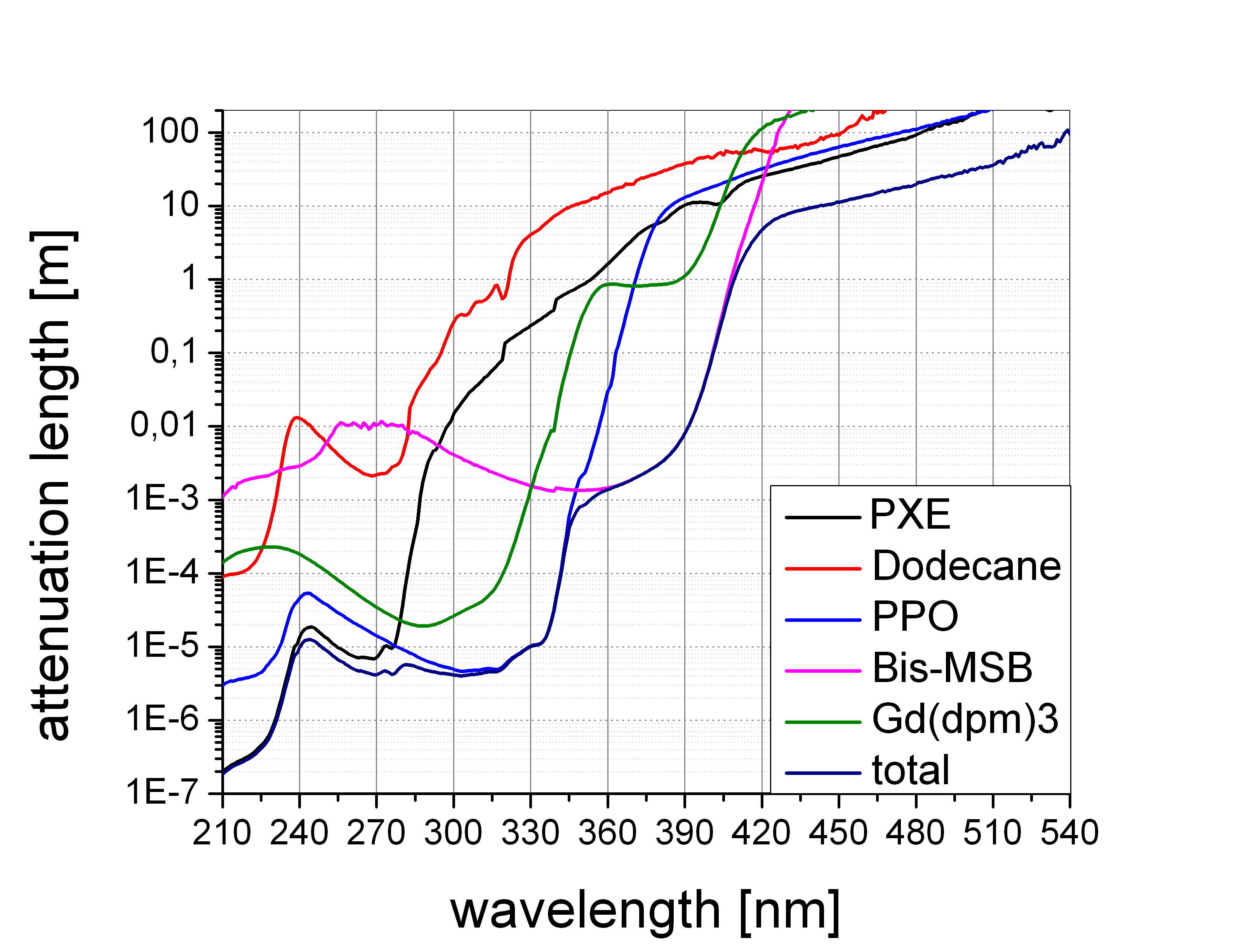}
    \label{LS_02}}
\end{center}
\caption[]{(a) Primary scintillator emission spectrum of the Double Chooz Target measured in a triangular cell. (b) Contributions of Target components to the total attenuation length. The absorption above 420~nm is dominated by optical impurities in the ingredients.
}
\end{figure}

Figure~\ref{LS_01} shows the scintillator emission spectrum for the Target as measured with a fluorimeter. The solvent molecules in a triangular scintillator cell were excited from the back using UV radiation. The light had to pass a few mm through the scintillator before the emission spectrum was recorded. This spectrum is used in the optical model of our simulation (see section~\ref{simulation}), since it also represents the emission of the light yield setup to determine the production of primary light after excitation by radioactive sources. The cells used in these measurements were 1\,cm in dimension. The farther light has to travel through scintillator the more the scintillator spectrum is shifted towards longer wavelengths. This is because the re-emission after bis-MSB absorption can only be at longer wavelengths due to energy conservation. The optical model of the DC simulation takes into account this wavelength shift of the emission spectrum as compared to the one shown in fig.~\ref{LS_01}.  

\subsection{Light absorption and re-emission}
In fig.~\ref{LS_02} the wavelength dependent attenuation length, as measured in a 10\,cm quartz cell using a spectrophotometer, is shown for the Target liquid. In addition, the contribution of the single components are presented as they were obtained from individual measurements of the molar extinction coefficients and then calculated at the given concentrations. Since the molar extinction coefficient varies over 10 orders of magnitude within the plotted wavelength range, each curve was determined by combining several measurements with varying concentrations of the component of interest. These data provide another important input for our optical model in the simulation. A wavelength dependent ``impurity'' spectrum was added to compensate for a slight discrepancy between calculated attenuation length and the measured values of the final mixture in the very transparent region above 420~nm. At this level there were significant batch-to-batch variations for the optical purity of the chemicals and small impurities could have been picked up from the surfaces of materials in contact with the liquid during  large scale production. 

As can be seen in fig.~\ref{LS_02} the absorbance above 420\,nm was dominated by the contributions of the o-PXE and the PPO. Since this wavelength region is quite far from the corresponding absorption peaks (269\,nm for PXE and 303\,nm for PPO) we assume that the absorbance was dominated there by non-fluorescent impurities in those components. The Gd-loading had negligible impact on the transparency for scintillation light thanks to the high optical purity of the Gd-complex. Below 420\,nm the bis-MSB starts to dominate. Whereas light absorbed above this wavelength was mainly lost, the light absorbed below was re-emitted with high probability around $400 - 450$\,nm corresponding to the wavelength region for which the PMTs were most sensitive. The quantum efficiency of bis-MSB is about 0.86~\cite{Buck:2015jxa}. Although there is no significant contribution to the scintillation light expected below 350\,nm this region is still of interest for Cherenkov light. The scintillator stability was checked in calibration campaigns and by investigating the time dependence of the Gd peak in the energy spectrum after spallation neutrons. Results are presented in section~\ref{perf_stab}. 

\subsection{Material compatibility}

To ensure optical stability over the long data taking period of the DC experiment, liquid/solid interface exposures with materials, such as containing vessel or calibration devices, in contact with liquids, in particular the Gd-loaded Target scintillator, should not spoil their chemical and optical stability. Special care was needed to meet this challenge. The long-term stability of the liquid scintillators developed for DC has been investigated by means of spectrophotometric techniques. The transmission of a collimated light beam through 10~cm of material was routinely measured. Dedicated 10~cm long HELLMA Suprasil$^{\rm TM}$ quartz cells, with low UV absorbance windows were used. 
Sensitivity was thus provided to detect any potential chemical evolution of the scintillator leading to an increase in the absorbance, i.e.,~to an optical degradation.

For any new scintillator sample, a 10~cm quartz cell was filled (100~mL) and the liquid was flushed with $N_2$ in order to purge oxygen, a potential hazard as regards the chemical stability of the scintillator. The cell was hermetically sealed through airtight stoppers and stored in darkness at room temperature. The transmission $T$ in the wavelength range $300~{\rm nm} < \lambda < 800~{\rm nm}$ was routinely measured, once to twice a month. Since pure quartz shows very low absorption in the optical wavelengths of interest, the spectrophotometric instrument calibration was usually referenced to the transmission $T$ of the beam in air, defined as $T = 100\%$. The spectrophotometric measurement was used not only to monitor the relative changes of the sample transmission, but also to determine the absolute wavelength-dependent attenuation length. For the latter case, the effect of  light losses due to reflections at the air-quartz-liquid and liquid-quartz-air interfaces were corrected. This was done by self-referencing the transmission spectra to the response around $\lambda=600~{\rm nm}$. No optical degradation was observed in the wavelength region $\lambda> 650$~nm. This gives a handle by which to correct small fluctuations in the offset of each scan with respect to the others, which are believed to be due to surface effects and instrumental systematics. This offset correction was of the order of a few tenths of percent, at most. In this region it can be safely assumed that light absorption is negligible. To improve the accuracy of this absolute measurement, the small absorption from the quartz windows was also corrected. The same experimental procedure was repeated with a twin sample stored at elevated temperature, typically $40^\circ{\rm C}$.

\begin{figure}[htpb!]
	\centering
	\includegraphics[width=.49\textwidth]{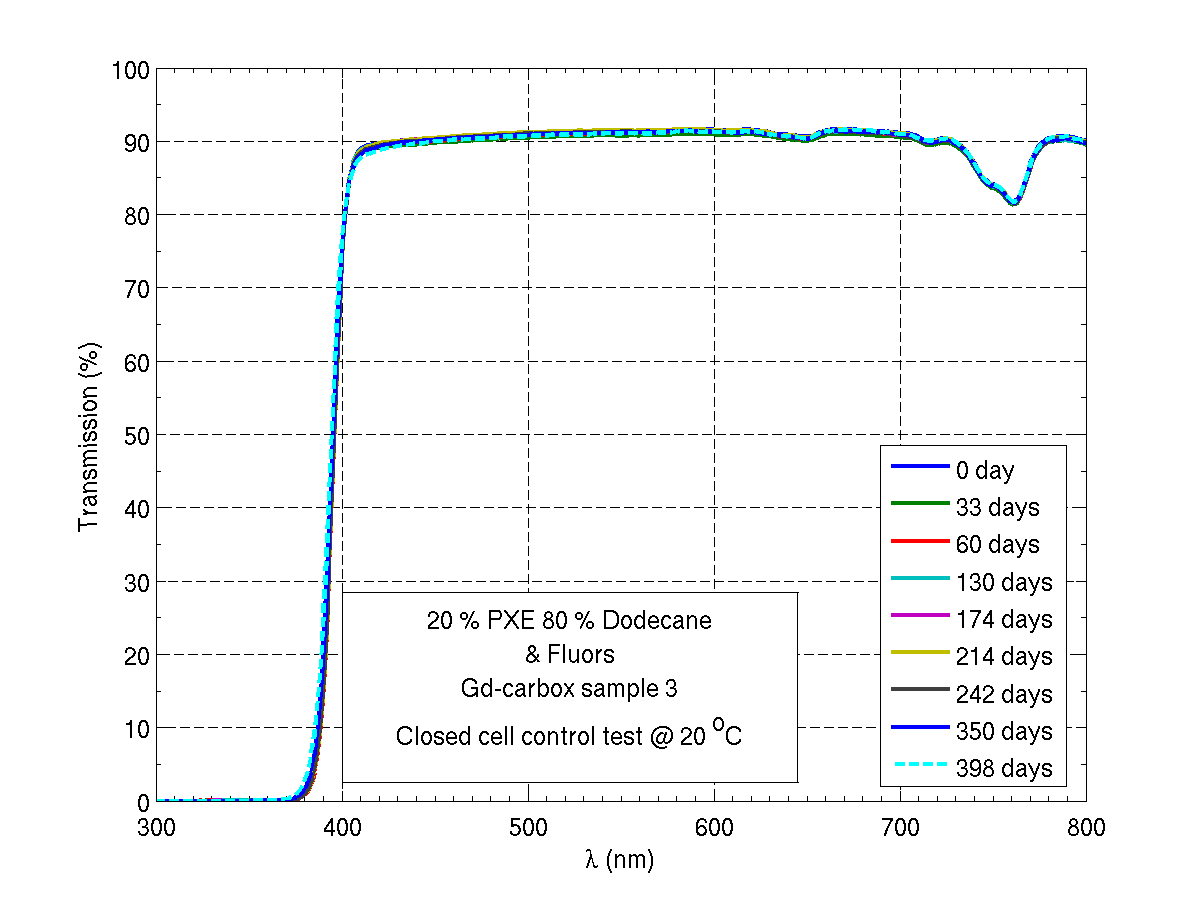}
	\includegraphics[width=.49\textwidth]{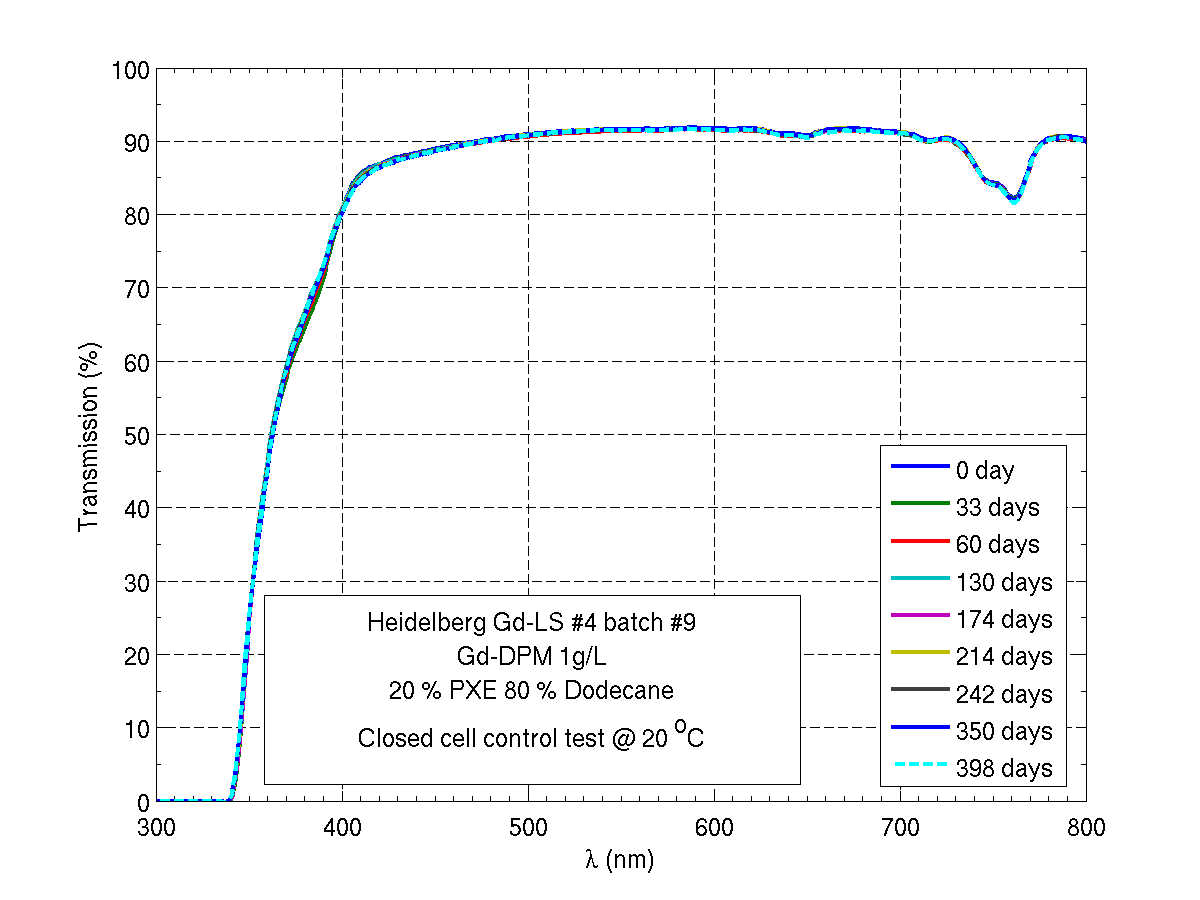}
    \caption{Light transmission ($T$) of Gd doped liquid scintillators. A Gd-CBX (top) and a Gd-BDK (bottom) prototype sample are shown, both stored at $20^\circ{\rm C}$.}
	\label{fig:GSHD}
\end{figure}

This was intended as an accelerated aging test, under the hypothesis that a change in the temperature influences just the dynamics of the chemical reactions. The overall systematic uncertainty on the absorbance (including the spectrometer stability and the planarity of the quartz windows) was proved to be at or below 0.5\%. Spectrophotometry thus provides a very powerful and precise tool to investigate the stability of scintillators. Fig.~\ref{fig:GSHD} shows the optical survey of two R\&D Gd-loaded scintillators synthesized in early 2005, a Gd-BDK and a Gd-carboxylate (Gd-CBX)~\cite{Ardellier:2006mn} test sample. Within the sensitivity of the instrument, no degradation was observed for 1~year of data taking.

The development of the measurement and certification protocols was a long, careful and stringent task. Room temperature was monitored and controlled within the range 14--18$^\circ{\rm C}$. Reference samples and material exposed samples were prepared and tested using 20~cells in parallel. Storage, transfer and measurement operations were investigated. Careful handling, cleaning and visual inspection were applied at each step. The materials in contact with liquids were classified into two main categories: permanent materials (acrylic, gluing, stainless steel vessels) and temporary materials such as calibration devices. Databases of all materials with their characteristics and tests were developed. The final material certification even required the use of a 500~l glove box operated under highly pure $N_2$ (6.0~grade). Dedicated preparation methods and protocols were developed with reference measurements and time evolution studies of liquids exposed to detector materials. The final certification protocol was then based on extrapolation scaling laws to full detector size and exposure time with safety margins. The effect induced by the acrylic glue on the liquid, which was unavoidable, was taken as the baseline threshold for all other materials to which more stringent criteria were thus applied.

All DC subsystems with exposure to liquid scintillators were tested, in total more than 100~materials, over a period of 6~years. A full analysis software package for material compatibility testing and evolution investigations was developed in Matlab$^{\rm TM}$. The most relevant systems in contact with the Gd-LS were made of fluorinated hydrocarbons (PFA, PVDF, PTFE) or acrylic, which were all classified as certified materials. According to the compatibility tests small surfaces or temporary exposure to stainless steel, nylon or Delrin (polyoxymethylene homopolymer) could also be allowed, e.g.,~for the calibration systems. For the other Gd-free liquids no restrictions on the use of stainless steel were applied. Among  rejected materials for contact with the IV liquid were a titanium dioxide coating, shrink tubing and the VM2000 adhesive foil.   

Thanks to all the care taken in liquid scintillator development and stability checking through material compatibility activities, both DC detectors demonstrated unprecedented stability over time with 7~years for the far detector and 4 years for the near one (see section~\ref{perf_stab}). In particular, there was no evidence for degradation neither of the liquids nor the acrylic materials such as crazing or optical deterioration. 

\section{Target and GC proton numbers}
To calculate the proton number $n_p$ in the Target and GC, two parameters need to be known: the absolute liquid mass $m$ and the relative H fraction $f_H$ in the scintillator molecules ($n_p=\frac{f_H\cdot m}{m_H}$). Whereas masses can be measured at the 0.1\% level and even below, the H fraction can be determined only at the 1\% level with standard technologies such as CHN elemental analysis. The situation of the Target is special since it was produced and mixed as a single batch that was later used to fill both detectors. Therefore the H fraction is identical between both detectors and the uncertainty in $n_p$ is estimated purely from the mass determination including temperature effects. Since temperature variations change the liquid density, the scintillator mass in the active volume is slightly temperature dependent. Distinct from the Target liquid, the GC scintillator was mixed separately for far and near detectors with slightly different chemical compositions, since it was originally not foreseen that this volume would be used as a neutrino target. Therefore the GC H fraction needed to be measured. Due to correlations in the uncertainty of the H fraction determinations in the samples, higher precision can be obtained for the H fraction ratio of (GC near)/(GC far) than for an absolute estimation.

For the Target the value of the H fraction can be directly calculated from the chemical composition, since the liquid consisted of well defined molecules with a known hydrogen content, whereas in the GC there was mineral oil containing many different molecules making a calculation unreliable. Therefore, the precision of the H fraction knowledge is better for the Target than for the GC. 

Neutrino interactions in the acrylic material contribute dominantly to the first energy bin of the positron spectrum around 1\,MeV, since the energy deposited in the active scintillating material is mainly from the 2x511\,keV positron annihilation gammas. The proton number of the acrylic material was considered in the simulations and the corresponding uncertainty can be neglected for the neutrino analyses.  

\subsection{Target}
Before detector filling, the Target liquid was stored in a dedicated tank in the neutrino laboratory. This tank had a volume of 12\,m$^3$ and was fully coated with PVDF on its interior. It stood on three feet located at 120$^\circ$ intervals on a 1.2\,m radius circle. These feet were used as interfaces with weight sensors (SB5-50kN-C3 from FLINTEC). Under each foot of the weighing tank one sensor with a measurement range of 5\,t was set and the three measurements were summed. The overall range of the weighing system, which was used for both detectors, was then 15\,t with an accuracy below 0.1~\%. 

The weights of the near and far detector Target liquid were measured as follows:
\begin{itemize}
\item Near detector Target mass: $8326.5\pm3.8$~kg
\item Far detector Target mass: $8291.5\pm7.3$~kg
\end{itemize}
To extract the proton numbers one has to take into account the hydrogen fraction of $0.1360\pm0.0004$. With these inputs the absolute proton number can be deduced:
\begin{itemize}
\item Near detector proton number: $(6.767\pm0.020)\cdot10^{29}$
\item Far detector proton number: $(6.739\pm0.021)\cdot10^{29}$
\end{itemize}

For the $\theta_{13}$ analysis with two detectors the fully correlated uncertainty of the H fraction cancels. Applying the numbers above, the near/far proton number ratio is $1.0042\pm0.0010$. 

\subsection{Gamma Catcher}

\subsubsection{Volume determination}
For the GC scintillator mass determination in the near detector, one has to take into account an observed leakage between GC and Buffer liquids. Depending whether the liquid exchange happened before or after the GC chimney level was reached during filling, the mass measurements could be biased by the amount of liquid leaked. Although there are indications that the leak was in the chimney area, the most reliable method to determine the GC liquid mass in the near detector is from the volume estimate of the vessel dimensions. This method is almost leak independent, since $f_H$ is very similar in the GC and Buffer liquids. To calculate the mass from the volume, the liquid density is also needed which is known to high precision. The leak had a minor impact on the data taking, since the characteristics of scintillation events in the Buffer allow to reject them with selection cuts. 

The detector temperature during the data taking period was $13.4\pm0.5^\circ$C in the near and $13.5\pm1.0^\circ$C in the far detector. Since the detector was filled at about 15$^\circ$C, a temperature correction is applied to the liquid density of 0.804 at 15$^\circ$C (see table~\ref{LS_T02}) using the thermal expansion coefficient $\alpha = 0.0006/$K extracted from density measurements between 15 and 25$^\circ$C. The mass can thus be calculated as the product of density and liquid volume.

The liquid volume in the GC was calculated based on information from the technical drawings as well as measurements. The Target volume is subtracted from the GC volume together with all other known elements placed in the GC as the acrylic support structure, filling tubes fixtures, etc. The geometry and density of the acrylic material is included in the simulation (see section~\ref{simulation}) and antineutrino interactions on Hydrogen nuclei in the acrylic are taken into account accordingly. For the Target an uncertainty of 2\,mm on height and diameter was estimated. For the GC slightly higher uncertainties of 5\,mm for the height and 2.6 (2.9)\,mm for the diameter of the near (far) detector vessels were taken. In the calculations a perfect cylinder model was assumed and corrections were applied to take into account thermal expansion effects. This way, volumes of 22.62\,m$^3$ (near detector) and 22.41\,m$^3$ (far detector) were determined with a 0.5\% relative uncertainty. 

The mass of the GC liquid can be determined directly using flowmeter data and weight measurements. Whereas flowmeter data are available for both detectors, a weight measurement was  conducted only for the near detector liquid. From the flowmeter data for the far detector we get a GC mass of 18.08\,t (15$^\circ$C) with a relative uncertainty of 0.7\%. In the near detector the flowmeter indicated a volume of 22.608\,m$^3$ in very good agreement with the calculated volume estimate. This corresponds to a mass of 18.18\,t. This mass is determined at a filling level 3\,cm above the start of the GC chimney. Moreover, the weight of the GC liquid transport truck was measured before and after  filling. Subtracting the liquid not filled into the detector or chimney we obtain a mass (volume) of 18.16\,t (22.581\,m$^3$). Although all volume calculations, flowmeter data and weight measurement agree at the per mille level, a rather large uncertainty of 1.7\% was set on the last two values to account for a possible bias from leakage. This value is obtained from the estimated amount of GC liquid identified in the Buffer taking into account UV-Vis and GC-MS measurements as well as LED calibration data. If we calculate the weighted mean of the volumes as reported above we obtain:

\begin{itemize}
\item Near detector GC volume: $22.615\pm0.106$m$^3$ (corresponding to $18.200\pm0.087$~t)
\item Far detector GC volume: $22.438\pm0.094$m$^3$ (corresponding to $18.056\pm0.079$~t)
\end{itemize}

\subsubsection{Hydrogen fraction measurements}
The hydrogen fraction in the GC samples was measured at BASF Ludwigshafen and TU M\"unchen (TUM) by combustion analysis (CHN). The uncertainties of the two measurements are similar and estimated to be 1\% making this the dominant part of the total proton number uncertainty in a single detector analysis. The main part of the systematic uncertainties is correlated between measurements in case they are performed on the same day using identical instruments, methods and calibration. Therefore the two chemically similar, but not identical, GC samples of the near and far detectors were measured in series within one sequence in both laboratories. The $f_H$ ratio between GC$_{near}$ and GC$_{far}$ could be determined to a precision at the 0.1\% level by performing multiple measurements.

For $f_H$, we use the average value of the TUM measurements, since a tenfold measurement was performed as compared to a sixfold measurement at BASF. For the GC samples we get a hydrogen ratio  $f_H =14.53\pm0.15$\% in the near and 
$f_H =14.58\pm0.15$\% in the far detector. If we use the standard deviation of the mean of the tenfold measurement as the uncorrelated part of the uncertainty, we get only 0.12\% (near) and 0.08\% (far) relative error as compared to the 1\% relative error including the correlated part. 

The $f_H$ measurements at BASF were 1\% higher, 
but the ratio of the hydrogen fractions between near and far detector agreed within 0.1\%. These results demonstrate that the estimate for the precision of the method at the 1\% level is reasonable. Moreover the results show that this uncertainty has a highly correlated part when comparing values obtained within a sequence of measurements done in one setup.  

\subsubsection{GC proton number}
Applying the values obtained above, the final results for the GC proton numbers in the two detectors are:
\begin{itemize}
\item Near detector proton number: $(1.580\pm0.018)\cdot10^{30}$
\item Far detector proton number: $(1.573\pm0.017)\cdot10^{30}$
\end{itemize}
This corresponds to a near/far ratio of $1.0045\pm0.0067$. In the uncertainty of this ratio the mainly correlated part of $f_H$ cancels, so the final uncertainty is dominated by the mass numbers.

\section {Gas and liquid operation systems}
\label{liquidandgas}

The high purity filling systems were designed to allow several filling modes, thermalization of the liquids (FD only) and various liquid operations under a nitrogen blanket. All systems were built as modules and were partly movable. 

\subsection{Nitrogen system}

To avoid scintillator degradation due to contact with oxygen or air humidity as well as for fire prevention, a nitrogen atmosphere was maintained in all detector and tank volumes during the filling and operation of the detector. A slight overpressure of up to 2~mbar over ambient pressure effectively prevented air from getting into the detector.

The nitrogen was provided by a 3000~liter liquid nitrogen plant (nitrogen quality 4.7) per experimental site in front of the tunnels to the neutrino laboratories. Gas lines to the detectors were made of stainless steel tubes (1/4 to 3/4~inch). Particle filters were installed close to the nitrogen plant and inside the neutrino laboratory before the gas entered the detector. The gas pressure was first reduced to about 4 bar before the N$_2$ entered the tunnel and then further reduced to $<5$~mbar inside the neutrino laboratory. The system was equipped with over- and underpressure protection systems at various levels. In addition, a subsystem used for N$_2$ flushing of the detector vessels before filling was operable at 150~mbar.  

\subsection{Liquid handling systems}
\label{sec:LH}
\begin{figure}[ht]
    \centering
    \includegraphics[width=0.8\textwidth]{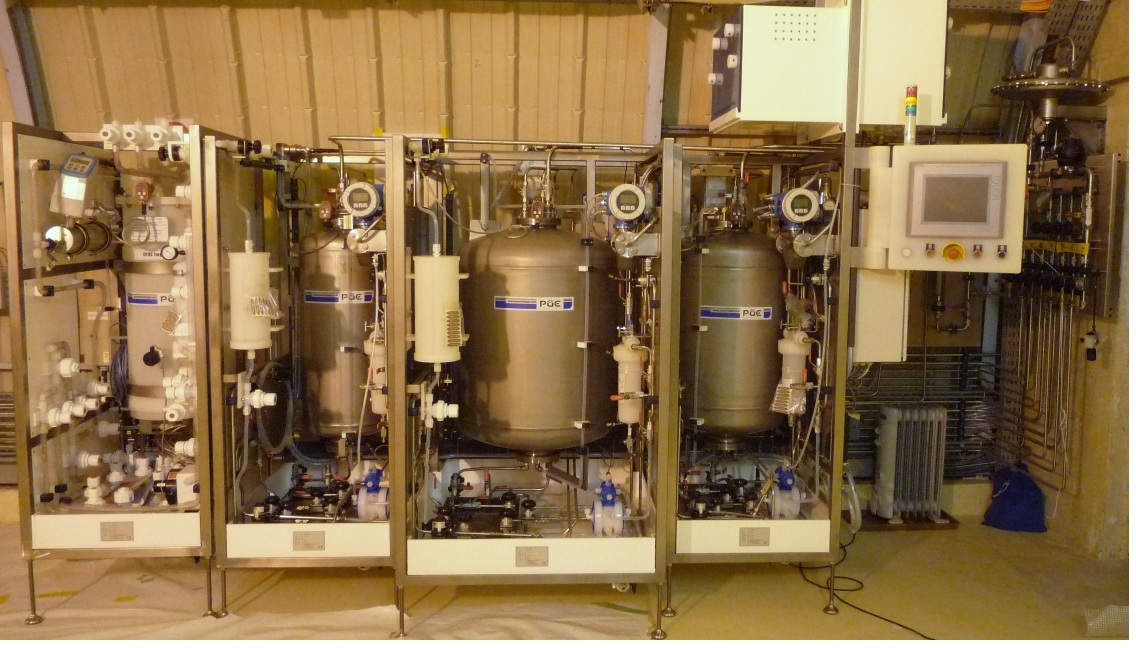}
    \caption{Detector Fluid Operating System installed in the underground laboratory of the far detector. From left to right are shown: Target, GC, buffer and veto liquid systems; the control unit including touch screen; the over-, underpressure safety and low pressure nitrogen boxes, and one of the gas system manifolds.}
    \label{fig:DFOS}
\end{figure}

The underground liquid systems as shown in fig.~\ref{fig:DFOS} for the far laboratory were designed as a specialized multipurpose tool in order to handle all detector liquids at all stages of detector life~\cite{PPDiss}\cite{MFDiss}. The systems were divided into four individual modules and located in close proximity to the detector. The modules had individual frames and worked independently of each other. Each frame was specifically made for its purpose in order to meet all needs for the individual liquids. The modules were equipped with intermediate tanks (100~liter scale) to allow decoupling of the filling system from the area in which the main part of the GC, buffer and IV liquids were stored, which was outside the tunnel. In the far detector, all tanks and tubing for the liquids other than the Target were made of stainless steel. In the near detector, the approximately 200~m lines connecting the laboratory to the truck unloading area outside were made of carbonated PFA. Pressure driven Teflon pumps were used for the filling of all volumes. All modules were equipped with high precision flow meters. 

\begin{figure}[ht!]
    \centering
    \includegraphics[width=0.8\textwidth]{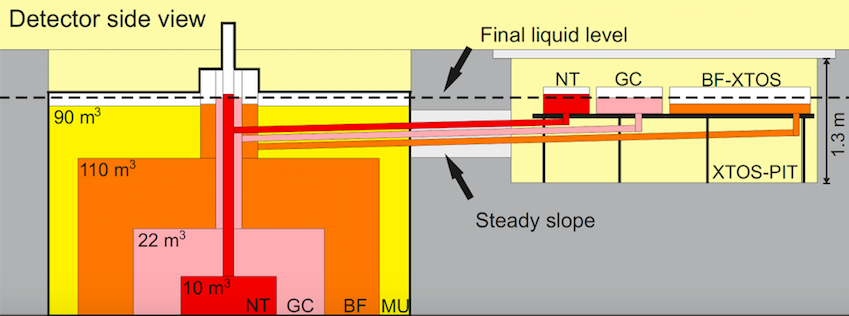}
    \caption{Detector side view for the far detector case including the connection between the detector chimneys and the expansion tank system (XTOS), made with 3/4-inch tubes.}
    \label{fig:XTOS}
\end{figure}

For level regulation of the Target, GC and Buffer when the volume of the detector liquids changed due to temperature fluctuations, expansion tanks were used as shown in fig.~\ref{fig:XTOS}. The 15~cm high tanks allowed the liquids to expand or shrink by several centimeters in height. The expansion tanks were located in pits next to the detector and connected via tubes to the detector chimney. The tank sizes were designed to allow for temperature fluctuations inside the detector at the level of 1\,K. No extra tank for the IV was needed since the liquid level in this volume was below the chimney of the vessel and therefore the liquid level variations were moderate. The expansion tanks for the GC and Buffer were made of stainless steel whereas a PVDF tank was used for the Target. The level inside the expansion tanks was regularly monitored by checking it on semi-transparent side tubes mounted to each tank. To minimize light leaks baffle boards were installed inside the expansion tanks to reduce the probability for photons to enter the inner detector volume to a negligible level. For the near detector these tanks were placed on balances and the weight was monitored as well. 

\subsection{Detector filling and monitoring}

During the filling of the detectors the stress on all vessels had to be minimized and all relevant parameters had to be kept as identical as possible. The specification on level differences of neighboring volumes during filling was to stay within less than 3~cm, so all volumes had to be filled in parallel. The pressure differences in the blanket system were kept below 0.5~mbar. Finally, the densities of all four detector liquids were matched and temperature differences were minimized. 

During filling several monitoring systems for the levels, pressures and temperatures were used~\cite{PPDiss}\cite{MFDiss}. Independent methods of level measurement were applied, all of them sensitive down to the mm level. The absolute and differential pressures in the gas volumes were measured with a precision of 0.1~mbar and temperature sensors were placed at several detector positions.

The design concept of the liquid level monitoring system was to have at least two completely independent measurements which measure the liquid level to  better than 1\,cm accuracy. In total, four different and independent systems for level measurement were chosen. Three of them monitored the absolute liquid level in the detector while the fourth system was focused only on differences of liquid levels.

The absolute liquid levels in the GC, Buffer and IV were determined during filling by a laser measuring the distance between the top lid of the detector and a PTFE cup floating on the liquid surface. The cylindrical cups were used as reflectors and kept in place by a tube with a slightly bigger diameter than the float itself. The lasers were mounted on an adjustable flange. The GC was a special case, since there was no straight pipe available directly connected to the bottom of the GC vessel. Therefore a tube connection  filled with liquid via the detector top was installed allowing communication between the liquid in the GC and a dedicated pipe for the laser measurement inside the veto vessel (this system was called the ``Loristube''). The second measurement in the outer three volumes was accomplished using hydrostatic pressure sensors. Each sensor head was put directly into the medium and measured the hydrostatic pressure created by the liquid above. 

The absolute liquid level in the Target was monitored with high accuracy by a distance measurement using a ``Tamago'' (Endress + Hauser, Proservo NMS 53). This gauging system is based on the principle of displacement measurement. A displacer is suspended on a measuring wire which is wound onto a finely grooved drum and lowered into the liquid from the top. As soon as the displacer touches the liquid its weight is reduced because of the buoyant force of the liquid. This weight change is recognized by the system and interpreted as liquid level. No additional hydrostatic pressure sensor was used in the Target volume. 
 
The differential liquid level between all volumes was measured by applying a common vacuum to suck up the four liquid levels (above the detector) into a ruled area. Since the densities in all the liquids were identical, the differential liquid levels remained the same at any absolute height. This system provided an important cross-check to make sure the liquid level difference between neighboring volumes was never more than the specified 3~cm.

The basic concept for the gas pressure monitoring was to monitor the absolute and differential gas pressures with precision better than 1\,mbar. One system monitored the absolute gas pressure; the other was focused only on differential pressures in the detector. These independent measurements provided the possibility to cross-check the pressure values in the detector. The sensors used for this system had an accuracy of 0.1\,mbar with a resolution of 0.01\,mbar. The system was installed on the top lid of the Inner Veto. In addition to these sensors, the atmospheric pressure was monitored and all information was collected at a monitoring PC in each neutrino laboratory. 

A temperature measurement system was designed to check the Target thermal stability during detector filling. It measured the temperature at five levels from the bottom to the top of the Target vessel. The temperature sensors were welded to stainless steel mounts separated by flexible elements. All material was enclosed in hermetically sealed PTFE hose. The cables ran out from the detector within the PTFE hose. This system as well as the level measurement systems for the Target were removed after filling was completed. There were several temperature sensors inside the GC, Buffer and IV volumes for continuous monitoring of the temperature in the detector. Temperature fluctuations inside the detector were within 1$^\circ$C over the full year. 

Temperature was also measured by  sensors inside the Buffer. Each detector was equipped with 12 magnetometers (Bmons) used to monitor magnetic fields as well as temperature. Two sensors each were attached to the bottom and top surfaces, with an additional eight on the vertical wall. These Bmons were mounted on acrylic plates approximately 30\,cm in length and 8\,cm in width. They were secured to mounting rails via two stainless steel bolts passing through each plate. The cable passage into the Buffer used the same penetrations as the PMT cables.

\section{Photomultiplier Tubes}

\subsection{\label{sec:OverView} Overview of the inner PMT system}
\begin{figure}
  \centering
  \includegraphics[width=0.8\textwidth]{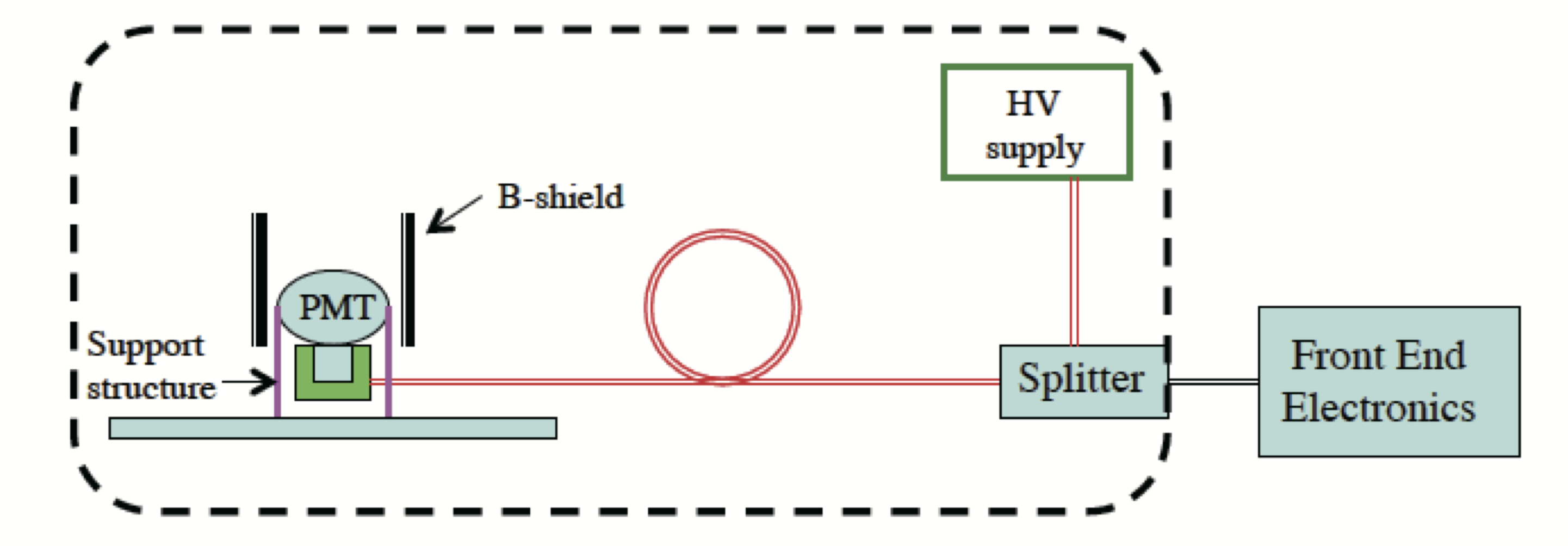}
 \caption{Schematic of Inner PMT system.}
 \label{fig:IPMT_System}
\end{figure}

The inner PMT system was one of the core parts of the DC detector, which detected the scintillation light and gave information about the energy and timing of the signals. Each DC detector used 390 low-background PMTs of 10~inch diameter, uniformly arranged around the interior of the cylindrical Buffer oil tank. The total photocoverage was $\sim$ 13\% and each PMT received from 0 to a few photoelectrons (PE) from the neutrino-induced signal, depending on its energy and vertex position. Figure~\ref{fig:IPMT_System} shows a unit of the PMT system, which included a PMT with its magnetic shield, support structure, high voltage (HV) supply,  HV/signal splitter, and cables.  
 
The PMTs were immersed in the transparent Buffer oil. A 22\,m long coaxial cable was used for both HV transfer and PMT signal readout. Therefore, a HV/signal splitter circuit was required between the HV supply, PMT and electronics.  
%
\subsubsection{\label{sec:PMT} The Photomultiplier}
\begin{figure}[htbp]
 \begin{center}
  \includegraphics[width=0.8\textwidth]{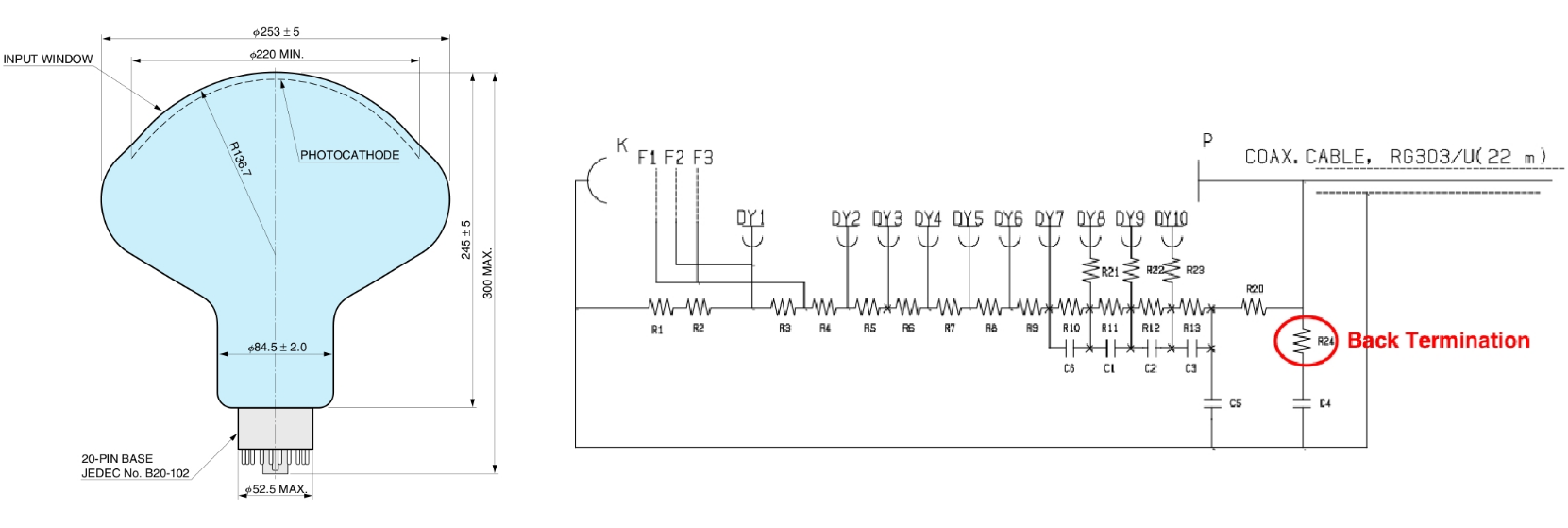} 
  \hspace{0.5cm}
  \caption{Schematic of the PMT structure (left) and circuit diagram of the base \cite{DCiPMT_JINST_2016_Abe}\cite{GermanPMTpaper}.}
  \label{fig:PMT_struc}
 \end{center}
\end{figure}

The PMT was a low-background type Hamamatsu Photonics K.K. (HPK) R7081 with oil-proof base. Figure~\ref{fig:PMT_struc} shows a schematic of the PMT structure and the circuit diagram of the PMT base. The diameter of the glass is 253\,mm and the length of the tube is 308~mm. The diameter of the photocathode area is at least 220\,mm. The average weight of the glass part is 1.06~kg to be compared to a total weight of 2.5\,kg. The cathode material is bi-alkali with about 25\,\% quantum efficiency (QE) for 420\,nm light. The QE spectrum matches well with the bis-MSB light emission spectrum. 
   
The PMTs were operated with positive HV. The signal was transmitted through the  coaxial HV cable with no DC decoupling capacitor in the base circuit. 
A back-termination resistor (R24) was connected to the anode. 
It reduced the amplitude of the signal by half but quickly damped multi-reflections of signals caused by impedance mismatch. The PMT base was encapsulated in a transparent epoxy resin covered by an acrylic housing. During the commissioning of the far detector we observed the emission of light around the base circuits of some PMTs. The rate of this light emission correlated with the applied high voltage and the ambient temperature. However, these light noise~\cite{DCiPMT_JINST_2016_Abe} events could be efficiently removed in the analysis. In the near detector each PMT base was covered with a black sheet to mitigate this light noise effect. As PMT cable, a Teflon jacketed cable with an impedance of 50~$\Omega$ (RG303) was used.  
 
Radioactive gamma-ray emission from the PMT glass was one of the dominant sources for the singles background counting rate of the DC detectors.
The glass of the DC PMTs has very low radioactivity because the sand from which it was made was carefully selected and was melted in a platinum coating pot in order to prevent contamination from the pot wall. The contaminations of radioactive elements were measured to be U:~13\,ppb, Th:~61\,ppb and $^{40}$K:~3.3\,ppb. The low background rates enabled the total neutron capture analysis for the DC $\theta_{13}$ measurement~\cite{DC_Nature}. 

\subsection{Inner PMT magnetic shields}
\label{shield}

An external steel shield surrounded the DC far detector. The steel pieces were demagnetized with the goal of obtaining a magnetic field inside the detector below 0.5 gauss. However, the effective magnetic field level and its direction once the detector was closed were unknown and expected to  differ for the far and near detectors, since the near detector did not have an external steel shield.

Passive individual PMT magnetic shields were designed to protect the PMTs from the unknown B-field present in the detector. Various shield materials and configurations  were tested \cite{paperciemat1}. 
The charge response of the Hamamatsu R7081 PMTs under magnetic fields was measured in a dedicated experimental setup \cite{paperciemat2}. A 0.5 mm thick cylindrical mu-metal shield of 275 mm height and 300 mm inner diameter was chosen. The shield extended beyond the highest point of the PMT by 5.5\,cm. 
Cylinders were spot-welded and heat treated at 1080$^\circ$C for three hours after fabrication. 

\begin{figure}[ht]
\centering
\includegraphics[width=0.48\textwidth]{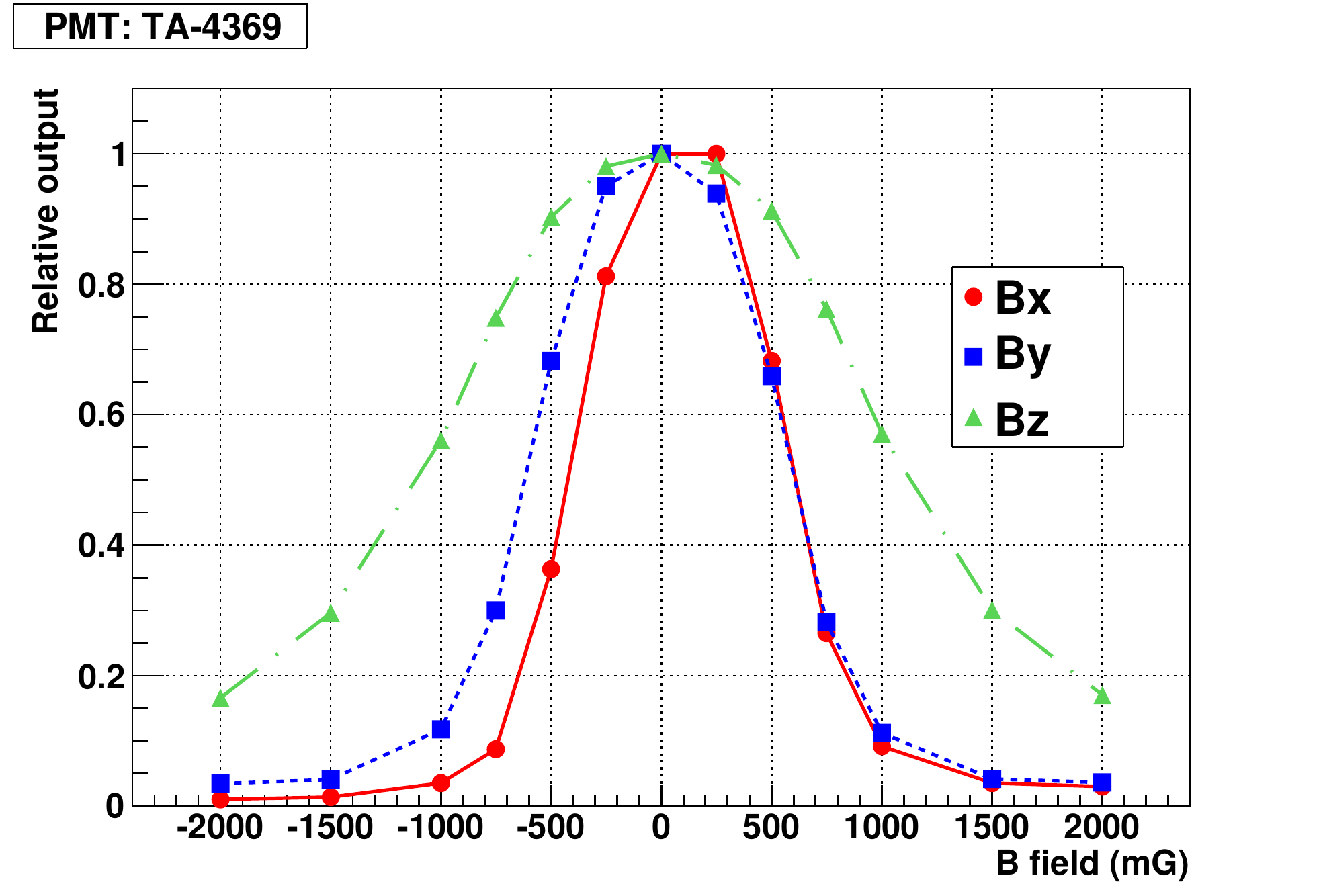}
\includegraphics[width=0.48\textwidth]{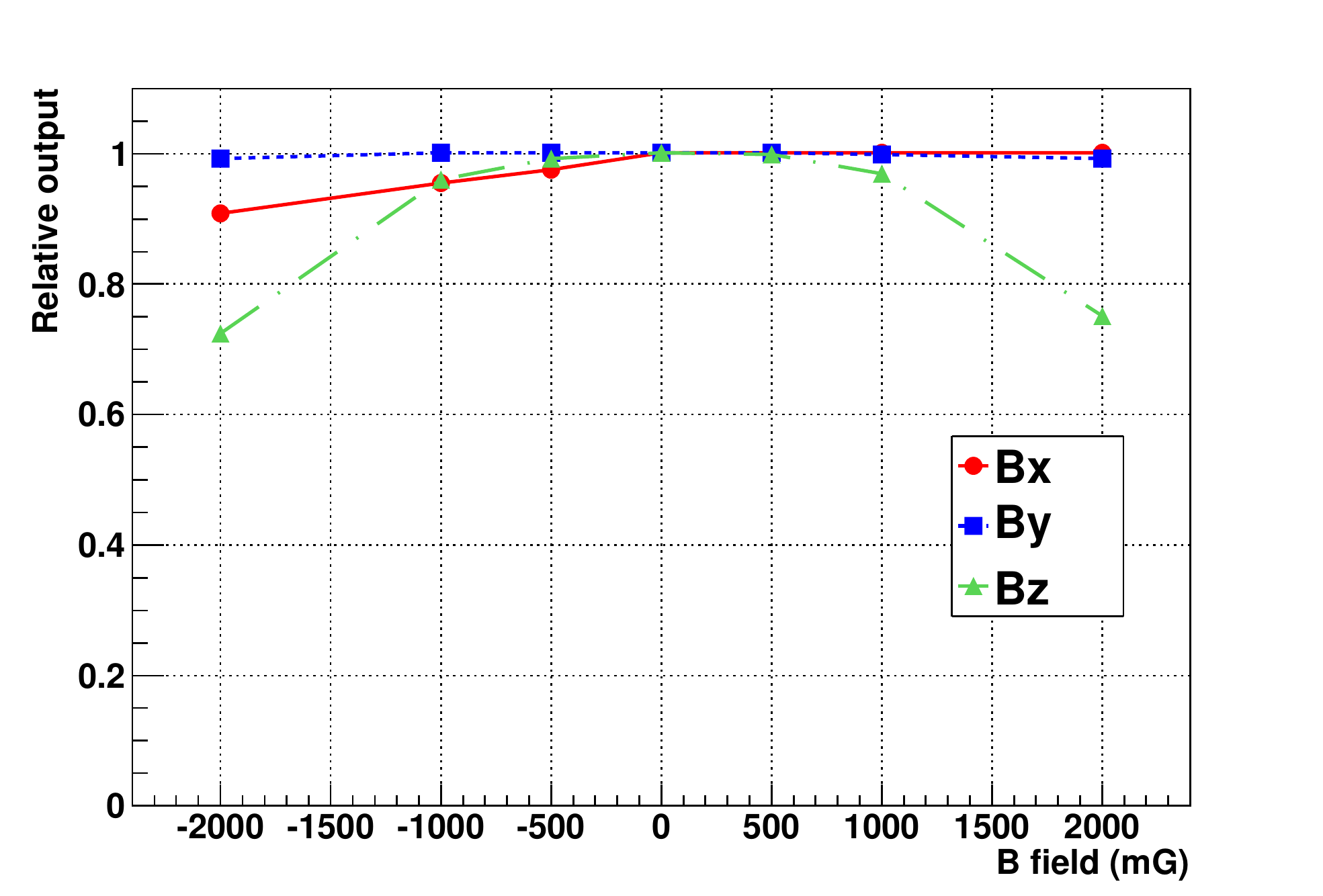} 
\caption{Relative charge response of the Hamamatsu R7081 PMT with respect to that at zero B-field for various values and directions (X and Y are the transverse directions and Z the longitudinal direction with respect to the PMT) of the magnetic field before (top) and after (bottom) the mu-metal shield installation.}
\label{shield_perf}
\end{figure}

Figure~\ref{shield_perf} shows the variation of the PMT charge response under various magnetic fields with and without the mu-metal shield. A B-field of 500\,mG applied transverse to the PMT reduces the PMT signal by 20-60\,\%. 
Using the magnetic shield, the loss of the PMT signal with respect to zero B-field is less than 10\% for 1~gauss external B-field, meeting our requirements.

A random sample of 10\% of the mu-metal shields were tested after fabrication to verify the expected performance of the shield in terms of shielding factor along the longitudinal axis of the cylinder. All the shields passed the quality tests and met specifications.

\subsection{The Inner PMT support structure}
\label{supp}

The PMT support structure was required to hold each PMT individually allowing its orientation towards the Target center with a precision better than 2$^\circ$ and a tolerance on  PMT center position better than 2~cm. The DC PMT support structure consisted of two parts: an acrylic structure holding the PMT and its magnetic shield, and a stainless steel structure affixing the acrylic one to the detector wall.

\begin{figure}[ht]
\centering
\includegraphics[height=4.3cm]{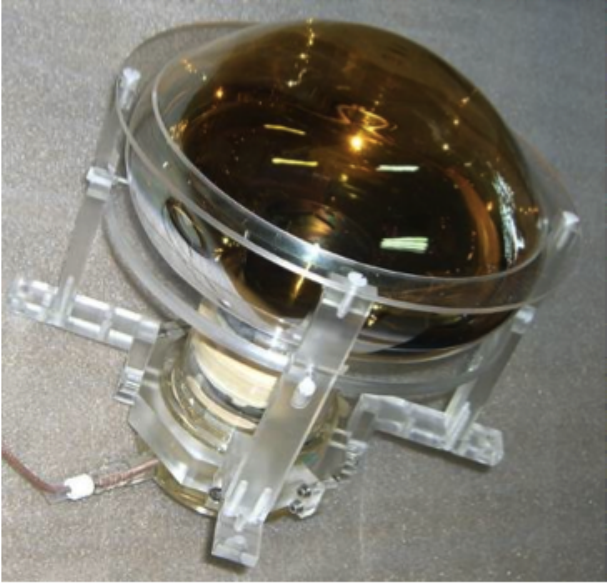} 
\includegraphics[height=4.3cm]{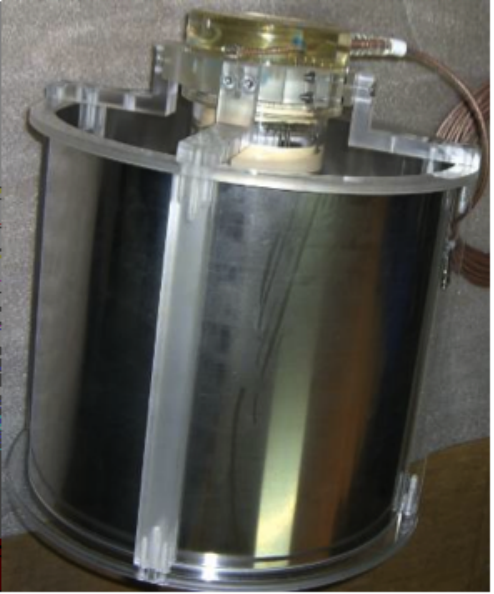}
\caption{PMT acrylic support structure (left) and assembly with mu-metal shield (right).}
\label{support}
\end{figure}

The acrylic structure clamped the PMT in three areas: just above and below its equator and at the oil-proof base. Figure~\ref{support} shows photos of the mechanical support assembled with the PMT. 
The material used was acrylic GS233, similar to that used for the Target and GC vessels, and compatible with the Buffer oil. 
For the  DC near detector PMTs, the acrylic base was covered with a black polyester film in order to reduce the light noise emission observed in the far detector PMTs. The acrylic PMT support structure was fixed to the Buffer walls by intermediate U-shaped stainless steel pieces attached to its bottom part. 
These pieces were attached to stainless steel vertical struts placed around the cylindrical Buffer walls and to stainless steel ring structures in the top and bottom Buffer lids. 

\begin{figure}[ht]
\centering
\includegraphics[width=0.7\textwidth]{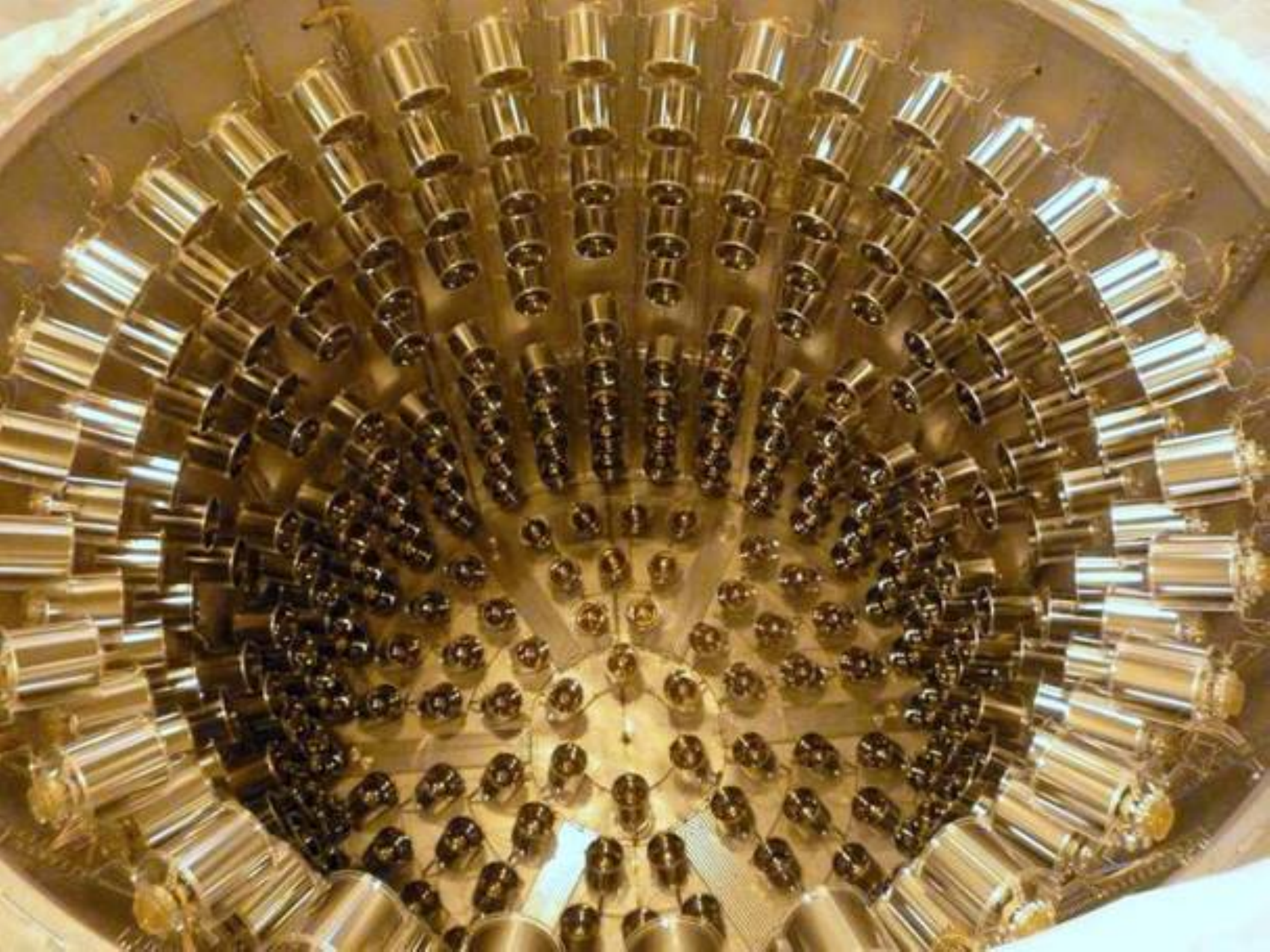}
\caption{Distribution of the PMTs inside the Buffer tank.}
\label{total}
\end{figure}

The position and orientation of the 390~PMTs were optimized for a homogeneous detector response. They were distributed in 10 rings on the wall with 15~PMTs in the two central rings and  30~PMTs per ring otherwise. The distributions on the bottom and top Buffer lids were the same with 60~PMTs divided in six 60$^\circ$ wedges. Figure~\ref{total} shows the final distribution of the PMTs installed in the detector. To assure uniform and identical detectors the PMT serial numbers and anode orientations with respect to the supporting frame were randomized. The final positions of all the PMTs after their installation in the Buffer were surveyed and registered in our database. Four survey photogrammetry targets were installed on the front acrylic ring of each PMT support and removed after the survey. All the materials passed exhaustive cleanliness procedures and radiopurity tests. 

\subsection{Inner Veto PMTs} 
\begin{figure}[ht!]
    \centering
    \includegraphics[width=0.4\textwidth]{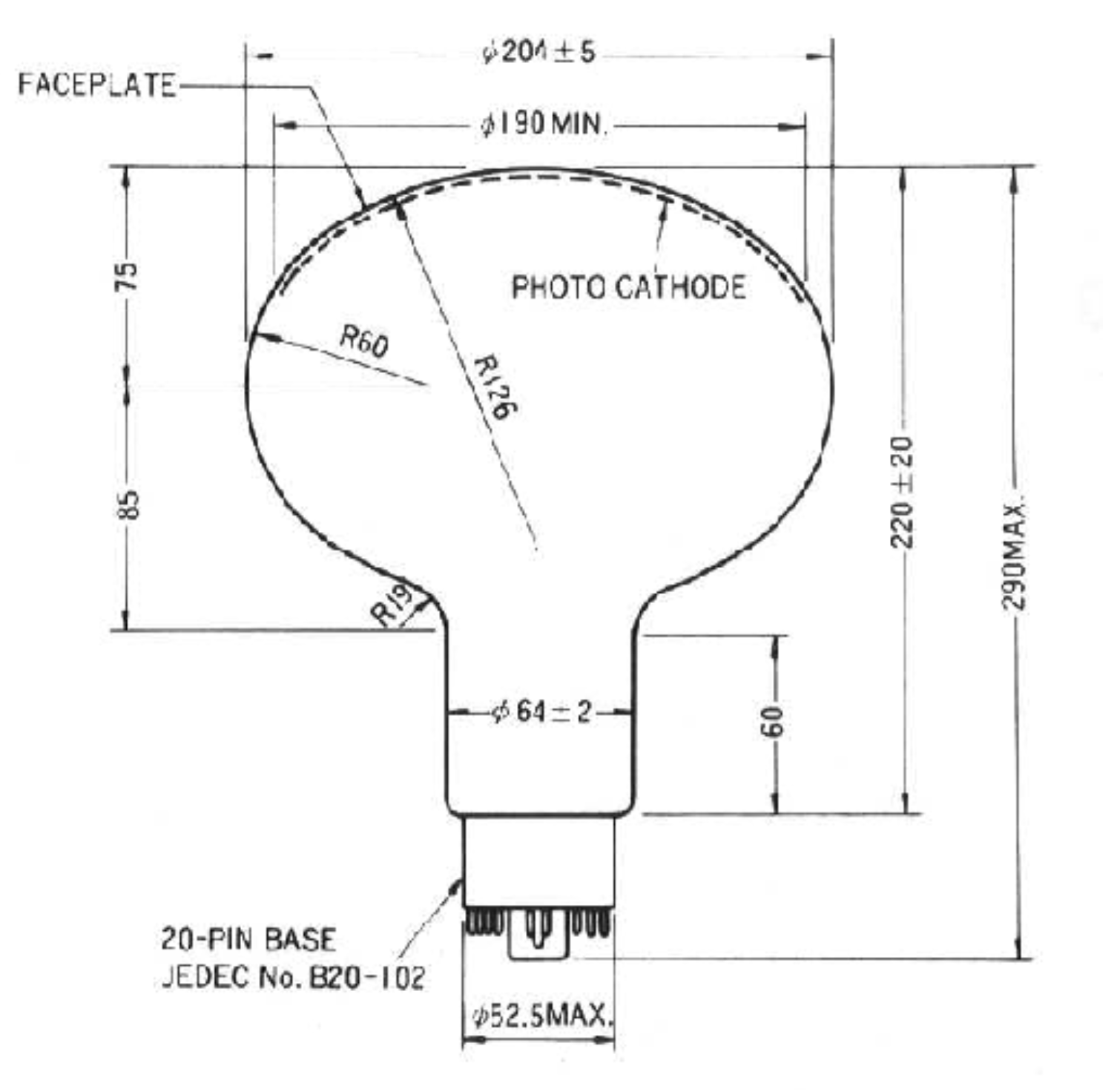}
    \caption{Dimensions and geometry of the Hamamatsu R1408 photomultiplier tube.}
    \label{fig:dc_veto_pmt_ham}
\end{figure}

The 78 (8-inch) IV PMTs per detector were selected from a larger sample of Hamamatsu R1408 tubes that were available from previous experiments (IMB, Super-Kamiokande). Figure~\ref{fig:dc_veto_pmt_ham} shows their dimensions and geometry. The selection was based on agreed test procedures and quality criteria.

\begin{figure}[ht!]
    \centering
    \includegraphics[height=6cm]{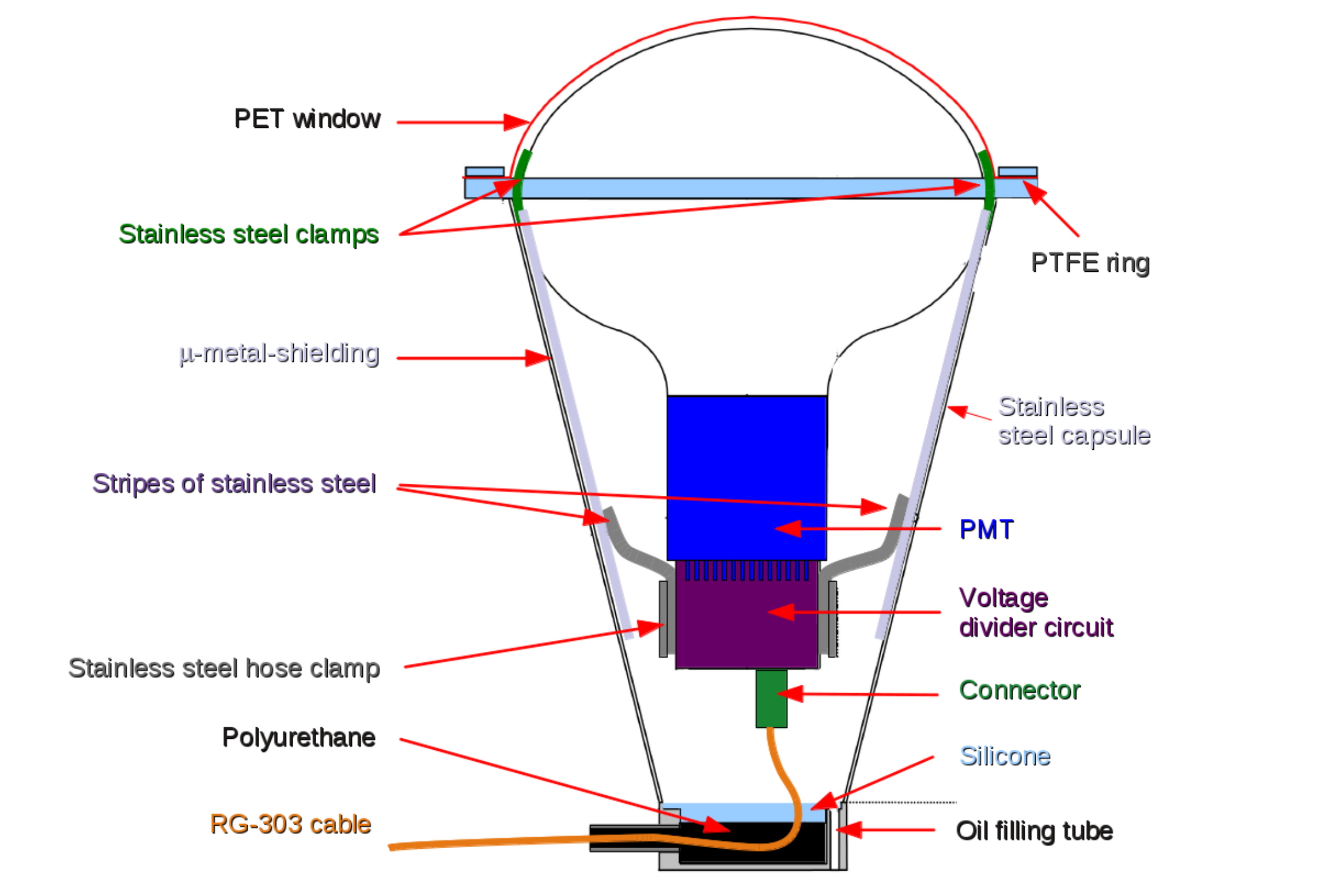}
    \includegraphics[height=6cm]{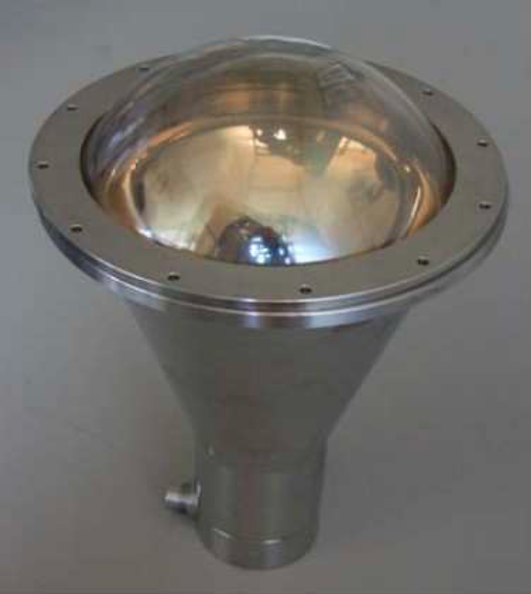}
    \caption{Stainless steel enclosure with a transparent PET window, an oil-tight cable feedthrough, mu-metal shielding, the Hamamatsu R1408 PMT, and the PMT base. The inner volume was filled with mineral oil.}
    \label{fig:dc_vet_pmt_encaps}
\end{figure}

Each PMT was encapsulated in a stainless steel enclosure with a transparent PET front window, filled with spectroscopy quality mineral oil matching the index of refraction of the scintillator (see fig.~\ref{fig:dc_vet_pmt_encaps}). A PTFE ring at the flange and a layer of polyurethane at the cable feedthrough sealed the capsule. Steel clamps near the flange of the PET window, and in the narrow part of the encapsulation where they were attached to the socket of the base PCB, held the PMT in place.

\begin{figure}[ht!]
    \centering
    \includegraphics[width=0.7\textwidth]{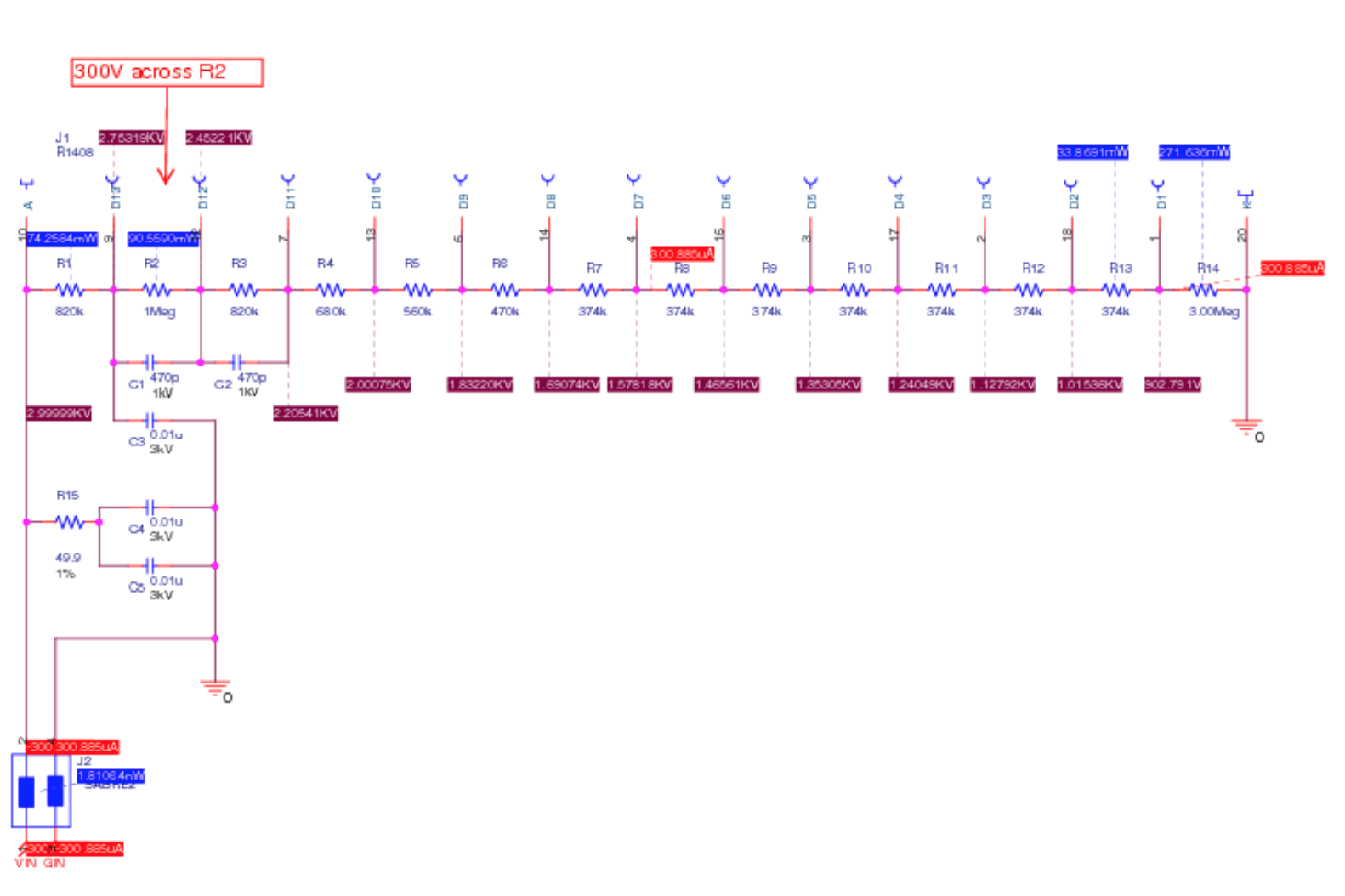}
    \caption{Schematic diagram of the inner veto PMTs' base circuit.}
    \label{fig:dc_base_schematic.png}
\end{figure}

A single RG-303 coax cable with 50\,$\Omega$ impedance connected to the base PCB (see fig.~\ref{fig:dc_base_schematic.png}) carried the high voltage and the PMT signal. At the other cable end, the same voltage splitters as for the inner detector were used. PMT magnetic shields to protect the PMTs from the residual magnetic field were fabricated as truncated cones of mu-metal (company Meca Magnetic) by spot-welding and thermal annealing. With a length of 200\,mm and diameters of 216\,mm at the top and 119\,mm at the bottom, they fit into the steel encapsulation such that the upper edge was at the equator of the PMT glass bulb. Laboratory measurements in a dedicated setup confirmed that the reduction in charge response from the expected remaining field at the IV PMTs' positions was less than 10\%~\cite{phdgreiner}.  

\subsection{High Voltage System for ID/IV}
\label{sec:HV_PMT}

As described in section~\ref{sec:OverView}, a single cable carried the HV and the PMT signal. Therefore, a custom circuit to split the signal from the HV was designed and fabricated. To reduce the noise from the HV power supply, there was a low pass filter at the HV input, and to reduce electromagnetic interference and crosstalk, each splitter circuit was enclosed in an individual aluminum box connected to ground.

\begin{figure} 
\begin{center}
\includegraphics[width=0.8\textwidth]{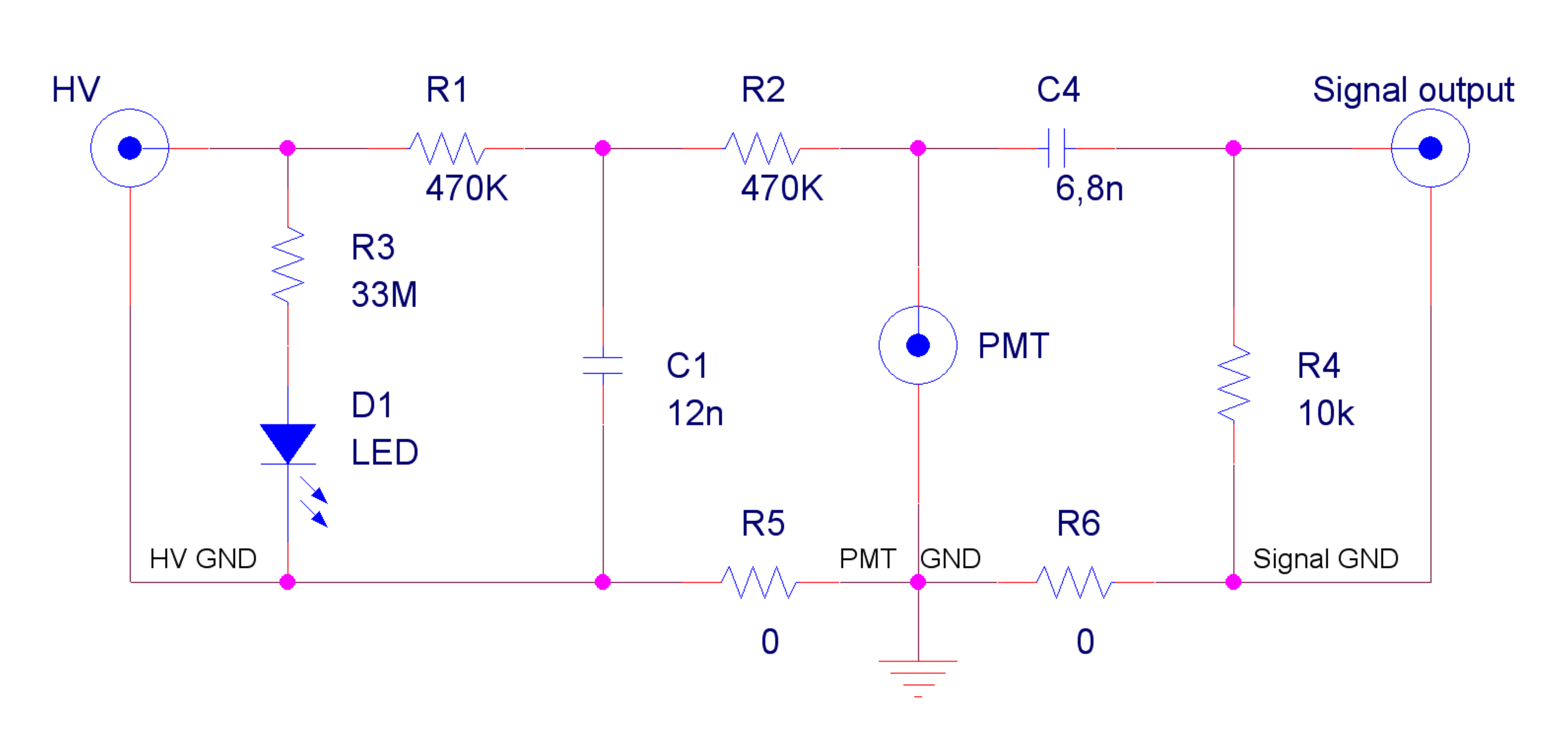}
\caption{Splitter circuit diagram. R5 and R6 are for on-site modification of ground connection resistance for possible ground-loop problems.}
\label{fig:HV:Splitter1}
\end{center}
\end{figure}

Figure~\ref{fig:HV:Splitter1} shows the splitter circuit diagram.
R1 and C1 form the low pass filter with a cutoff frequency of 28~Hz.
C4 separates the signal and the HV and R2 prevents this signal from going to ground through C1. Finally R4 prevents  voltage spikes if the splitter is  disconnected from the Front-End Electronics.

The HV power-supply modules used in DC (CAEN A1535P, see section~\ref{sec_elec}) have hardware and software limits for output voltage.
The HV module is able to monitor its output voltage with 0.5~V precision.
To ensure consistency between the monitored and actual output voltage,
the output voltage was measured by a voltmeter at the splitter input.
Moreover,  module calibration was performed using a special module developed by CAEN.
This module has the same structure as the HV module, but with voltmeters installed in place of HV chips.
The deviation between monitored and output voltage was found to be within
0.2\% after calibration.
Further details on this system are given in~\cite{DCiPMT_PhysPro_2012_Sat}.

\subsection{PMT calibration}
\label{sec:IMPT_Tests}
In total 803 PMTs were tested and characterized to obtain a knowledge of their behavior and signal responses. The results were also used to improve the Monte Carlo simulations of the DC detectors. The calibrations were performed by two independent groups.

The German test facility at MPIK was able to calibrate 30~PMTs simultaneously. 
The PMT array was placed inside a Faraday-shielded dark room. 
Each PMT was surrounded by a cylindrical $\mu$-metal shield. 
To avoid crosstalk, all analog and digital electronics were placed outside the Faraday cage. 
The PMT cables were fed through a hole in the Faraday cage walls and connected to splitter boxes to separate the signal and HV. 
An LED triggered, scintillator-filled quartz ball served as a central light source. 
In addition, each PMT could be illuminated individually by an LED system. 
More details and results can be found in~\cite{GermanPMTpaper}\cite{pmt_QT}\cite{DCiPMT_JINST_2013_Has}.

The test system used by the Japanese groups had two steps at different facilities. 
The first test step, performed in Japan, used an individual PMT evaluation system. 
A picosecond laser pulser and 7 LEDs along a rotational arm were used as light sources. 
A light source having about 430\,nm wavelength, nearly equal to that of our scintillator, was chosen. 
The second testing step was performed at MPIK to check for possible damage due to PMT transport. 
This facility deployed 8 optical fibers from one LED to test 8 PMTs simultaneously. More details and results can be found in~\cite{DCiPMT_NIMA_2012_Mat}.
\begin{figure}[htbp]
\begin{center}
\includegraphics[width=0.6\textwidth]{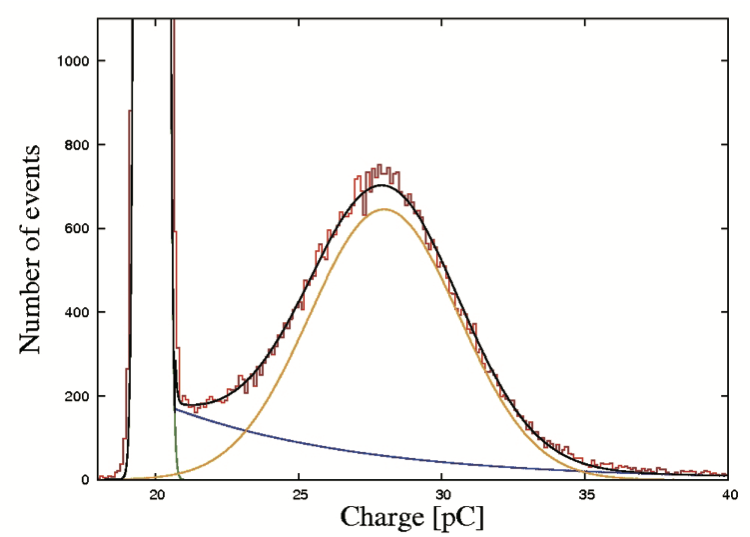}
\caption{Typical single photoelectron charge response~\cite{GermanPMTpaper}. Pedestal and SPE peaks are fit by two gaussians, adding an exponential to describe the valley region.}
\label{fig:SPE-spectrum}
\end{center}
\end{figure}

\begin{figure}[htbp]
\begin{center}
\includegraphics[width=0.5\textwidth]{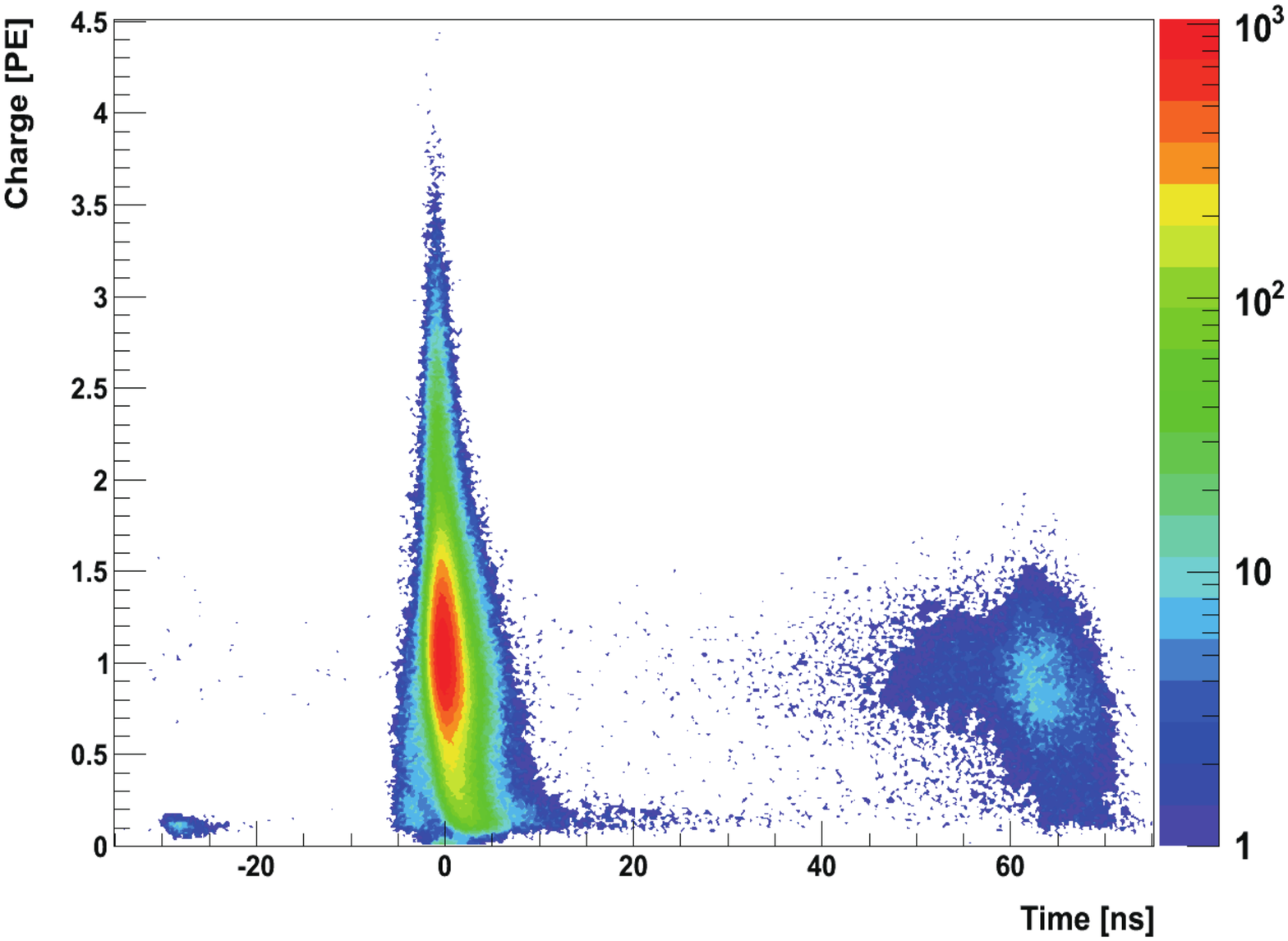}
\includegraphics[width=0.45\textwidth]{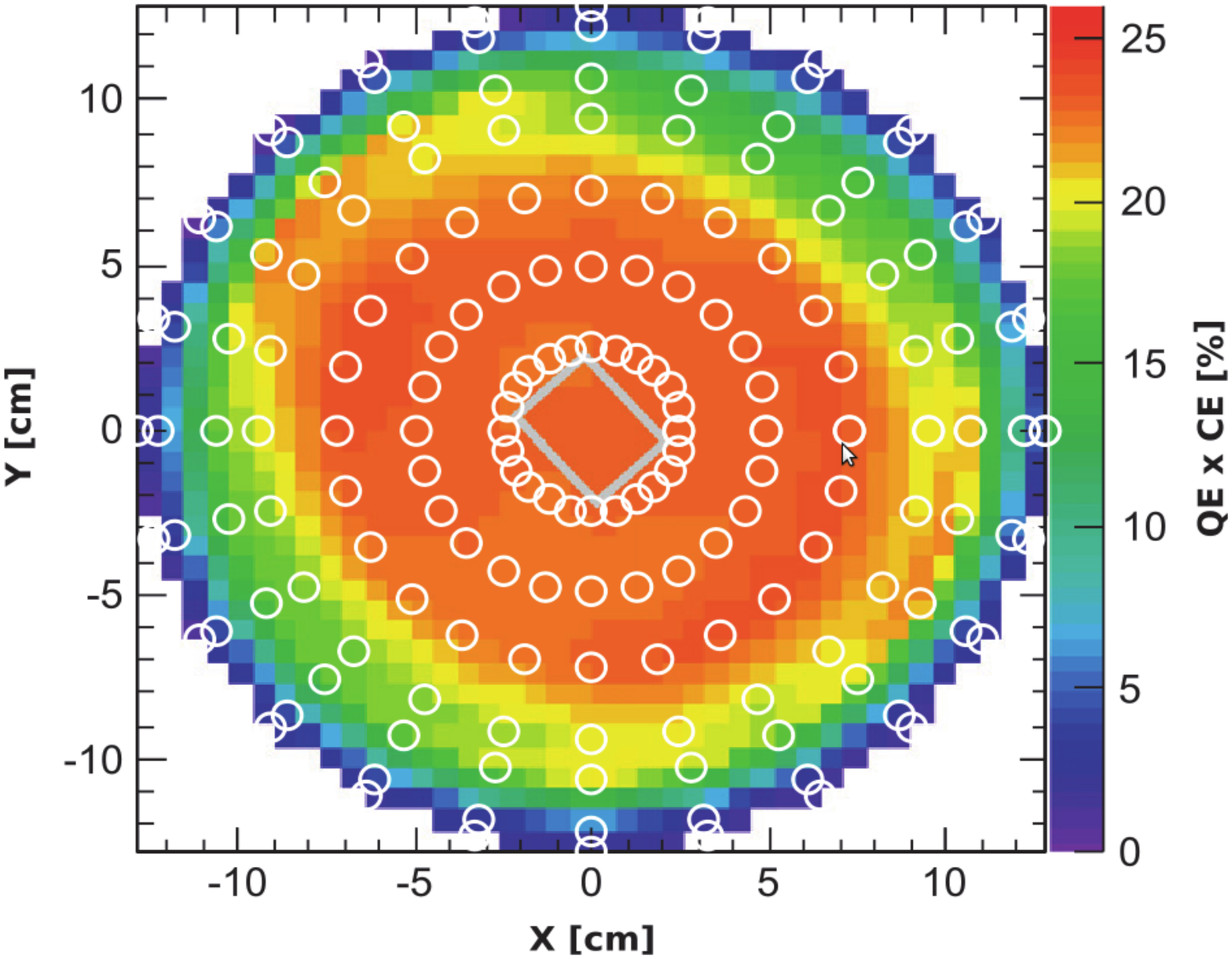}
\caption{Transit time versus charge plotted event by event \cite{pmt_QT} (top). 
One can identify main pulses around $t = 0$\,ns with a transit time spread of $\approx 3\,$ns, pre-pulses around $t=-25$\,ns and late-pulses up to $t=70$\,ns. The fraction of single PE in the measured population was about 95\%. The efficiency map (bottom) is shown for a typical PMT surface \cite{DCiPMT_NIMA_2012_Mat}. The efficiency values are the product of two components, the quantum efficiency (QE) and the collection efficiency (CE) of the conversion electrons.}
\label{fig:TT-QE}
\end{center}
\end{figure}

The calibrations included HV scans to determine the nominal operating HV value at a gain of $10^7$, single PE response regarding energy resolution and peak-to-valley ratio. Figure~\ref{fig:SPE-spectrum} shows that the typical peak-to-valley ratio was greater than 3.5.
The transit time distribution including pre- and late-pulse probabilities is shown in fig.~\ref{fig:TT-QE} together with an efficiency map for the PMT's photocathode. This efficiency map was also measured by using a standard PMT provided by Hamamatsu HPK as reference. Finally, the calibration program included a determination of dark count events, which are caused by thermal emission of electrons from the photocathode. The measured average rate per PMT of 2200~counts/s was well below the DC specification of 8000~counts/s.~\cite{GermanPMTpaper}
About 100 PMTs were measured in both systems to cross-check the results in order to rule out systematics between the different calibrations.

\section{Electronics and DAQ}
\label{sec_elec}

\begin{figure}
\begin{center}
\includegraphics[width=0.72\textwidth]{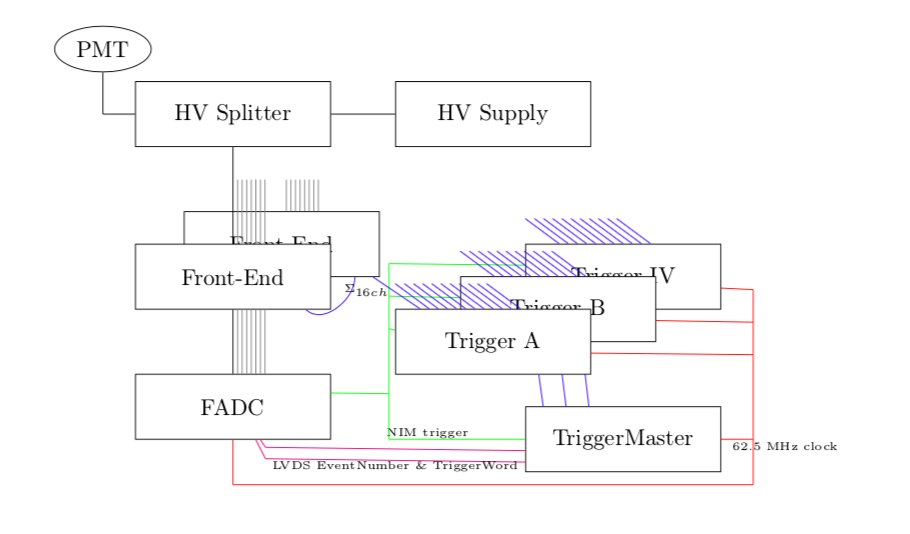}
\caption{Double Chooz simplified electronics scheme.}
\label{fig:electronics}
\end{center}
\end{figure}

Figure~\ref{fig:electronics} shows the DC electronics scheme for both the 390 inner detector PMTs and the 78 IV PMTs per detector. Identical electronics was used for each detector, and each detector was operated independently.

The HV of each PMT was supplied by a CAEN A1535P High Voltage supply unit, allowing each inner detector PMT to be operated at a gain of $10^7$.  Each PMT signal was decoupled from the HV cable by the HV splitter (see section~\ref{sec:HV_PMT}), amplified by the Front-End electronics (see section~\ref{fee}) and sampled synchronously at 2 ns by waveform digitizers or an FADC system (see section~\ref{fadc}). The Front-End electronics also summed and shaped groups of up to 16~PMTs forming output signals whose amplitude was proportional to the charge observed per group. These signals were fed into the DC trigger system (see section~\ref{trigger}) which generated a global trigger if threshold conditions on the summed amplitude of all groups and multiplicity conditions were met. As well as sending a NIM trigger signal, the trigger system also sent information on the events in the form of two 32-bit LVDS words: the first contained the EventNumber and the second contained information about the trigger known as the TriggerWord.

Both the Trigger and FADC Systems ran via VME bus, located within 6 VME crates, with each crate controlled by a single-board computer (MVME3100 or MVME7100 PowerPC) running Debian Linux with Data Acquisition software written in Ada.

\subsection{Front-End Electronics}
\label{fee}
The Front-End Electronics (FEE) used an 8-channel custom-made NIM module which amplified the PMT signals, matched their dynamic range to the FADC, and also provided dedicated signals for the trigger system.

For the first period of running, with far detector only, the FEE operated with a gain of $\sim$\,7, such that after amplification the mean single photoelectron (PE) amplitude was $\sim$\,35\,mV with an analog RMS noise level of $\sim$\,1.2~mV at the output of the amplifier. During the first phase of running (FD-I), a nonlinearity in the reconstructed charge of the digitized PMT pulses was observed and found to be due to an insufficiency of the input analog noise level compared to digitization step size of the FADC~\cite{FADC}. The FEE gain was therefore doubled for the two-detectors phase and the noise level increased to $\sim$1.6\,mV. Both effects reduced the observed nonlinearity.

A dedicated output for the trigger system was also provided. First, a stretched and summed analog signal from the 8 input channels was put out for each NIM module. A summing circuit then allowed chaining two Front-End NIM modules together, resulting in a summed and stretched signal for up to 16 input channels. These signals went to the trigger system to allow a trigger decision based on the deposited energy for each group of channels.

\subsection{Flash ADC}
\label{fadc}

The eight amplified outputs of each Front-End amplifier were digitized by an 8-Channel waveform digitizer developed specifically for the experiment in partnership with CAEN, which is commercially available as model Vx1721 (VME64x) or V1721~\cite{CAENV1721} (legacy VME). It is based on an 8-channel, 8-bit, 500\,megasamples per second FADC~\cite{FADC_chip}.   Dedicated in-house firmware allowed the synchronous running of these cards, with a flexible readout which suffered no deadtime in part due to the implementation of the fastest VME64x transfer protocol, 2eSST, at a maximum speed of 320\,$\mbox{MiByte/s}$. Additionally, geographical addressing is supported, which is important for dealing with a large number of cards.  Futher detail on the hardware, firmware and performance of this card can be found in~\cite{FADC}.  

A total of 49 FADC cards were used to digitize the inner detector PMT signals, with a further 10 used for the IV PMT signals. In order to balance the loading on each crate, 4 VME crates dedicated to the FADC signals each held 12 or 13 cards, while one crate holding the trigger system also contained 6 cards receiving IV PMT signals. 

The dynamic range of the Vx1721 is 1\,V, such that a typical ADC step is $\sim$ 4\,mV. A typical single PE signal had an amplitude $\sim$ 8 (16) ADC counts (depending on running phase). Typically, the number of PE per channel over the energy range of interest in DC was low, $\sim$~1~PE per PMT, for an energy deposit of 2\,MeV in the inner detector, so channels did not saturate. The duration of the scintillation signals was less than 200\,ns, such that a waveform of 256\,ns was long enough to encompass the signal while providing sufficient baseline to calculate the pedestals.

The use of FADC was one of the unique features of DC as a $\theta_{13}$ experiment. It enabled the collaboration to get unique results on several topics such as pulse shape analysis to reject background events related to fast neutrons~\cite{DoubleChooz:2015mfm}, event classification based on the spectral analysis of scintillation light~\cite{DoubleChooz:2017cqd} and a measurement of ortho-positronium formation~\cite{DoubleChooz:2014iyf}.

\subsection{Neutrino Data Acquisition}
\label{daq}
The Neutrino DAQ system included two software packages: the Read Out Processor (ROP) and the EventBuilder. The ROP software ran on each single-board computer controlling a VME crate. Commands were sent from the EventBuilder to the ROPs to configure, initialize, begin and end acquiring data. During acquisition, each ROP sent data to the EventBuilder, which combined the data from each ROP to form a complete event. The EventBuilder processed the data using rapid online algorithms, determined whether to store the complete event or not, and wrote the event to disk. Additionally, information about the run and the data collected was recorded periodically via a ROOT macro, allowing the shifters to identify technical problems rapidly. 

\subsubsection{Read Out Processor}
\label{rop}
The ROP software was written in Ada with the GNAT 4.3.2 compiler from Ada Core Technologies, running on a single core MVME3100 PowerPC~\cite{mvme3100} for the data acquisition of the Near Detector and a faster dual-core  MVME7100 PowerPC~\cite{mvme7100} for the Far Detector system.  The home areas of the PowerPCs were stored on an external server, such that the home areas on all the PowerPCs were identical.

An Ada VME library was developed providing a  low level interface to the VME driver for the PCI--VME bridge Tsi148, distributed by Emerson-Network-Power. High level interfaces were provided to manage master and slave windows on VMEBus, and allowing VME DMA transfer. In addition, interfaces to the Linux pread, pwrite, ioctl, select and pselect system calls were provided.

The ROP software comprised two tasks; the first communicated  with the EventBuilder and the second with the VME cards held within its crate. The ROP configured the cards contained within its crate, polled for triggers, and retrieved data from the cards (FADC and/or Trigger cards). It packed multiple events into blocks and sent them to the EventBuilder via Ethernet. 

In order to maintain deadtimeless operation at trigger rates of $\sim$500\,Hz, the ROP had to transfer data efficiently from the cards' buffers to memory.  For this, the ROP heavily used chained Direct Memory Access (DMA) transfer for both reading and writing data to multiple cards, allowing the CPU to do other operations such as writing data to the EventBuilder while the transfer was in progress.  Block transfer of the waveforms from the FADCs was made using the double edged Source Synchronized Transfer (2eSST) giving transfer speeds of 320~MB/s (fastest). For the trigger system, Block Transfer (BLT) was used giving transfer speeds of 40~MB/s. 

\subsubsection{EventBuilder}

The EventBuilder software ran on a DELL PowerEdge R610. It communicated with the Run Control server, sent commands to the ROPs, received data from the ROPs, assembled the data, performed simple analyses on the waveforms, and zipped and wrote the data and run-wise monitoring information to disk.

After data were collected from each ROP, the waveforms were analyzed online in 5 parallel processes. The online analysis served two main purposes. The first was to provide reduced channel-wise data to be used by the shifters to determine whether the detector was functioning normally or not. The second was to reduce the quantity of data written for light noise~\cite{DCiPMT_JINST_2016_Abe} events at the far detector and muon events at the near detector.

The combination of the online data reduction and parallelized zipping of the data was necessary to reduce the quantity of data written to disk, transferred offsite and eventually analyzed. 

\subsubsection{Online Data Analysis} 
\label{minidata}
Waveform analysis was performed on each channel by the EventBuilder, determining:

\begin{itemize}
\item Waveform Pedestal
\item Charge --- using the full integration window
\item Pulse Start Time at which the negative pulse crossed threshold (2 ADC counts below the Pedestal)
\item Time Over Threshold 
\item Minimum ADC value of waveform
\end{itemize}
 
For pulses with amplitudes larger than 6 ADC channels (3 times the Threshold setting), the following variables dedicated to the muon reconstruction were calculated:

\begin{itemize}
\item Interpolated Rise Time ($T_{90} - T_{10}$)
\item Interpolated 50\% Start Time $T_{50}$
\end{itemize}

These variables were written to disk for all events.
With these channel-wise variables, event-wise variables were built and used to determine the event type on-the-fly for monitoring purposes and for subsequent data reduction. 

\subsubsection{Online Data Reduction}
The FADC waveforms comprised the dominant data written to disk. In order to reduce the quantity of data written to disk, for certain types of events only the online analyzed variables were stored (i.e.,~the FADC waveforms were removed). To make a significant reduction in the data volume, FADC waveforms were not written for 
\begin{itemize}
\item Light noise at the far detector --- determined by the event-wise variable that measured the spread of the inner detector pulse start-times (RMS\_TStart $>$70). This online cut removed waveforms from approximately two thirds of light noise events (175 Hz).
\item Muons at the near detector --- determined by the IV muon bit from the trigger system. Waveforms were removed for $\sim 190$\,Hz of muons and  stored for only about 1\% of the muons to perform dedicated studies.
\end{itemize}

At the end of data taking, the far detector wrote a 13\,GB/hourly run for a trigger rate of 470\,Hz, and the near detector wrote a 9\,GB/hourly run for a trigger rate of 275\,Hz.

\subsubsection{Data handling and transfer}

The DC Collaboration relied on the database software MySQL for the collection, handling and distribution of various data in real time. These included detector monitoring data, metadata regarding the processing 
and storage of physics data files, and metadata regarding simulation data files, as well as performance data for several detector subsystems. Several instances of MySQL servers were implemented in different locations to provide a fast and fail-safe data handling system~\cite{PhD_Schopp}.
For the two-detector phase, a system built from seven servers was used to provide the database infrastructure. At each detector site, two servers were installed. These were the master servers of the laboratories and their backup servers. The remaining three servers were located at a computing centre (centre de calcul of the French IN2P3). Two servers duplicated the data from the laboratories and allowed access locally at the computing centre. The third one was used as a standalone server for data produced at the computing center. Looking at the long-term performance of the system, we find an overall good performance with an average daily downtime of less than 1.5\,sec.

After the online data reduction, binary data written to disk were sent to the IN2P3 computing center in Lyon, France. The data transfer system was developed in Python using iRODS~\cite{iRods} and the MySQL database.
Event reconstruction, such as vertex reconstruction and energy
calibration, was performed and combined two data files from the neutrino DAQ and OV DAQ. Original and reconstructed data files were stored in the High-Performance Storage Systems (HPSS) in Lyon.

\subsection{Trigger}
\label{trigger}
The trigger initiated the readout with a trigger decision based on the light detected by the PMTs. It was optimized for reliability and high efficiency. For each of the two detectors, there were two largely independent trigger chains monitoring their inner detector volumes. Each one used only half of the 390 PMTs with independent hardware. The PMTs associated to the two chains A and B were interleaved with each other reducing the spatial variation over the Target volume. The decisions of the two chains were eventually combined with each other and additional triggers into a global ``OR'' of all positive decisions. This redundant system had two important advantages. On the one hand it increased  reliability: In the case of a failure of one of the chains the events were still recorded. The failure would have been detected by the online monitoring and offline software even if only a small fraction of events were affected and appropriate measures could be taken. On the other hand it allowed measurement of trigger efficiencies by comparison of the trigger decision of one chain against that of the other.

The trigger decision was based on the analogue output of the PMTs. The complexity of the system was reduced by grouping typically 13 PMTs in the FEE. The trigger decision was derived from the analogue sum of these groups. Each of the two independent chains handled 16 input groups. Each input from a group was discriminated at two different levels. In addition the signal of the 16 groups was summed into a total  representing the energy deposition in the central detector. This total energy was discriminated at four different levels representing thresholds for muon-like events, neutron capture, and neutrino interactions. The fourth, very low threshold was used for trigger studies, but only a small fraction of these events could be recorded due to rate limitations. The trigger decision was based on a combination of the total energy and the number of fired groups. Requiring more than one fired group (typically 3) largely reduced the sensitivity to noise or sparks on individual PMTs. All thresholds as well as the logical combination of the discriminator outputs were fully programmable.

A third identical hardware chain was used to trigger on light in the IV system detected by the 78~PMTs. No redundancy was implemented for the IV system. Typically six veto-PMTs formed one input group. The trigger decision from the two redundant chains of the inner detector and the decision from the veto system were combined into a global trigger decision in the trigger master board. It could handle eight independent triggers from each of the three chains plus seven additional external trigger inputs. In addition it created a so-called random trigger internally. It was derived from the internal clock and fired at a fixed, programmable frequency. The trigger master board recorded all trigger decisions in a 32-bit trigger word. It was  possible to prescale every trigger, i.e.,~to accept only every $n$-th positive decision, with a programmable scale-factor $n$. A trigger mask was applied to the trigger word afterwards, excluding triggers currently not in use. The remaining bits of the trigger word were ``OR''-ed into the final trigger decision. Upon a positive trigger the master board sent a trigger bit to all other subsystems, initiating the readout. It also sent the trigger word and an event number which was the number of triggers since the start of a run. The information in the trigger word was used by some of the subsystems to adjust the amount of data recorded according to the event type. In parallel the trigger initiated its own readout, recording the input status, the input rates, and several internal registers. 

Each trigger chain as well as the master board operated on an internal clock. Those clocks were synchronized to an external clock module. It provided a stable clock and time stamps. There was the possibility to link the master clock to a GPS signal. This possibility was fully tested but never used. The clock was fanned out to all subsystems for their synchronization. Internally the trigger  operated on a $32\,\mathrm{ns}$ clock. The inputs were discriminated and the results processed at the rate of this clock. A new trigger decision was produced every $32\,\mathrm{ns}$. Each trigger initiated the readout for a readout window which was longer than $32\,\mathrm{ns}$. If the trigger decision was still fulfilled at the end of the readout window, a so-called follow-up trigger was issued to continue the readout for another readout window. No relevant information was lost. 

A complete trigger for each of the two detectors consisted of 3 trigger boards handling the two redundant chains for the inner volumes and the veto, a trigger master board, a clock module (either the one at the near or the far detector was in use at any time), and a number of fanout cards for clock, trigger decision, trigger word and event number. All the hardware was implemented in the VME standard. The trigger logic was implemented in field programmable gate arrays (FPGA: Xilinx, Virtex-II, model XC2V500). Each trigger included a microcomputer with an Ethernet connection to the data acquisition for the programming and control of the trigger and its readout. The trigger efficiency was about 50\% at 0.3\,MeV and reached $> 99.9$\% at 0.5\,MeV with a negligible uncertainty. This was well below the minimum energy of a neutrino signal, two gamma rays with total 1.022\,MeV from the positron annihilation after the IBD interaction. 
As observed in data taken during FD-I, the inner detector PMTs show significant ringing after there were large signals produced by crossing muons. These pulses produce a number of additional triggers at characteristic times, between 200\,ns and 1\,$\mu$s, after the parent muon event, which can be explained by the interplay between the PMTs, Front-End electronics, and trigger system. Additionally, muon events cause a long undershoot on the stretcher signal from the Front-End Electronics, affecting the time period from 15 to 50\,$\mu$s after the parent muon. This undershoot results in a blinding of the trigger system in this time period.

To reduce the artificial triggers and recover from the inefficiency of the trigger system after muon events, a threshold raise condition was implemented in the firmware of the trigger master card. For the near and far detector running phase, the trigger master was configured such that after an event exceeds the high group multiplicity threshold on the IV trigger board, i.e.~a detector crossing muon, the trigger master effectively raises the inner detector trigger threshold, by emitting triggers only from events that pass the neutron energy threshold ($\sim 4$\,MeV). This threshold raise condition continues for 50\,$\mu$s and then the trigger master returns to normal working condition. More details on the trigger systems can be found in~\cite{triggerpaper}\cite{PhDAnselm}.

\subsection{Outer Veto DAQ}
\label{sec:OVDAQ}
 The Outer Veto DAQ ran independently of the neutrino detector DAQ system. Events from the two DAQ streams were merged offline using time stamps. The trigger master board produces a synchronisation signal which allows the OV to synchronize its clock. This synchronization signal is sent every 68.72\,s (0.015\,Hz). The OV DAQ system consisted of two software components: the Read Out Processor (ROP) and the OV EventBuilder. Each daisy chain of MAPMT (Multi-Anode PMT) boards was connected to one USB card. Each USB card was read out and controlled by an individual software ROP processor. The ROPs ran on a single multi-core computer controlling each individual USB board and consequently each MAPMT daisy chain. Commands were sent from the EventBuilder to the ROPs to configure, initialize, begin, and end acquiring data. During acquisition, each ROP sent data to the OV EventBuilder, which combined the data from each ROP to form a complete event based on timestamp.

 The OV EventBuilder processed the data using rapid online algorithms, determined if an event met coincidence requirements and was above a certain threshold; if so, the complete event was written to disk. Calibration data were also saved and used offline to calibrate the MAPMT. Additionally, information about the run and event builder quality monitoring were written periodically and at the beginning and end of each run, and stored in a MySQL database table. Both Event Builder and ROPs stored and retrieved calibration data from the MySQL database. A series of perl scripts monitored the data quality and hardware time responses automatically, including HV values for each of the MAPMT, determining the operation of the OV DAQ system.

 The OV DAQ system was controlled by a TCP/IP interface to the Double Chooz central run control and to the online monitor system. The OV online monitor system was a ROOT-based framework; it monitored power consumption of MAPMT boards, MAPMT responses, performance of individual ROPs, data disk throughput, computer CPU occupancy and performance of the OV event builder.

\subsection{Control and Monitoring}

Online systems to control and monitor the detectors and DAQ system were placed on the local area network inside the Chooz nuclear power plant facility and were limited in access from outside the control room via the internet. Four types of common software tools were developed, for run control, online data monitoring, information notification, and process control, written in C++. The graphical interfaces of these tools were initially implemented in Java for compatibility with the various software environments of shifters, and connected to each server via an SSH gateway.

In order to reduce the number of tasks required of  shifters, the run control GUI kept a sequential list,  used to take data according to a series of reserved run configurations. The various DAQ states were changed and repeated automatically.  Process control was able to restart several DAQ processes simultaneously in the proper sequence.

Since 2015, no one was required to be onsite for DAQ control. Instead,  shifters  connected to the system remotely from around the world. These systems 
provided an automated DAQ environment in order to simplify shifter duties. The DAQ system ran continuously, even when no one was connected to the shifter tool, and  each new run in the sequence started automatically.
Moreover, a Web-based application using Web-socket~\cite{WebS} enabled simplified access to the control tool. This could be accessed by mobile devices, such as smartphones, and there was no need to keep its window open.

\section{Detector simulation}
\label{simulation}

Simulation studies provided an important input for the optimization of the detector design in the early phase of the DC experiment. Moreover, they are an important tool for the modeling of the IBD signal prediction and the data analysis. Antineutrino events are generated over the full detector volume and the simulation results are then compared with the experimental data. The IBD signal in DC consists of a prompt positron followed by a delayed neutron capture. Therefore, a proper tracking of the electromagnetic and neutron interactions at low energies ($E<10$\,MeV) is essential for reproducing the observed data. In DC this task is managed by the DCGLG4sim package, which implements DC geometry and materials and the optical model of scintillator and PMTs. The DCGLG4sim software can be described as an extension of Geant4 (Geant4.9.2.p02~\cite{Geant4_1}\cite{Geant4_2}) for liquid scintillator antineutrino detectors and, more specifically, for DC.  Another software package is used to simulate the readout, implementing all DC electronic systems from the PMTs to the waveform digitizers. The output data have the same format as experimental data,  allowing the implementation of possible detector imperfections, such as dead channels. The simulation also includes custom models for scintillation processes, the photocathode optical surface, and thermal neutrons.

\subsection{Geometry and materials}

\begin{figure}[h!]
    \centering
    \includegraphics[width=0.7\textwidth]{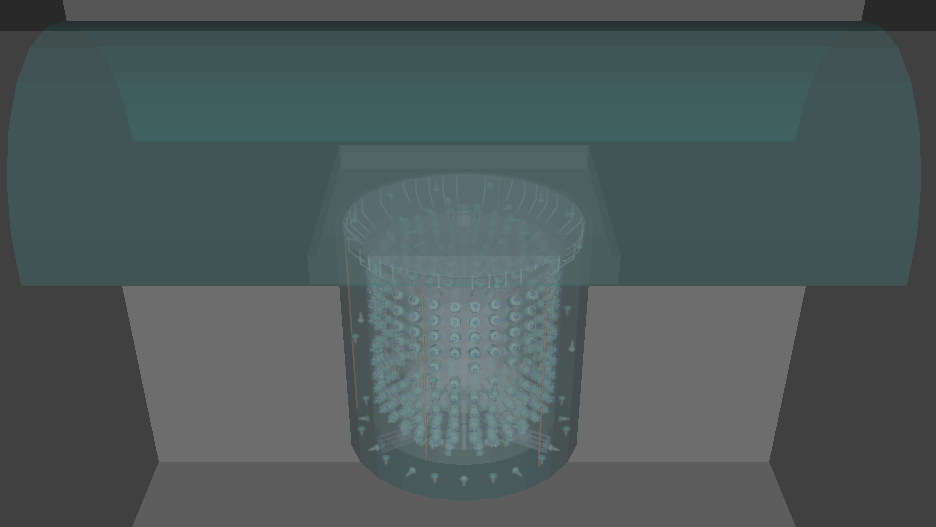}
    \includegraphics[height=6.5cm]{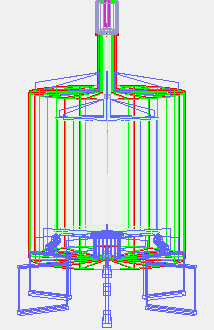}
    \caption{Visualization of the Double Chooz far detector geometry using Geant4 as it was used for the neutrino event generator. In the figure on the top each detector volume, the PMTs and also the cavern where the laboratory was built can be seen. The figure on the bottom shows the acrylic vessels including substructures.}
    \label{fig:MC_det}
\end{figure}

A detailed description of the detector geometry and material properties is part of the simulation. It includes acrylics vessels, Buffer and IV tanks, external shielding, and its immediate surroundings. The PMTs, including supports and mu-metal shields, are simulated as Geant4 physical volumes. During the detectors' construction, a photographic survey that gives sub-millimeter accuracy was performed for the orientation and position of all PMTs, tank walls and support placement. These values were implemented in the simulation. Figure~\ref{fig:MC_det} shows the detector representation in the simulation. Measured or calculated values of the molecular composition and densities of the liquids and acrylic vessels are used to calculate the interaction cross-sections and energy losses in non-instrumented materials. 

\subsection{Optical Model}
The optical model is an important ingredient to achieve a precise energy calibration in the experiment. In particular for the spectral analyses in DC a good understanding and control of the energy scale with small systematic uncertainties is essential. 

After a charged particle (mainly electrons or positrons) deposits its energy in the simulation volume, this energy is converted into photons via the scintillation and Cherenkov processes. The Target and GC scintillators' properties were measured before  detector filling and are implemented in the simulation. These measurements include: wavelength-dependent attenuation lengths, light yield, emission spectra, quantum yields and time distributions (see section~\ref{sec:liquids}). Ionization quenching based on Birks' law is also taken into account. Quenching parameters for electrons and alpha particles were separately determined by laboratory measurements~\cite{CADiss}\cite{Aberle:2011zz}. The fine-tuning of some of the parameters such as the light yield was done with calibration data, using in particular radioactive sources deployed within the detector. The final input parameters for the DC simulation as regards  light production are 8152 (8320) photons/MeV for the Target (GC). Such a light level is consistent (within 5\%) with the estimates from laboratory studies using small scale setups before and after the DC data taking period.

The photons produced by the ionizing particles are then tracked and propagated by Geant4. Absorption and re-emission at longer wavelengths are included in the optical model. Reflectivity of surfaces and refraction indices are also taken into account. Polarization dependent reflection and refraction are simulated at the boundaries between dielectric materials. At metal surfaces such as the Buffer wall or PMT mu-metal shields, optical photons can be absorbed or reflected according to specular and diffuse reflectivity parameters. Photons incident on the PMT optical surface (defined by a mathematical model for a thin, semitransparent surface) are handled by a dedicated PMT model that simulates the position-dependent collection efficiency based on measurements performed before installation~\cite{DCiPMT_NIMA_2012_Mat}. If the photon is absorbed by the photocathode, a PE is generated based on the measured probability (quantum efficiency of the PMT). For each simulated event, the hit time of each PE and the PMT channel numbers are aggregated and passed to the readout simulation.

\subsection{Neutron Interactions}
The Geant4 version used in the DC simulation does not include the effects of molecular bonds on neutron elastic scattering. Molecular binding energies are typically of order 1~eV. Thus, we expect these effects to become important for neutrons with similar energies, which is well above the energy at which neutrons become thermal (25~meV). To overcome this issue, DC developed a transport package that takes into account hydrogen molecular bonds for neutron elastic scattering, based on~\cite{Granada_neutrons}. In this package, molecular C-H bonds were modeled which are similar to the ones found in the DC liquid components. The dominant fraction of the DC liquids consists of alkane-like chemical structures with mainly CH$_2$ groups. Finally, a radiative capture model with improved final state gamma modeling is also taken into account. The neutron capture time is better reproduced with these modifications than with the default Geant4, specifically for short times, where thermal processes dominate.

\subsection{Muons}
The atmospheric muon flux at the detector positions and its corresponding angular distributions are estimated using the MUSIC simulation package~\cite{muonMC}. This code simulates the propagation of muons through matter taking into account  energy loss due to ionization, pair production, Bremsstrahlung and scattering processes. As main inputs a description of the overburden profile including its composition as well as the initial muon energy and direction at ground level are implemented in the code.

The simulated muon flux at the detectors is in good agreement with the measured fluxes of $(3.64 \pm 0.04)\cdot10^{-4}\,\text{cm}^{-2}$\,s$^{-1}$ for the near detector and $(7.00 \pm 0.05) \cdot 10^{-5}$\,cm$^{-2}$\,s$^{-1}$ for the far one~\cite{DoubleChooz:2016sdt}. Moreover, with the simulated muon rates it was possible to estimate the rate of stopping muons in the DC detectors. From those numbers probabilities for muon captures on light isotopes could be estimated~\cite{DoubleChooz:2015jlf}.

\subsection{Readout System Simulation}
The Geant4-based simulation of DC returns as output the time each photon strikes the photocathode of each PMT, producing a PE and its charge. The Readout System Simulation (RoSS) converts this information into a format identical to that of the raw detector data. This conversion takes into account the response of the elements associated with  detector readout, namely the PMTs, FEE, FADC and the trigger system. 

The first step of the readout simulation relies on the measured probability density function (PDF) to empirically characterize the response to each single PE as measured by the full readout chain. A dedicated setup consisting of one readout channel was built to measure most of the necessary PDFs and to tune the design of the full readout chain. The channel-to-channel variations, such as gain, baseline, noise and single-PE widths are taken into consideration in the simulation. 

For each event, the measured data format consists of trigger information and a waveform for each PMT representing the digitized response recorded by the FADC. This is done by summing each PE waveform of each PMT to form the PMT's signal, which is converted to a digital waveform using 2\,ns time bins. In this way, the simulation models nonlinearity effects as observed in data. After calibration, data and simulation agree to better than 1\%.

\section{Calibration systems}
\subsection{Overview}
The calibration systems were designed to determine and validate the relative antineutrino detection efficiency between the near and far detectors to the sub-percent level. At the same time the absolute as well as the relative energy scales 
were calibrated with various low-activity radioactive sources. 

To minimize systematic uncertainties the basic idea in DC was to deploy the same radioactive gamma and neutron sources in both detectors. A wide range of source deployment positions along the central z-axis inside the Target and in the upper half of the GC volume were accessible through the calibration systems. The comparison of the detector response in different calibration campaigns allowed monitoring of the detector stability. 

Full calibration campaigns, including a $^{252}$Cf neutron source and several gamma sources ($^{68}$Ge, $^{137}$Cs, $^{60}$Co), were carried out about once per year. Between campaigns, the detectors, in particular the PMT gains, were calibrated using a dedicated multi-wavelength LED system.

\subsection{Glove Box} 
\label{glovebox}

The calibration of the Target volume required deployment of sources in a clean environment with a dry nitrogen atmosphere. Thus it was necessary to have a glove box (GB) interface with an associated clean room on top of the detector. Based on measurements of the thorium, uranium and potassium concentrations of the surrounding rock performed in the context of the Chooz experiment, a specific activity of 0.5\,Bq/g for dust suspended in the air was estimated. With the requirement that dust from the GB should not cause the singles rate in the Target to increase by more than 1\,Hz, a cleanliness of class ISO 6 sufficed for the GB interior. For the external clean room surrounding the GB ISO 9 conditions were specified. 

The volume of the GB needed to be large enough that sources could be easily manipulated, and the deployment system  assembled and disassembled safely. The  hermetically sealed chambers, made of stainless steel with an acrylic viewing window, were about 90\,x\,70\,x\,60\,cm$^3$ in size. The GBs in both detectors were connected to the general nitrogen gas system (see section~\ref{liquidandgas}) for purging. They were equipped with oxygen monitors for monitoring the progress of purging, flagging leaks and assuring a low oxygen level of less than 50\,ppm in the detector gas blanket during deployment. While the detector interface was open to the detector, the GB systems needed to be light tight. Feedthroughs for laser fibers and control and power cables were hermetic to prevent air incursion. To bring sources into the detector, a transfer airlock was required; sources were placed into this box through an external door, then the box was purged, after which the operator opened an internal door and brought the source inside the detector interface area. 

\subsection{Target z-axis deployment system}
\begin{figure}[h!]
    \centering
    \includegraphics[width=0.48\textwidth]{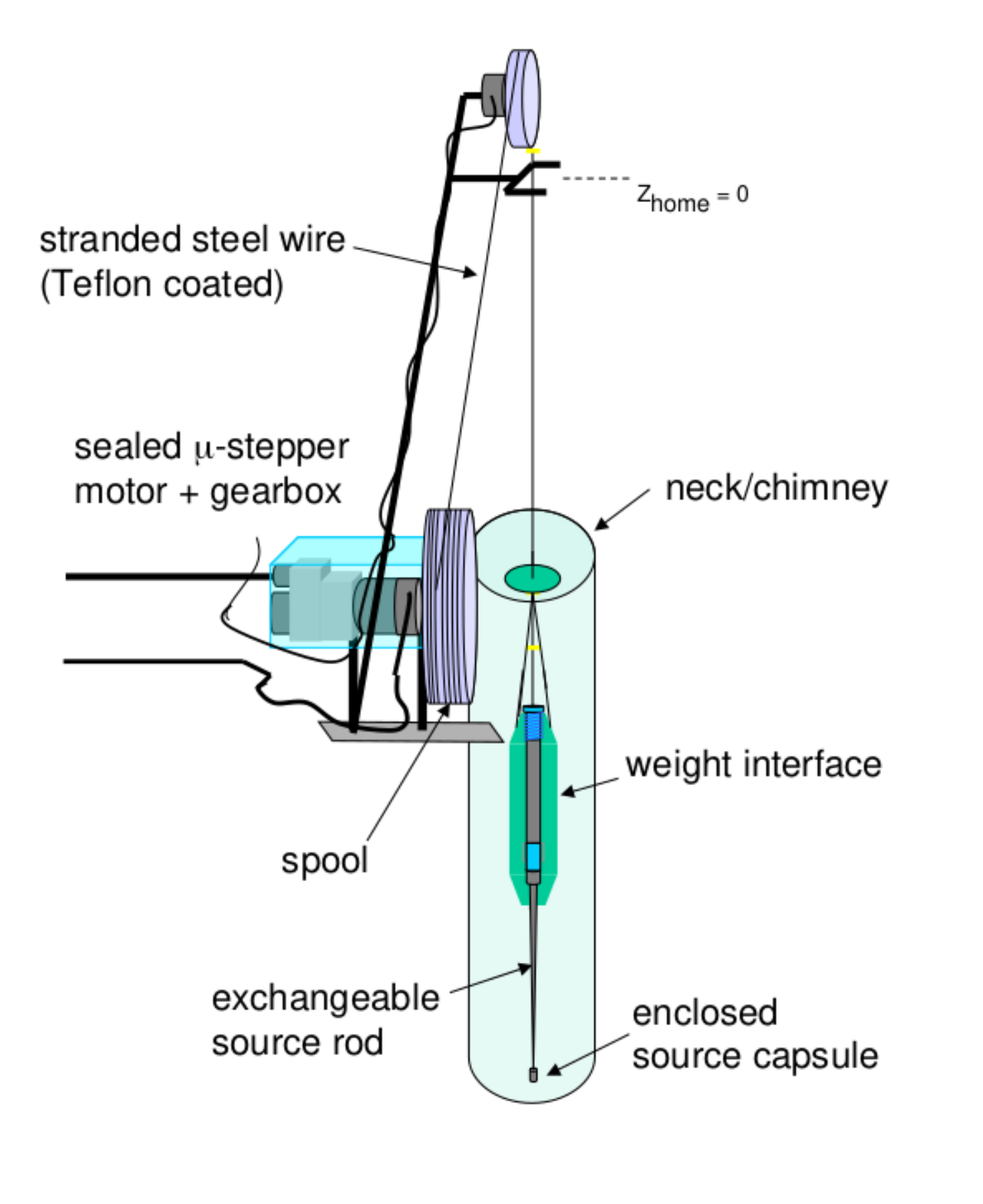}
    \includegraphics[width=0.4\textwidth]{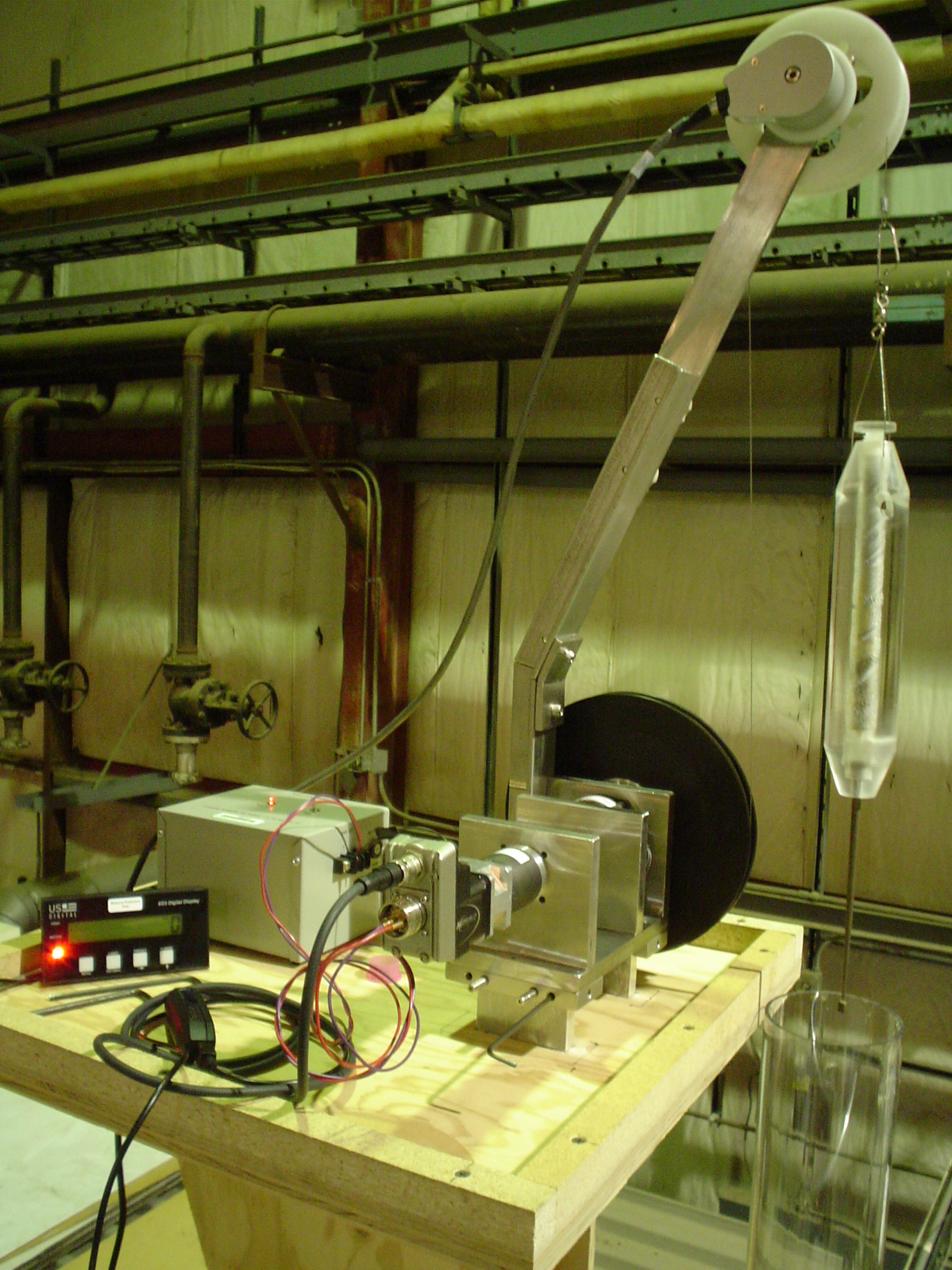}
    \caption{On the top the basic components of the z-axis system are illustrated. The picture on the bottom shows the far detector system on a test bench before installation in the glove box.}
    \label{fig:zaxis}
\end{figure}

With the z-axis calibration system (see fig.~\ref{fig:zaxis}) it was possible to deploy calibration sources along the central vertical axis of the Target through the detector chimney, extending down from the GB. The calibration sources were interchangeable and the deployment systems utilized by the near and far detectors were identical. The automated z-axis system could accommodate standardized rods, which contained a gamma or untagged neutron source with sizes of a few mm.

The system was specified to position sources with an absolute uncertainty  less than 1.5\,cm. The materials and geometry of the z-axis system were chosen to minimize uncertainties in the corrections for shadowing and absorption, while still guaranteeing  safe deployment. All components of the z-axis deployment system coming into direct contact with the Gd-LS were checked for material compatibility. Even components exposed just to the Gd-LS vapors were tested.

To protect the detector from oxygen and radio-contamination the z-axis system was operated in a nitrogen environment inside the GB. Deployment of the calibration system was required not to affect detector performance. Therefore, the system was largely automated and could be operated with the PMTs on. Electromagnetic noise was minimized during operation to reduce impact on data while the microstepper motor was running. The motor drove a reel equipped with a line (stainless-steel wire-rope with a diameter of 1/32$^{\prime\prime}$) at whose end the radioactive sources were attached. The wire rope was fixed to a transparent acrylic weight interface. The sources were mounted manually to this weight interface inside the GB with the deployment neck towards the detector closed. The stainless steel source rod was hooked into the body of the acrylic weight and held in place with a bolt from the top. The setup was optimized for a deployment speed of the order of 1\,cm/sec. A motor friction brake was added to guarantee that the load was still held in the case of a power failure. The system  was controlled via LabVIEW. Infrared cameras were mounted inside the GB for observation and control of the deployment.

\subsection{Guide tube calibration}
\begin{figure}[h!]
    \centering
    \includegraphics[width=0.51\textwidth]{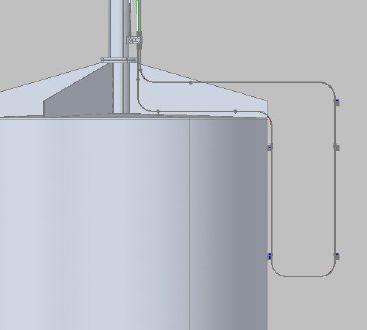}
    \includegraphics[width=0.4\textwidth]{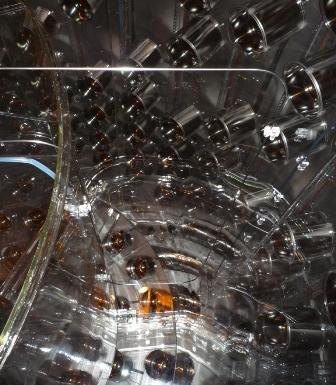}
    \caption{The guide tube system used to calibrate the detector in the top half of the GC.}
    \label{fig:GT}
\end{figure}

The guide tube (GT) was a calibration system allowing insertion of radioactive sources into the GC in the region above the top lid of the Target and along part of the cylindrical surfaces of the acrylic vessels as shown in fig.~\ref{fig:GT}. The identically designed GTs for the two detectors were systems of nested tubes. The wire and source capsule were entirely contained within the innermost Teflon tube running the full length of the system. Inside the GC this Teflon tube was contained within a stainless steel (SS304) tube, which was permanently attached to the Target lid ribs, the exterior Target wall and the internal GC wall at several points by glued acrylic fixtures. The sources were pushed by a wire through the GT. The sizes of the tubes were minimized to reduce shadowing/absorption effects, imposing strict dimensional constraints on the sources. The outer diameter of the 12\,m long SS304 tube was 5.2\,mm and the inner diameter of the Teflon sleeve was 3.0\,mm.

The GT formed a loop with each end terminating inside a sensor box at the top of the GC. A first inductive sensor defined the zero position of the wire. A second sensor close to the end of the GT was installed as a reference point to check the reproducibility of the source deployments. In an additional tube the wires were directed from the sensor box through the GC flange to the calibration clean tent (the same one as used for the GB). Outside the GC flange, the flexible Teflon tubing ran within black vinyl tubes to assure light tightness.

In the control box in the clean tent, the wire passed between two pulleys, one driven by a motor. A wheel encoder provided precise information on the wire position within the guide tube. Moreover, this component helped to keep the wire well aligned. A National Instruments USB-6212 DAQ module was optimized for fast sampling rates. It connected to a calibration laptop via the USB port and the system control was again via LabVIEW. 

\subsection{Light Injection system}
A multi-wavelength LED fiber system, called the inner detector light injection (IDLI) system, was used to inject controlled pulses of light into the inner detector from a set of fixed points.
The IDLI system consisted of a rack-mounted control box, six boxes with LEDs (each ``pulser box'' providing eight channels), optical fibers and diffuser disks. The design of the IDLI system built on and extended the broad concepts developed for calibrating the MINOS detectors~\cite{Adamson:2002ze,Adamson:2004mh}.
Within the detector, optical fibers were routed inside Teflon tubes and  terminated at custom holders attached to PMT mounts.
There were 46 injection points and a pair of fibers was routed to each injection point with one serving as a backup.
Of the 46 injection points, 32 (20 on the side wall, 6 on the top and 6 on the bottom) were equipped with diffuser disks to increase the number of illuminated PMTs.
The light exiting the diffuser disks was distributed with an opening angle of 22$^\circ$.
The remaining 14 injection points were not equipped with diffuser disks but directly illuminated the detector providing a significantly narrower beam with an opening angle of only 7$^\circ$.
These narrow beams were directed in parallel to the top and bottom of the cylindrical tank at 7 levels on the side wall allowing illumination of the Buffer, GC and Target volumes. PMMA fibers were used for the 32 diffused channels while quartz fibers were used for the 14 narrow beam channels. 

\begin{figure}[htb]
    \centering              
    \includegraphics[width=0.8\textwidth]{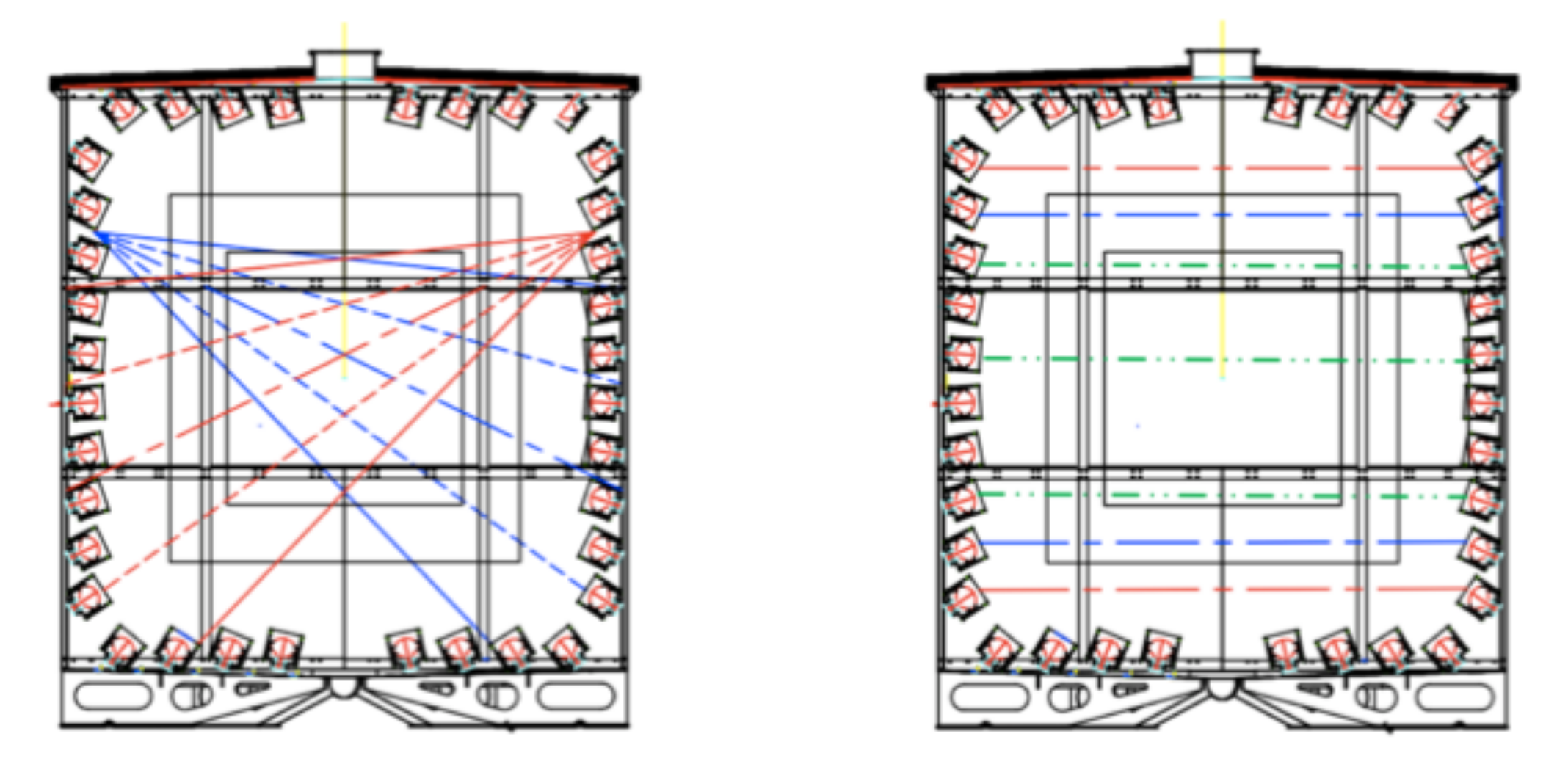}
    \caption{Schematic view of light paths inside the detector for 2 of the 32 wide beams with diffuser disks (left) and narrow beams without diffuser disks (right).}
    \label{fig:IDLI_injection}
\end{figure}

The fibers exited the detector via fluid-tight flanges alongside PMT cables and extended about 1~m before being terminated by an SMA connector. 
A total of 92 jacketed PMMA fibers about 30~m long and with 1~mm core diameter and SMA connectors on both ends ran outside the detector from the detector top to the electronics hut where the control and LED pulser boxes were located. Figure~\ref{fig:IDLI_injection} shows a schematic view of the light paths inside the detector.

Inside the LED pulser boxes the fiber ends were coupled to blue and UV LEDs. Each LED pulser box had 8~channels, each with 3~LEDs of 385, 425 and 475~nm wavelength and a PIN photodiode to monitor the light intensity.

Flash rate, light intensity, pulse width and choice of wavelength were controlled remotely by the control box via an RS232 interface to a computer in the DC online system.
The IDLI system provided a rectangular timing signal, which was divided into two. One part was fed into the trigger system to create an external trigger, and the second part was recorded by FADC.

Calibration data with the IDLI system were taken regularly by the automated calibration run sequence to measure the PMT and readout electronics gains, time offsets and stability in time. 
Gain calibration corrected for nonlinearity arising from readout electronics and charge reconstruction (see section~\ref{fadc}) by using data taken with various light levels from under 1\,PE up to more than 10\,PE. The time offset of each PMT was measured from the time offset of the signal relative to the external trigger provided by the IDLI system, accounting for the distance between the light injection point and the illuminated PMT. The stability of the time offset was  shown to be within the accuracy of the IDLI system measurement, which was estimated to be 0.5~ns by a study using the MC simulation.

For calibration of the IV PMTs, a pulsed light source (LEDs with wavelengths of 365\,nm and 470\,nm) was used to measure the gains and time offsets of the PMTs on a regular basis. The light entered the IV through an LED-fiber system (IVLI). Fiber ends were attached to the PMT cases a few cm from the PET windows.

\section{Radiopurity}

\label{radio_sec}
Accidental background was partly induced by radioactivity in the materials within and around the detector. Therefore, it was an important task to keep the amount of radioimpurities in the detector as low as possible. This required thoroughly screening all detector parts and materials, and, eventually, selecting only those materials fulfilling the requirements of radiopurity for the final detector configuration. Only in a few cases, e.g.,~for the detector liquids, was it possible to reduce the intrinsic radioactivity by some purification process.
The maximally allowed mass concentrations for radioimpurities strongly depended on the position of the investigated detector part and its mass; so no general upper limit could be established. The design goal for the DC far detector was no more than $\sim$\,0.8 accidental coincidences per day in the Target, which translates into a singles rate below $\sim$\,20\,Hz in the energy window for the prompt event, [0.7,\,12]\,MeV. With this constraint and the help of dedicated Monte Carlo simulations, limits on the radioimpurities for every detector component were obtained. 

The screening was carried out by direct gamma spectroscopy with a variety of germanium detectors in underground laboratories. Among them were the large HPGe detector for non-destructive radioassay at Saclay~\cite{cea} and the GeMPI detector at Gran Sasso~\cite{gsasso} with a sensitivity of about 10\,$\upmu$Bq/kg for U and Th. In addition, neutron activation analyses were performed for key components of the inner detector: the acrylics for Target and GC vessels as well as the wavelength shifter PPO~\cite{Aberle:2011ar}. The irradiations were done at the FRM II research reactor in Garching, Germany with a thermal neutron flux of $1.63\cdot 10^{13}$ cm$^{-2}\,$s$^{-1}$, with subsequent gamma spectroscopy in the TUM underground laboratory in Garching~\cite{Hofmann}.

The germanium detector is well suited for the measurement of highly energetic gammas, so samples from the outer parts of the DC detectors, such as the IV (including the vessel steel and PMT glass) and  shielding steel, were investigated with this technique. These parts of the detector could induce singles events in the Target only by emitting highly penetrating gamma rays. All the samples were investigated for their content of $^{238}$U, $^{232}$Th, and $^{40}$K; in such cases as the shielding steel samples $^{60}$Co was also of interest. The results of these measurements are shown in table~\ref{Tab:radio}. The values for steel samples of IV and buffer vessels and shielding were below the established limit of 1\,ppb for the three isotopes. The PMT glass and cavern rock were the main sources of gamma ray background. The PMT glass was made from low activity sand using a platinum coated furnace to reduce contamination. Radioactivity of the glass samples was measured during development of the low activity glass and production of the PMTs~\cite{merlin}. The average measurements were 13\,ppb, 61\,ppb and 3.3\,ppb for $^{238}$U, $^{232}$Th and $^{40}$K, respectively, assuming radio-equilibrium, which are much smaller than those of regular PMT glass. 

The U and Th concentrations in the scintillators can be estimated via a bismuth--polonium (BiPo) coincidence analysis assuming equilibrium in the decay chains. However, radon (Rn) diffusion or emanation from other detector materials also contributes in the BiPo analysis and the equilibrium might be broken. Indications for such effects are supported by the observation of locally higher BiPo rates around the chimney area and the hydrostatic liquid level sensors. Therefore the estimated mass concentrations should be interpreted as upper limits of the intrinsic liquid radiopurity. The results obtained in the Target (GC) were 0.4 (1.2)$\cdot 10^{-14}$~g/g U and 27.3 (1.8)$\cdot 10^{-14}$~g/g Th~\cite{MFDiss}, well within the DC specifications. The higher Th concentration in the Target could be explained by a small Th contamination identified by Ge-spectroscopy in some of the Gd-complex samples.  

A contribution to the correlated background rate in the Double Chooz detectors could be from ($\alpha$,n) reactions in the scintillators. Here, the prompt event is mimicked by the $\alpha$-particle and the delayed event by the capture of the neutron produced in this reaction. The dominant channel producing $\alpha$-particles in the DC detectors is expected to be from $^{210}$Po decays similar as in Borexino~\cite{Borexino:2008dzn} or KamLAND~\cite{KamLAND:2008dgz}. For this isotope the secular equilibrium in the uranium decay chain is typically broken due to the long half-life of $^{210}$Pb. The $\alpha$ energy of 5.4\,MeV in this decay is quenched by a factor of more than 10 in the LS. Therefore, the majority of these events is outside our prompt energy window. This is also the case for the other $\alpha$-decays in the detectors. From a conservative estimate of the $^{210}$Po-concentrations in DC and applying a conversion from $^{210}$Po-decays to ($\alpha$,n) reactions from KamLAND~\cite{KamLAND:2008dgz}, we obtain a negligible background contribution of $(3.7\pm0.5)\cdot 10^{-3}$ events per day~\cite{MFDiss} inside the Target liquid.

Dedicated purification steps for the liquid scintillators were carried out to meet the radiopurity requirements~\cite{Aberle:2011ar}. The organic solvents that were chosen were already tested for radiopurity in previous low energy neutrino experiments. Impurities in the Gd powder were mostly eliminated by sublimation. The possibility of such a purification method is one of the advantages of using metal-$\beta$-diketone (BDK) complexes to get the metallic Gd dissolved in the organic liquids. Traces of unstable europium (Eu) isotopes were also found in some of the Gd-complex samples. The main source for radioimpurities in the inner scintillator liquids was expected to be from potassium in the PPO. To minimize its contribution {\em Neutrino Grade} PPO from Perkin-Elmer (Netherlands) was used. For this quality a special synthesis procedure for removal of potassium is applied. 

\begin{table*}
\caption{\label{Tab:radio}
Measurements of the concentration of radioimpurities in the detector elements contributing most to the singles event background.}

\vspace{0.3 cm}
\centering
\begin{tabular}{|cccc|}
\hline
Element & Mass [g] & Isotope & Concentration [ppb] \\
\hline
\multirow{3}{2cm} {PMT Glass} & \multirow{3}{2cm} {390 $\times$ 1200} & $^{40}$K & 3.3   \\
                                &                                     & $^{238}$U & 13  \\
                                &                                     & $^{232}$Th & 61  \\
\hline
\multirow{1}{2cm} {Rock}      & \multirow{1}{2cm} {2.4$\cdot$10$^{8}$} &  $^{232}$Th & 5000  \\
\hline
\multirow{3}{2cm} {Buffer Tank} & \multirow{3}{2cm} {4.5$\cdot$10$^{6}$} & $^{40}$K & $<$0.1 \\
                                &                                     & $^{238}$U & $<$0.5  \\
                                &                                     & $^{232}$Th & 1.1  \\
\hline
\multirow{2}{2cm} {IV Tank (barrel)} & \multirow{2}{2cm} {1$\cdot$10$^{7}$} & $^{40}$K & $<$0.1  \\
                                &                                     & $^{232}$Th & $<$1  \\  
\hline
\multirow{2}{2cm} {IV Tank (lid and bottom)} & \multirow{2}{2cm} {5$\cdot$10$^{6}$} & $^{40}$K & $<$1 \\
                                &                                     & $^{232}$Th & 4.7  \\
\hline
\multirow{3}{2cm} {Target \& GC acrylics} & \multirow{3}{2cm} {1.88$\cdot$10$^{6}$} & $^{40}$K & $<$0.1   \\
                                &                                     & $^{238}$U & $<$0.1  \\
                                &                                     & $^{232}$Th & $<$0.1  \\
\hline
\end{tabular}
\centering 
\end{table*}

Monte Carlo simulations were carried out to investigate the rate and energy spectrum of the singles events induced by the identified radioimpurities that determine the trigger rate of the DC detector. The full geometry of DC detector was included, using the results of the radiopurity measurements as input for the simulation.

The full spectrum of decay particles was considered for the simulation of $^{40}$K in the inner parts of the detector (Target and GC liquid, as well as acrylics), including the branching ratios of $\beta$ and EC decay; for  simulation of the effect of radioimpurities in the outer parts of the detector (such as the shielding steel or components of the IV) on the Target only highly energetic gammas were simulated. 
The rates obtained for the decay chains of uranium and thorium are consequently dominated by the highly energetic gamma lines of $^{214}$Bi (1.12\,MeV and 1.76\,MeV) and $^{208}$Tl (2.6\,MeV). 

The singles event rate for energy depositions in the Target and GC in the range [0.7,\,12]\,MeV obtained from the simulation of the main elements was about 5\,Hz, well below the design goal. The main sources were PMTs and cavern rock with 2\,Hz each, while small concentrations of $^{40}$K in Target and GC acrylics and liquid scintillators could have an impact in the singles rate ($\sim 1$\,Hz). The uncertainties on the measured $^{40}$K concentrations for these elements, consisting in upper limits in most of the cases, did not allow a precise prediction of their contribution to the singles event rate.
The rate measured in the same energy window in the FD at the beginning of data taking was 7.7\,Hz, in good agreement with the prediction considering the existence of the additional light noise background.
The comparison of the simulated spectra to the real detector data allowed an additional energy calibration and, in some cases, attribution of the sources of radioactivity-induced background.

\section{Detector stability}
\label{perf_stab}
One of the main concerns in the design phase of the DC experiment was the long-term stability of the Target Gd-scintillator. Stability aspects of the Gd-scintillator are twofold: the first is related to chemical stability and homogeneity of the Gd-complex within the Target volume, the second to the optical stability, in particular of the scintillator absorption length. The former case can be monitored by looking at the neutron detection efficiency, the latter by studying the detector resolution and the absolute number of photoelectrons measured by the PMTs.   
The production of the Gd-scintillator happened in the first half of 2010, about one year before the start of DC data taking in the far detector. From 2011 to the end of 2017 there are almost 7 years of detector data available. Calibration campaigns with radioactive sources deployed inside the detectors were performed about once per year.      

\subsection{Neutron detection efficiency}
To evaluate the neutron detection efficiency in the Gd analysis three factors were studied:

\begin{equation}\label{Formula:DelayedEfficiency}
\varepsilon_{\text{delay}} = \varepsilon_{n-\text{capture}} \times \varepsilon_{\text{cut}} \times \varepsilon_{\text{spill}}
\end{equation}

\noindent where the Gd-fraction $\varepsilon_{n-\text{capture}}$ studies the relative abundance of neutron captures on Gd-nuclei in the Target scintillator, $\varepsilon_{\text{cut}}$ evaluates the efficiency of the IBD selection cuts (energy, time and vertex) and $\varepsilon_{\text{spill}}$ studies border effects between the Target and the GC in the IBD selection due to neutron mobility. A $^{252}$Cf neutron source in the center of the detector was used to determine $\varepsilon_{n-\text{capture}}$. To be selected as a Cf neutron event the following conditions had to be fulfilled: 
\begin{itemize}
\item Prompt event (fission gammas):
\begin{itemize}
\item 4 $< E_{\text{vis}}<$ 30\,MeV
\item at least  1.5\,ms after the last trigger event
\end{itemize}
\item Delayed events (neutron captures):\\
More than 1 event could occur (multiplicity $>$ 1) with the following properties:
\begin{itemize}
\item 0.5 $<E_{\text{vis}}<$ 25\,MeV
\item 0.5 $<\Delta t <1000\,\mu$s
\end{itemize}
\end{itemize}

The Gd-capture events had energies mainly distributed between 3.5 and 10\,MeV (Gd-capture peak and its tail). The H-captures had visible energies distributed in the range [1.3, 3.5]\,MeV, where the lower energy cut for the hydrogen captures was chosen in order to exclude the region with correlated background. The fraction of neutron captures on Gd nuclei is defined as the ratio of events in two subsamples:
\begin{equation}\label{Formula:GdFractionDefinition}
f_{Gd} = \dfrac{{N} (3.5 < E_{\text{delayed}} < 10\, \text{MeV})}{{N} (1.3 < {E}_{\text{delayed}} < 10\, \text{MeV})}\,.
\end{equation}

With this definition the energy spectrum is divided into events due to H- or Gd-captures. The Gd-fraction is rather independent of energy scale or optical effects. To first order, it is just determined by the ratio of Gd to H nuclei in the liquid. Therefore, it was an adequate parameter to monitor the chemical stability and solubility of the Gd-complex in the scintillator. It should be noted that the small amount of carbon captures happening around 5\,MeV was included in the Gd-capture region. 

\begin{figure}
\begin{center}
	\includegraphics[height=60mm]{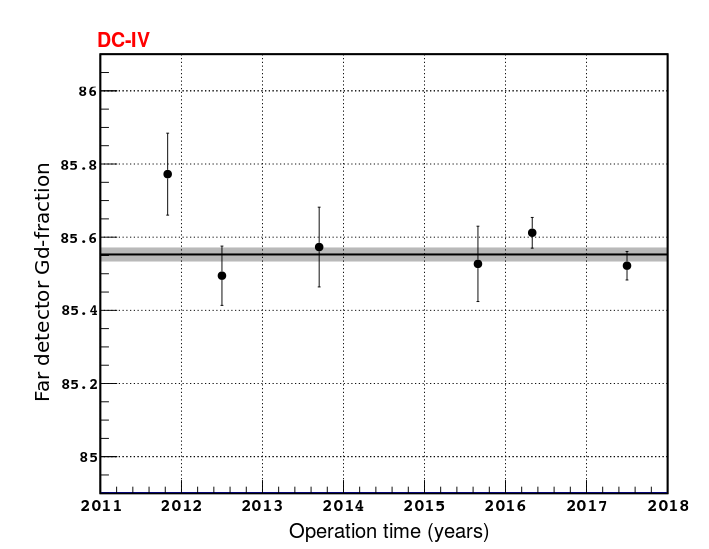}
\end{center}
\caption[]{Gd-fraction stability in the far detector over 7 years of data taking.
\label{GdFD}}
\end{figure}

In fig.~\ref{GdFD} the stability of the Gd-fraction is shown, measured at the center of the Target volume. For the last two calibration points the Cf source was kept longer inside the scintillator and the statistical uncertainty could therefore be significantly reduced. Over the full data taking period there is no hint of any loss of solubility of the Gd-complex or formation of layers with different Gd concentrations. Additional $^{252}$Cf calibration points along the z-axis also showed  consistent values for the neutron detection efficiency over time. Similar results were obtained for the near detector with a shorter data taking period. 

\subsection{Light response stability}
\begin{figure*}
\begin{center}
	\includegraphics[width=1\textwidth]{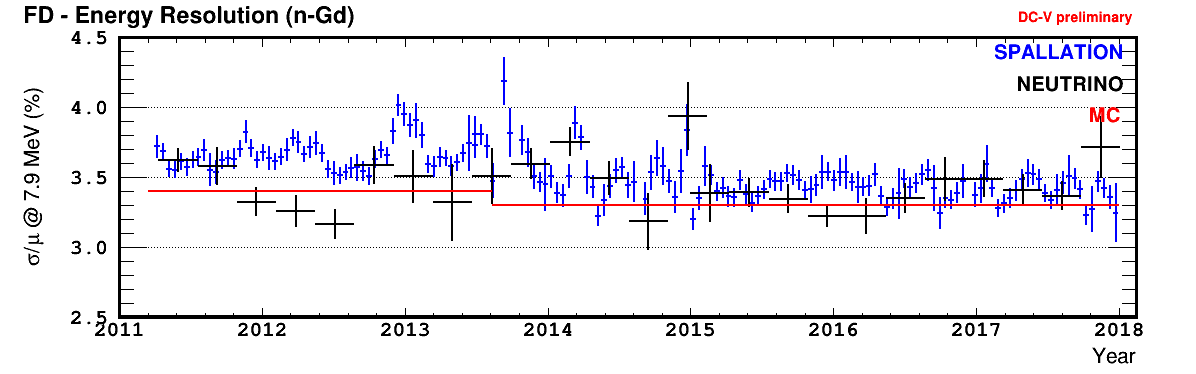}
\end{center}
\caption[]{Stability of energy resolution at about 8\,MeV (Gd-peak) in the Double Chooz far detector over the 7 years of data taking. The blue data points correspond to the capture of spallation neutrons, the black points to neutron captures from IBD candidate events and in red the energy resolution as predicted by the simulation is shown.
\label{fig:resovst}}
\end{figure*}

The stability of the energy scale in the oscillation analysis of DC was demonstrated to be controlled down to a level of 0.4\% (0.2\%) before (after) the electronic gain upgrade introduced in section 6.1. These errors are estimated using an independent data set (Gd and H n-captures from IBD signal) from the one used to determine uniformity and stability corrections (Gd and H n-capture by muon spallation). The calorimeter performance was a consequence of the excellent stability of optical properties of both the Target and GC scintillator liquids, as well as an appropriate application of drift corrections related to variations of the number of active PMT channels, PMT gains and reconstructed charge. In particular, stability of the PE and strongly linked PMT charge response is noteworthy, reflected in a steady energy resolution over the 7 years of data taking as illustrated in fig.~\ref{fig:resovst}. At the beginning of FD data taking, some noisy PMTs had to be turned off. After an electronics upgrade in 2013 and due to the an improved strategy of light noise reduction~\cite{DCiPMT_JINST_2016_Abe}, PMTs could be switched on again. With these optimization steps the energy resolution slightly improved as seen in the step of the otherwise flat red MC expectation line in fig.~\ref{fig:resovst}. From these DC data there is no hint for any scintillator degradation, neither concerning the light yield nor the attenuation length. 

\begin{figure*}
\begin{center}
	\includegraphics[width=1\textwidth]{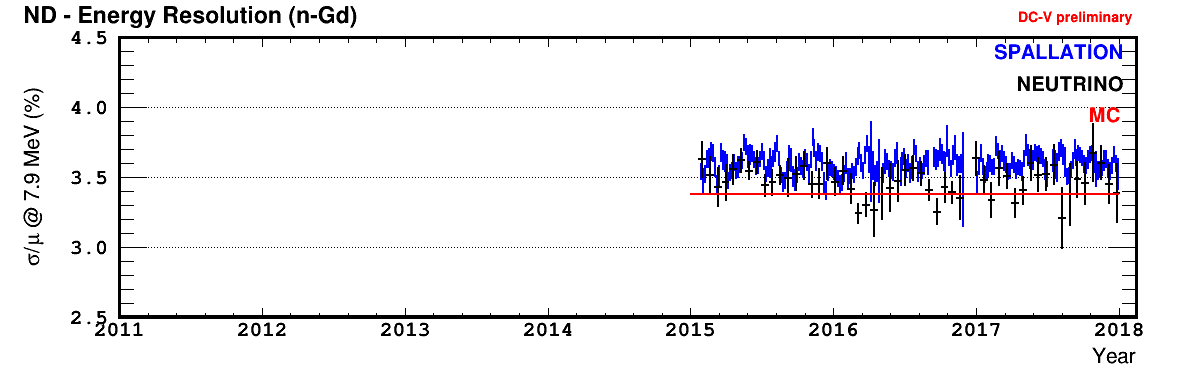}
\end{center}
\caption[]{Stability of energy resolution at about 8\,MeV (Gd-peak) in the Double Chooz near detector over the 3 years of data taking. The blue data points correspond to the capture of spallation neutrons, the black points to neutron captures from IBD candidate events and in red the energy resolution as predicted by the simulation is shown.
\label{fig:resovst_b}}
\end{figure*}

A surprising observation was even a small improvement of the detected PE level in the Target of the Near Detector with an asymptotic increase stabilizing to +2\% relative to the initial level over the 3~years. This increase in the light response could be interpreted as an improvement of the transparency of the Target scintillator due to the settling of dust particles or the floating up of micro-bubbles produced during the filling operation. Optical simulations with a scintillator attenuation length of about 5\,m ($>430$\,nm) show a remarkable agreement with data regarding the detected PE over the full data taking period. Moreover, the non-uniformity of response over the full sensitive volume as well as the energy resolution (taking into account both stochastic and non-stochastic contributions), as shown in fig.~\ref{fig:resovst} and \ref{fig:resovst_b}, are well modeled by the MC.

\section{Summary} 
The Double Chooz experiment demonstrated technology capable of observing reactor neutrinos and met its goal of measuring the neutrino mixing angle $\theta{_{13}}$.
Two underground detectors were built and operated, one near detector to monitor the emitted flux at about 400\,m from the reactor cores and one far detector at 1.05\,km to measure the oscillation effect. The detectors had to be as identical as possible to profit from cancellation of correlated systematic uncertainties in the comparison of the measured antineutrino spectra. Major design goals related to long term stability and radiopurity were met and partially exceeded. In particular the long-term stability of the Gd-loaded Target scintillator was important in Double Chooz, both to ensure a sufficiently long running time (several years), and to avoid systematics due to a possibly different evolution of the liquids in the two detectors. 

\begin{acknowledgements} 
We thank the company EDF (``Electricity of France’'), the European fund FEDER, the Région Grand Est (formerly known as the Région Champagne-Ardenne), the Département des Ardennes and the Communauté de Communes Ardenne Rives de Meuse. We acknowledge the support of the CEA, CNRS/IN2P3, the computer centre CC-IN2P3 and LabEx UnivEarthS in France; the Max Planck Gesellschaft, the Deutsche Forschungsgemeinschaft DFG, the Transregional Collaborative Research Centre TR27, the excellence cluster `‘Origin and Structure of the Universe’' and the Maier-Leibnitz-Laboratorium Garching in Germany; the Ministry of Education, Culture, Sports, Science and Technology of Japan (MEXT) and the Japan Society for the Promotion of Science (JSPS) in Japan; the Ministerio de Economía, Industria y Competitividad (SEIDI-MINECO) under grants FPA2016-77347-C2-1-P and MdM-2015-0509 in Spain; the Department of Energy and the National Science Foundation in the United States; the Russian Academy of Science, the Kurchatov Institute and the Russian Foundation for Basic Research (RFBR) in Russia and the Brazilian Ministry of Science, Technology and Innovation (MCTI), the Financiadora de Estudos e Projetos (FINEP), the Conselho Nacional de Desenvolvimento Científico
e Tecnológico (CNPq), the São Paulo Research Foundation (FAPESP) and the Brazilian
Network for High Energy Physics (RENAFAE) in Brazil.

Furthermore we acknowledge the valuable contributions of C.~Bauer, N.~Danilov, C.~Hagner, T.~Hayakawa, A.~Milsztajn, D.~Motta, Y.~Krylov 
and H.~Watanabe. 

\end{acknowledgements}



\end{document}